\documentclass[11pt,a4paper]{article}

\usepackage{ifpdf}
\usepackage{latexsym}
\usepackage{amssymb}
\usepackage{amsmath}
\usepackage[authoryear, round]{natbib}
\usepackage{graphicx}

\newtheorem{theorem}{Theorem}[section]

\newtheorem{remark}[theorem]{Remark}
\newtheorem{example}[theorem]{Example}
\newtheorem{proposition}[theorem]{Proposition}
\newtheorem{definition}[theorem]{Definition}
\newtheorem{corollary}[theorem]{Corollary}

\numberwithin{equation}{section}

\begin{document}
\title{Estimating discriminatory power and PD curves when the number of defaults is small}

\author{%
Dirk Tasche, Lloyds Banking Group\footnote{%
The opinions expressed in this paper are those of the author and do not necessarily reflect views of Lloyds Banking Group.}}

\date{}
\maketitle

\begin{abstract}
The intention with this paper is to provide all the estimation concepts and techniques
that are needed to implement a two-phases approach to the parametric estimation of probability of
default (PD) curves. In the first phase of this approach, a raw PD curve is estimated
based on parameters that reflect discriminatory power. In the second phase of the approach,
the raw PD curve is calibrated to fit a target unconditional PD. 
The concepts and techniques presented include a discussion of different definitions
of area under the curve (AUC) and accuracy ratio (AR), a simulation study on the performance
of confidence interval estimators for AUC, a discussion of 
the one-parametric approach to the estimation of PD curves by \citet{VanDerBurgt}
and alternative approaches, as well as a simulation study on the performance of
the presented PD curve estimators. The topics are treated in depth in order to
provide the full rationale behind them and to produce results that can be implemented
immediately. 
\end{abstract}


\section{Introduction}
\label{sec:intro} 

In the current economic environment with its particular consequence of rising credit default rates all 
over the world, at first glance it might not seem very appropriate to look after estimation issues 
experienced in portfolios with a small number of defaults. However, low default estimation issues
can occur quite naturally even in such a situation:
\begin{itemize}
	\item It is of interest to estimate \emph{instantaneous} discriminatory power of a score function
	or rating system. ``Instantaneous'' means that one looks only at the defaults and survivals that occurred
	in a relatively short time period as one year or less. In typical wholesale portfolios 
	that represent the scope of a rating system the number of borrowers does not exceed 1000. As a consequence,
	the number of defaults observed within a one year period might well be less than 20.
	\item Similarly, when estimating forward-looking point-in-time (PIT) conditional probabilities of
	default per score or rating grade \emph{(PD curve)}, it makes sense to construct the estimation 
	sample from observations in a relatively short time period like one or two years
	in order to capture the instantaneous properties of a potentially rather volatile object. The
	previous observation on the potentially low number of defaults then applies again.
\end{itemize}
The topics dealt with in this paper are closely related to these two issues. The intention with
the paper is to clarify conceptual issues of how to estimate discriminatory power 
and PD curves, to provide ready-to-use formulas for the related concepts, and to look at
low-default related performance issues by means of simulation studies.

The paper is organised as follows:
\begin{itemize}
	\item \textbf{Section \ref{sec:concepts}:}
	We introduce the concept of a two-phases approach to the calibration of score functions and
	rating systems and present a simple probabilistic model that is appropriate as a
	framework to discuss the two phases in a consistent manner. 
	Additionally, we introduce some technical notation for further use within the paper.
	The focus in this paper will be on
	on the \emph{estimation phase} whose technical details are studied in sections \ref{sec:framework},
	\ref{sec:aspects}, and \ref{sec:parametric}. For the purpose of reference, technical
	details on the \emph{calibration phase} are provided in appendix~\ref{sec:calibration}.
	\item \textbf{Section \ref{sec:framework}:} The estimation of discriminatory power of a score function
	or a rating system represents an important part of the estimation phase. In particular, when sample sizes
	are small, it is therefore crucial to have a clear and consistent view on 
	which definition of discriminatory power should 
	be used. This question is discussed in depth in section \ref{sec:framework}.
	\item \textbf{Section \ref{sec:aspects}}: We replicate simulation studies by \citet{Engelmannetal03a, Engelmannetal03b}
	with some refinements in order to investigate how accurate confidence interval calculations with different 
	methods for discriminatory power are. In contrast to the studies by \citet{Engelmannetal03a, Engelmannetal03b},
	the current study is arranged in such a way that the true value of discriminatory power is known. Additionally,
	even smaller sample sizes are considered and the results are double-checked against exact results from 
	application of the Mann-Whitney test.
	\item \textbf{Section \ref{sec:parametric}:} \Citet{VanDerBurgt} suggested 
	fitting cumulative accuracy profile (CAP) curves with a one-parameter family of exponential
	curves in order to derive conditional probabilities of default by taking the derivatives of the 
	estimated curves. \Citeauthor{VanDerBurgt} believes that this approach 
	is appealing in particular for low default portfolios where the defaulter sample size is small.
	In section \ref{sec:parametric}, we investigate the suitability of 
	\citeauthor{VanDerBurgt}'s approach by applying it to a situation
	where the conditional score distributions are known and, hence, the conditional probabilities of default
	can be calculated exactly. Performance of \citeauthor{VanDerBurgt}'s estimator is compared to the performance 
	a the logit estimator and two modifications of it. While it turns out that there is no uniformly 
	best estimator, the results also demonstrate the close relationship between discriminatory power as 
	measured by area under the curve or accuracy ratio and parametric PD curve estimators.
	A consequence of this, however, is high sensitivity of parametric estimates of PD curves to 
	poor specifications of discriminatory power.
	\item \textbf{Section \ref{sec:conclusions}:} Conclusions.
\end{itemize}

\section{Basic concepts and notation}
\label{sec:concepts}

Within this paper, we assume that there is a score function or a rating system that can be deployed 
to inform credit-related decisions. We do not discuss the question of how such score functions or rating
systems can be developed. See, e.g., \citet{Engelmann&Rauhmeier} for some approaches
to the development of rating systems. Instead, in this paper, 
we look at the questions of how the power of such score functions or
rating systems can be assessed and how probability of default (PD) estimates \emph{(PD curves)}
associated with score values or rating grades can be derived. 

In sub-section \ref{sec:approach} we present a general concept for the calibration of score functions and rating
systems which is based on separate estimation of discriminatory power and an unconditional
probability of default. In sub-section \ref{sec:model} a simple probabilistic model is introduced that will help
in the derivation of some ideas and formulas needed for implementation of the concept. 
Moreover, in sub-section \ref{sec:notation} we recall for further reference some properties of
distribution functions and some notation related to such functions.


\subsection{Estimation phase and calibration phase}
\label{sec:approach}

In this sub-section, we introduce the concept of a two-phases approach to the calibration of 
a score function or a rating system: The first phase is the \emph{estimation} phase, 
the second phase is the \emph{calibration and forecast} phase. 

\subsubsection{Estimation} 
\label{sec:estimation}

The aim here is to estimate conditional PDs per score (or grade) and
	the discriminatory power of the rating system (to be formally defined in sections \ref{sec:framework} and
	\ref{sec:discontinuous})
	from a historical sample of scores or rating grades associated with borrowers 
	whose solvency states one period after the scores were 
	observed are known. The composition of the sample is not assumed to be representative of current or future portfolio 
	composition. In particular, the proportion of defaulters and survivors in the sample may differ
	from proportions of defaulters and survivors expected for the future. The estimated conditional PDs therefore
	are considered \emph{raw PDs} and have to be \emph{calibrated} before being further used.
	The estimation sample could be the development sample of the rating system or a
	validation sample.
	
	In the following, we will write $x_1, \ldots, x_{n_D}$ when talking about a sample 
	of scores or rating grades of defaulted borrowers and $y_1, \ldots, y_{n_N}$ when talking about
	a sample of surviving borrowers. In both these cases, the solvency state of the borrowers one
	period after the observation of the scores is known. In contrast, we will write 
	$s_1, \ldots, s_n$ when talking about a sample of scores 
	of borrowers with unknown future solvency state.	
	
\subsubsection{Calibration and forecast} 
\label{sec:forecast}

The aim here is to calibrate the raw PDs from the estimation
	step in such a way that, on the current portfolio, 
	they are consistent with an unconditional PD that may be different to the 
	unconditional PD of the estimation sample. This calibration exercise is needed because for the borrowers in the
	current portfolio scores (or rating grades) can be determined but not their future solvency states.
	Hence direct estimation of conditional PDs with the current portfolio as sample is not possible.
	We will provide the details of the calibration under the assumption that the \emph{conditional} score
	distributions (formally defined in \eqref{eq:cond_dist} below) 
	that underlie the estimation sample and the conditional score distributions of
	the current portfolio are the same. This assumption is reasonable if the estimation sample 
	was constructed not too far back in time 
	or if the rating system was designed with an intention of creating a through-the-cycle (TTC)
	rating system. 
	
	As mentioned before, the unconditional PDs of estimation sample and current portfolio
	may be different. This will be the case in particular if a point-in-time (PIT) calibration of 
	the conditional PDs is intended, such that the PDs can be used for forecasting future default rates.
	But also if a TTC calibration of the PDs is intended (such that no direct forecast of default
	rates is possible), most of the time the TTC unconditional PD will be different to the realised
	unconditional PD of the estimation sample. Note that the \emph{unconditional} score distributions
	of estimation sample and current portfolio can be different but, on principle, are linked together
	by equation \eqref{eq:unconditional} from sub-section \ref{sec:model}.
	
	The question of how to forecast the unconditional PD is not treated in this paper. An  
	 example of how PIT estimation of the unconditional
	PD could be conducted is presented by \citet[][section III]{Engelmann&Porath}. 
	Technical details of how the calibration of the conditional PDs 
	can be done are provided in appendix \ref{sec:calibration}.


\subsection{Model and basic properties}
\label{sec:model}

Speaking in technical terms, in this paper we study the joint distribution and 
some estimation aspects of a pair $(S, Z)$ of real random variables. The variable $S$ is 
interpreted as the \emph{credit score} (continuous case) or
\emph{rating grade}\footnote{%
In practice, often a rating system with a small finite number of grades is derived from a
score function with values on a continuous scale. This is usually done by mapping
score intervals on rating grades. See \citet[][section 3]{Tasche2008a} for
a discussion of how such mappings can be defined. Discrete rating systems are preferred by
practitioners because manual adjustment of results (overrides) is feasible. Moreover,
results by discrete rating systems tend to be more stable over time.} (discrete case) 
observed for a solvent borrower at a certain point in time.
Hence $S$ typically takes on values on a
continuous scale in some open interval $I \subset \mathbb{R}$ or on a discrete
scale in a finite set $I = \{1, 2, \ldots, k\}$. 

\textbf{Convention:} Low values of $S$ indicate low creditworthiness (``bad''), 
high values of $S$ indicate high creditworthiness (``good'').

The variable $Z$ is the \emph{borrower's state of solvency} one
observation period (usually one year) after the score was observed. $Z$ takes on values in
$\{0,1\}$. The meaning of $Z=0$ is ``borrower has remained solvent'' (solvency or survival), 
$Z=1$ means ``borrower has become insolvent'' (default). 
We write $D$ for the event $\{Z=1\}$ and $N$ for the event $\{Z=0\}$.
Hence 
\begin{equation}
	D \cap N = \{Z=1\} \cap \{Z=0\} = \emptyset, \quad D \cup N = \text{whole space}.
\end{equation}
The marginal distribution of the state variable 
$Z$ is characterised by
the \emph{unconditional probability of default} $p$ which is defined as
\begin{equation}\label{eq:PDunconditional}
	p = \mathrm{P}[D] = \mathrm{P}[Z =1] \in [0,1].
\end{equation}
The joint distribution of $(S, Z)$ then can be specified by
the two conditional distributions of $S$ given the states of $Z$ or
the events $D$ and $N$ respectively. In particular, we define 
the conditional distribution functions
\begin{equation}\label{eq:cond_dist}
	\begin{split}
	 F_N(s) &= \mathrm{P}[S \le s\,|\,N] = \frac{\mathrm{P}[\{S \le s\}\,\cap N]}{1-p}, \quad s \in I,\\
	 F_D(s) &= \mathrm{P}[S \le s\,|\,D] = \frac{\mathrm{P}[\{S \le s\}\,\cap D]}{p},\quad s \in I.
	\end{split}
\end{equation}
For the sake of an easier notation we denote by
$S_N$ and $S_D$ random variables with distributions $\mathrm{P}[S \in \cdot\,|\,N]$ and
$\mathrm{P}[S \in \cdot\,|\,D]$ respectively.
In the literature, $F_N(s)$ sometimes is called \emph{false alarm rate} 
while $F_D(s)$ is called \emph{hit rate}.

By the law of total probability, the distribution function $F(s) = \mathrm{P}[S\le s]$ of the 
marginal (or unconditional) distribution of the score $S$ 
can be represented as
\begin{equation}\label{eq:unconditional}
	F(s) = p\,F_D(s) + (1-p)\,F_N(s), \quad \text{all\ } s.
\end{equation}
$F(s)$ is often called \emph{alarm rate}.

The joint distribution of the pair $(S, Z)$ of score and borrower's state one period later
can also be specified by starting with the unconditional distribution $\mathrm{P}[S \in \cdot\,]$
of $S$ and combining it with the \emph{conditional probability of default} $\mathrm{P}[D\,|\,S] =
1-\mathrm{P}[N\,|\,S]$. Recall that in general the conditional probability 
$\mathrm{P}[D\,|\,S] = p_D(S)$ can 
be characterised\footnote{We define the indicator function
$\mathbf{1}_M$ of a set $M$ by $\mathbf{1}_M(m) = \begin{cases}
	1, & m \in M,\\
	0, & m \notin M.
\end{cases}$} by the property \citep[see, e.g.][section 4.1]{Durrett}
\begin{equation}\label{eq:def_condPD}
	\mathrm{E}[p_D(S)\,\mathrm{1}_{\{S\in A\}}]\ = \ \mathrm{P}[D\cap \{S\in A\}],
\end{equation}
for all Borel sets $A \subset \mathbb{R}$.
It is well-known (Bayes' formula) that equation \eqref{eq:def_condPD} implies closed-form
representations of $\mathrm{P}[D\,|\,S=s] = p_D(s)$ in two important
special cases:
\begin{subequations}
\begin{itemize}
	\item $S$ is a discrete variable, i.e.\ $S \in I = \{1, 2, \ldots, k\}$. Then
\begin{equation}\label{eq:pd_discrete}
		\mathrm{P}[D\,|\,S = j] \ =\ \frac{p\,\mathrm{P}[S = j\,|\,D]}{p\,\mathrm{P}[S = j\,|\,D] +
		(1-p)\,\mathrm{P}[S = j\,|\,N]}\,, \quad j \in I.
\end{equation}
	\item $S$ is a continuous variable with values in an open interval $I$
	such that there are Lebesgue densities $f_N$ and $f_D$ 
	of the conditional distribution functions $F_N$ and $F_D$ from \eqref{eq:cond_dist}. Then	
\begin{equation}\label{eq:pd_continuous}
		\mathrm{P}[D\,|\,S = s]\ =\ \frac{p\,f_D(s)}{p\,f_D(s) + (1-p)\,f_N(s)}\,,\quad s \in I.
\end{equation}
\end{itemize}
\end{subequations}
A closely related consequence of equation \eqref{eq:def_condPD} is the fact that $p$, $F_N$, and $F_D$
can be determined whenever the unconditional score distribution $F$ and the conditional 
probabilities of default $\mathrm{P}[D\,|\,S]$ are known. We then obtain
\begin{subequations}
\begin{equation}
	p  = \mathrm{E}\bigl[\mathrm{P}[D\,|\,S]\bigr] = \begin{cases}
	\sum_{j=1}^k \mathrm{P}[D\,|\,S = j]\,\mathrm{P}[S = j], & \text{$S$ discrete}\\[1ex]
	\int_I \mathrm{P}[D\,|\,S = s]\,f(s)\,d s, & \text{$S$ continuous with density $f$}.
\end{cases}
\end{equation}
If $S$ is a discrete rating variable, we have for $j\in I$
\begin{equation}
\begin{split}
\mathrm{P}[S = j\,|\,D]  &= \mathrm{P}[D\,|\,S = j]\,\mathrm{P}[S = j] / p,\\
\mathrm{P}[S = j\,|\,N]  &= \bigl(1-\mathrm{P}[D\,|\,S = j]\bigr)\,\mathrm{P}[S = j] / (1-p).
\end{split}
\end{equation}
If $S$ is continuous score variable with density $f$, we have for $s\in I$
\begin{equation}
\begin{split}
f_D(s) & = \mathrm{P}[D\,|\,S = s]\,f(s) / p,\\
f_N(s) & = \bigl(1-\mathrm{P}[D\,|\,S = s]\bigr)\,f(s) / (1-p).
\end{split}
\end{equation}
\end{subequations}


\subsection{Notation for distribution functions}
\label{sec:notation}

At some points in this paper we will need to handle distribution functions and their inverse 
functions. For further reference we list in this subsection the necessary notation and 
some properties of such functions:
\begin{itemize}
	\item A (real) distribution function $G$ is an increasing and
	right-continuous function $\mathbb{R} \to [0,1]$ with $\lim\limits_{x\to -\infty} G(x) = 0$ and
	$\lim\limits_{x\to \infty} G(x) = 1$.
	\item Any real random variable $X$ defines a distribution function $G = G_X$ by 
	$G(x) = \mathrm{P}[X \le x]$.
	\item \textbf{Convention:} $G(-\infty) = 0$ and $G(\infty) =1$.
	\item Denote by $G(\cdot - 0)$ the left-continuous version of the distribution function 
	$G$. Then $G(\cdot - 0)\le G$ and 
$G(x - 0) = G(x)$ for all $x$ but countably many $x \in \mathbb{R}$ because $G$ is non-decreasing.
	\item For any distribution function $G$, the function $G^{-1}$ is its \emph{generalised inverse} or
\emph{quantile function}, i.e.
\begin{subequations}
\begin{equation}
	G^{-1}(u) = \inf\{x \in \mathbb{R}:\, G(x) \ge u\}, \quad u \in [0,1].
\end{equation}
In particular, we obtain
\begin{equation}\label{eq:infty}
	-\infty = G^{-1}(0) < G^{-1}(1) \le \infty.
\end{equation}
\end{subequations}
	\item Denote by $\varphi(s)$ the standard normal density and by $\Phi(s)$ the standard normal distribution
	function.
\end{itemize}

\section{Discriminatory power: Theory}
\label{sec:framework}

\citet[][section 8.1]{Hand97} described ROC curves as follows: ``Often the two degrees of freedom
[i.e.\ the two error types associated with binary classification] are presented simultaneously for
a range of possible classification thresholds for the classifier in a \emph{receiver operating
characteristic (ROC) curve}. This is done by plotting true positive rate (sensitivity) on 
the vertical axis against false positive rate (1 - specificity) on the horizontal axis.''

Translated into the notation introduced in section \ref{sec:concepts}, for a fixed score
value $s$ seen as threshold the true positive rate
is the hit rate $F_D(s)$ while the false positive rate is the false alarm rate $F_N(s)$. 
In these terms, CAP (Cumulative Accuracy Profile) curves \citep[not mentioned by][]{Hand97} 
can be described as a plot of 
the hit rates against the alarm rates across a range of classification thresholds.
If all possible thresholds are to be considered, these descriptions formally can 
be expressed in the following terms.

\begin{definition}[ROC and CAP]\label{de:ROC&CAPcurves}
Denote by $F_N$ the distribution function $F_N(s) = \mathrm{P}[S_N\le s]$ of the scores
conditional on the event ``borrower survives'', by $F_D$ the distribution function 
$F_D(s) = \mathrm{P}[S_D\le s]$ of the scores
conditional on the event ``borrower defaults'', and by $F$ the unconditional distribution
function $F(s) = \mathrm{P}[S \le s]$ of the scores. 

The \emph{Receiver Operating Characteristic (ROC)} of the score function then is defined as the
graph of the following set $\mathrm{gROC}$ (``g'' for graph) of points in the unit square:
\begin{subequations}
\begin{equation}\label{eq:ROCset}
	\mathrm{gROC} \ =\ \bigl\{\bigl(F_N(s), F_D(s)\bigr):\ s \in \mathbb{R}\cup\{\pm \infty\}\bigr\}.
\end{equation}
The \emph{Cumulative Accuracy Profile (AUC)} of the score function is defined as the
graph of the following set $\mathrm{gCAP}$ of points in the unit square:
\begin{equation}\label{eq:CAPset}
	\mathrm{gCAP} \ =\ \bigl\{\bigl(F(s), F_D(s)\bigr):\ s \in \mathbb{R}\cup\{\pm \infty\}\bigr\}.
\end{equation}
\end{subequations}
\end{definition}
Actually the point sets $\mathrm{gROC}$ and $\mathrm{gCAP}$ can be quite irregular (e.g.\ if one of
the involved distribution functions has an infinite number of discontinuities and the set of 
discontinuities is dense in $\mathbb{R}$). In such a case it would be physically impossible to plot 
on paper a precise graph of the point set. In most parts of the following, therefore, we will focus
on three more regular special cases which are of relevance for theory and practice:
\begin{enumerate}
	\item $F$, $F_N$, and $F_D$ are smooth, i.e.\ at least continuous. This is usually a reasonable 
	assumption when the score function takes on values on a continuous scale.
	\item The distributions of $S$, $S_N$ and $S_D$ are concentrated on a finite number of points. This
	is the case when the score function is a rating system with a finite number (e.g.\ seven or seventeen
	as in case of S \& P, Moody's, or Fitch ratings) of grades.
	\item $F$, $F_N$, and $F_D$ are empirical distribution functions associated to finite samples of scores
	on a continuous scale. This is naturally the case when the performance of a score function is analysed 
	on the basis of non-parametric estimates.
\end{enumerate}
In the smooth situation of 1) the sets $\mathrm{gROC}$ and $\mathrm{gCAP}$ are compact and connected such that
there is no ambiguity left of how to draw a graph that -- together with the x-axis and the vertical line through
$x=1$ -- encloses a region of finite area. In situations 2) and 3), however, the sets $\mathrm{gROC}$ and 
$\mathrm{gCAP}$ consist of a finite number of isolated points and hence are unconnected. While this, in a certain sense,
even facilitates the drawing of the graphs, the results nonetheless will be unsatisfactory when it comes 
to a comparison of the discriminatory power of score functions or rating systems. Usually, therefore, in such 
cases a certain degree of interpolation will be applied to the points of the sets $\mathrm{gROC}$ and $\mathrm{gCAP}$
in order to facilitate their visual comparison. We will discuss in section \ref{sec:discontinuous}
the question of how to do best the interpolation to satisfy some properties that are desirable from a statistical
point of view.

Before, however, in section \ref{sec:GenObsCont} we have a closer look on the properties of ROC graphs in smooth 
contexts. These properties then will be used as a kind of yardstick to assess the appropriateness of interpolation approaches
to the discontinuous case in section \ref{sec:discontinuous}.


\subsection{Continuous score distributions}
\label{sec:GenObsCont}

In this subsection, we will work most of the time on the basis of one of the following two assumptions.

\textbf{Assumption N:} The distribution of the score $S_N$ conditional on the borrower's survival
is continuous, i.e.
\begin{equation}\label{eq:Ncontinuous}
	\mathrm{P}[S_N = s] = 0\quad \text{for all\ }s.
\end{equation}
\textbf{Assumption S:} The unconditional distribution of the score $S$ is continuous 
(and hence by \eqref{eq:unconditional} so are the distributions of $S_N$ and $S_D$), i.e.
\begin{equation}\label{eq:continuous}
	\mathrm{P}[S = s] = 0\quad \text{for all\ }s.
\end{equation}
Additionally, the following technical assumption is sometimes useful.

\textbf{Assumption:} 
\begin{equation}\label{eq:smaller}
	F_D^{-1}(1) \le F_N^{-1}(1).
\end{equation}
This is equivalent to requiring that the essential supremum of $S_D$ is not greater than
the essential supremum of $S_N$. Such a requirement seems natural under the assumption that
low score values indicate low creditworthiness (``bad'') and high score values indicate
high creditworthiness (``good'').

As an immediate consequence of these assumptions we obtain representations
of the ROC and CAP sets \eqref{eq:ROCset} and \eqref{eq:CAPset} that are more convenient for
calculations.
\begin{subequations}
\begin{theorem}[Standard parametrisations of ROC and CAP]\label{pr:functionROC}\ \\
With the notation of definition \ref{de:ROC&CAPcurves} define the functions ROC and CAP by
\begin{align}\label{eq:ROC}
	\mathrm{ROC}(u) & = F_D\bigl(F_N^{-1}(u)\bigr), \quad u \in [0,1],\\
\label{eq:CAP}
	\mathrm{CAP}(u) & = F_D\bigl(F^{-1}(u)\bigr) = 
	F_D\bigl((p\,F_D(\cdot)+(1-p)\,F_N(\cdot))^{-1}(u)\bigr), \quad u \in [0,1].
\end{align}
For \eqref{eq:CAP}, assume $p > 0$ (otherwise ROC and CAP coincide).
Under \eqref{eq:Ncontinuous} (assumption N) then we have
\begin{equation}\label{eq:ROC_graph}
	\bigl\{\bigl(u, \mathrm{ROC}(u)\bigr): u \in [0,1]\bigr\} \ \subset\ 
	\mathrm{gROC}.
\end{equation}
If under \eqref{eq:Ncontinuous} (assumption N), moreover, 
the distribution of $S_D$ is absolutely continuous with respect to the distribution
of $S_N$ (i.e.\ $\mathrm{P}[S_N \in A]=0 \ \Rightarrow\ \mathrm{P}[S_D \in A]=0$), then\footnote{%
The absolute continuity requirement implies that $F_D$ is constant on the intervals 
on which $F_N$ is constant.} ``$=$'' applies also
to \eqref{eq:ROC_graph}:
\begin{equation}\label{eq:sets_equal}
	\bigl\{\bigl(u, \mathrm{ROC}(u)\bigr): u \in [0,1]\bigr\}\ =\ 
	\mathrm{gROC}.
\end{equation} 
Equation \eqref{eq:continuous} (assumption S) implies
\begin{equation}
	\bigl\{\bigl(u, \mathrm{CAP}(u)\bigr): u \in [0,1]\bigr\} \ =\ 
	\mathrm{gCAP}.\label{eq:CAP_graph}	
\end{equation}
\end{theorem}
\end{subequations}
\textbf{Proof.} Note that \eqref{eq:infty} implies in general
\begin{subequations}
\begin{equation}
	0 =  \mathrm{CAP}(0) = \mathrm{ROC}(0). \label{eq:null}
\end{equation}
For $p > 0$, we have $\bigl\{s: p\,F_D(s)+(1-p)\,F_N(s) \ge 1 \bigr\} \subset \bigl\{s: F_D(s) \ge 1\bigr\}$
and hence 
\begin{align}
	\mathrm{CAP}(1)  & = F_D\bigl((p\,F_D(\cdot)+(1-p)\,F_N(\cdot))^{-1}(1)\bigr) \notag\\
					& \ge F_D\bigl(F_D^{-1}(1)\bigr) \notag\\
					& \ge 1 \notag\\
	 \Rightarrow\quad \mathrm{CAP}(1) & = 1. \label{eq:eins}
\end{align}	
Additionally, if \eqref{eq:smaller} holds  -- which is implied by the absolute continuity assumption -- 
we obtain
\begin{align}
	\mathrm{ROC}(1)  & = F_D\bigl(F_N^{-1}(1)\bigr) \notag\\
					& \ge F_D\bigl(F_D^{-1}(1)\bigr) \notag\\
					& \ge 1 \notag\\
	 \Rightarrow\quad \mathrm{ROC}(1) & = 1. \label{eq:zwei}
\end{align}	
\end{subequations}
Now, by  \eqref{eq:Ncontinuous} (assumption N) we have $F_N\bigl(F_N^{-1}(u)\bigr) = u$ and by 
\eqref{eq:continuous} (assumption S) we have $F\bigl(F^{-1}(u)\bigr) = u$ \citep[see][section 21.1]{VanDerVaart}. 
This implies \eqref{eq:ROC_graph} and
``$\subset$'' in \eqref{eq:CAP_graph}. \\
Assume that distribution of $S_D$ is absolutely continuous with respect to the distribution
of $S_N$. For $s\in \mathrm{R}$ let $s_0 = F_N^{-1}\bigl(F_N(s)\bigr)$. By continuity of $F_N$
then we have $F_N(s_0) = F_N(s)$, and by absolute continuity of $S_D$ with respect to $S_N$ we also
have $F_D(s_0) = F_D(s)$.  This implies ``$=$'' in \eqref{eq:ROC_graph} because on the one hand
\begin{equation*}
	\bigl(F_N(s), F_D(s)\bigr) \ =\ \bigl(F_N(s), F_D(s_0)\bigr)  \ =\ 
	\bigl(F_N(s), \mathrm{ROC}\bigl(F_N(s)\bigr) 
	\ \in\ \bigl\{\bigl(u, \mathrm{ROC}(u)\bigr): u \in [0,1]\bigr\},
\end{equation*}
and on the other hand for $s = \pm \infty$ we can apply \eqref{eq:null}, \eqref{eq:eins}, and
\eqref{eq:zwei}.\\
The ``$=$'' in \eqref{eq:CAP_graph} follows from the fact that $S_D$ by \eqref{eq:unconditional} is
always absolutely continuous with respect to $S$. \hfill $\Box$

\begin{remark}\label{rm:standard}\ \\
A closer analysis of the proof of theorem \ref{pr:functionROC} shows that a non-empty difference
between the left-hand and the right-hand sides of \eqref{eq:ROC_graph} can occur only if there are
non-empty intervals on which the value of $F_N$ is constant. To each such interval on which $F_D$ is not constant
there is corresponding piece of a vertical line in the set $\mathrm{gROC}$ that has no counterpart in the
graph of the function $\mathrm{ROC}(u)$. Note, however, that these missing pieces are not relevant with
respect to the area below the ROC curve because this area is still well-defined when all vertical pieces
are removed from $\mathrm{gROC}$. In this sense, in theorem \ref{pr:functionROC} the absolute continuity
requirement and equation \eqref{eq:sets_equal} are only of secondary importance.
\end{remark}
In view of theorem \ref{pr:functionROC} and remark \ref{rm:standard}, 
we can regard ROC and CAP curves as graphs for functions \eqref{eq:ROC} and \eqref{eq:CAP} respectively, as
long as \eqref{eq:Ncontinuous} and  \eqref{eq:continuous} apply. This provides a convenient way to dealing analytically 
with ROC and CAP curves. In section \ref{sec:discontinuous} we will revisit the 
question of how to conveniently parametrize the point sets \eqref{eq:ROCset} and \eqref{eq:CAPset} in the 
case of score distributions with discontinuities. 

In this section, we continue by looking closer at some well-known properties of ROC and CAP curves.
In non-technical terms the following proposition \ref{pr:powerless} states: 
The diagonal line is the ROC and CAP curve of powerless rating systems (or score functions).
For a perfect score function, the ROC curve is essentially the horizontal line at level 1 while the CAP curve is 
made up by the straight line $u \mapsto u/p, u < p$ and the horizontal line at level 1.
\begin{subequations}
\begin{proposition}\label{pr:powerless}
Under \eqref{eq:Ncontinuous} (assumption N), in case of a powerless classification system (i.e.\ $F_D = F_N$) 
we have
\begin{equation}\label{eq:powerless}
	\mathrm{ROC}(u) = u = \mathrm{CAP}(u), \quad  u \in [0,1].
\end{equation}
In case of a perfect classification system\footnote{%
Note that in case of a perfect classification system the distribution of $S_D$ is not
absolutely continuous with respect to the distribution of $S_N$ as it would be required 
for \eqref{eq:sets_equal} to obtain.} (i.e.\ there is a score value $s_0$ such that $F_D(s_0) =1, F_N(s_0)=0$)
we obtain without continuity assumption that 
\begin{align}
	\mathrm{ROC}(u) & = \begin{cases}
	0, & u = 0,\\
	1, & 0 < u \le 1,
\end{cases} \label{eq:perfect_ROC}\\
\intertext{and, if $p > 0$ and $F_D$ is continuous,}  
	\mathrm{CAP}(u) & = \begin{cases}
	u/p, &  0 \le u < p, \\
	1, & p \le u \le 1.
\end{cases} \label{eq:perfect}
\end{align}
\end{proposition}
\end{subequations}
\textbf{Proof.} For \eqref{eq:powerless}, we have to show that 
\begin{equation}\label{eq:part}
	F_D\bigl(F_D^{-1}(u)\bigr) = u, \quad u \in [0,1].
\end{equation}
This follows from the continuity assumption \eqref{eq:Ncontinuous} \citep[see][section 21.1]{VanDerVaart}.

On \eqref{eq:perfect_ROC} and \eqref{eq:perfect}: Observe that $F_D(s_0) =1, F_N(s_0)=0$ for some $s_0$ implies 
\eqref{eq:smaller}. By 
\eqref{eq:null}, \eqref{eq:eins} and \eqref{eq:zwei}, therefore, we only need to consider the case $0 < u < 1$.
For $u > p$ we obtain
\begin{align}
	& F(s_0) = p\,F_D(s_0) + (1-p)\,F_N(s_0) = p  \label{eq:p} \\ 
	\Rightarrow\quad  & F^{-1}(u) \ge s_0\notag\\
	\Rightarrow\quad & F_D\bigl(F^{-1}(u)\bigr) = 1.\notag
\end{align}
This implies \eqref{eq:perfect_ROC} (with $p=0$), in particular, and \eqref{eq:perfect} for $u > p$. 
For $u < p$, equation \eqref{eq:p} implies $F^{-1}(u) < s_0$. By left continuity of $F^{-1}$,
we additionally obtain $F^{-1}(u) \le s_0$ for $u \le p$.
But
\begin{equation*}
	F(s) = p\,F_D(s) + (1-p)\,F_N(s) = p\,F_D(s), \quad s \le s_0.
\end{equation*}
Hence for $u \le p$
\begin{align*}
	 & F^{-1}(u)  = \inf\{s: p\,F_D(s) \ge u\} = F_D^{-1}(u/p)\\
	 \Rightarrow\quad & F_D\bigl(F^{-1}(u)\bigr) = F_D\bigl(F_D^{-1}(u/p)\bigr) = u/p.
\end{align*}
The last equality follows from the assumed continuity of $F_D$. \hfill $\Box$

By theorem \ref{pr:functionROC}, 
in the continuous case \eqref{eq:Ncontinuous} and \eqref{eq:continuous}, the common notions of AUC (area under the curve) and
AR (accuracy ratio) can be defined in terms of integrals of the ROC and CAP functions \eqref{eq:ROC}
and \eqref{eq:CAP}. Recall that the accuracy ratio commonly is described in terms like these: ``The quality of
a rating system is measured by the accuracy ratio AR. It is defined as the
ratio of the area between the CAP of the rating model being validated
and the CAP of the random model [= powerless model], and the area between the CAP of
the perfect rating model and the CAP of the random model'' \citep[][page 82]{Engelmannetal03a}.

\begin{definition}[Area under the curve and accuracy ratio]\label{de:AUC&AR}\ \\
For the function ROC given by \eqref{eq:ROC} we define the \emph{area under the curve} AUC by
\begin{subequations}
\begin{equation}\label{eq:AUC}
\mathrm{AUC} = \int_0^1 \mathrm{ROC}(u)\, d u.	
\end{equation}
For the function CAP given by \eqref{eq:CAP} we define the \emph{accuracy ratio} AR by
\begin{equation}\label{eq:AR}
	\mathrm{AR} = \frac{\int_0^1 \mathrm{CAP}(u) - u\, d u}{1- p/2 - 1/2} = 
	\frac{2\int_0^1 \mathrm{CAP}(u)\, d u -1}{1- p}.
\end{equation}
\end{subequations}
\end{definition}
In the continuous case \eqref{eq:continuous} (assumption S) AUC and AR are identical up to a constant
linear transformation, as shown by the following proposition.

\begin{proposition}[AUC and AR in the continuous case]\label{pr:AR_AUC}\ \\
If the distribution of the score function conditional on default is continuous then
\begin{equation*}
	\mathrm{AR} = 2\,\mathrm{AUC} - 1.	
\end{equation*}
\end{proposition}
\textbf{Proof.} Denote by $S'_D$ a random variable with the same distribution $F_D$ as $S_D$ but independent of $S_D$. 
Let $S_N$ be independent of $S_D$. Observe that $F_N^{-1}(U)$ and $F_D^{-1}(U)$ have 
the same distribution as $S_N$ and $S_D$ if $U$ is uniformly
distributed on $(0,1)$.
By the definition \eqref{eq:AR} of AR and Fubini's theorem, therefore we obtain
\begin{align}
	\mathrm{AR} & = \frac{2}{1-p} \left(p\,\mathrm{P}[S_D \le S'_D] + (1-p)\,\mathrm{P}[S_D \le S_N] - 1/2\right)\label{eq:ARgeneral}\\
	& = 2\,\mathrm{P}[S_D \le S_N] - 1\notag\\
	& = 2\,\mathrm{AUC} - 1.\notag
\end{align}
In this calculation, the fact has been used that $1/2 = \mathrm{P}[S_D \le S'_D]$ because the distribution of $S_D$ is assumed 
to be continuous. \hfill $\Box$

As the ROC curve does not depend on the proportion $p$ of defaulters in the population, proposition \ref{pr:AR_AUC} in
particular shows that AR does not depend on $p$ either. The following corollary is an easy consequence of propositions 
\ref{pr:powerless} and \ref{pr:AR_AUC}. It identifies the extreme cases for classification systems. A classification system
is considered poor if its AUC and AR are close to AUC and AR of a powerless system. It is considered powerful if 
if its AUC and AR are close to AUC and AR of a perfect system.

\begin{corollary}\label{co:AUC_AR}
Under \eqref{eq:Ncontinuous} (assumption N), in case of a powerless classification system (i.e.\ $F_D = F_N$) 
we have
\begin{subequations}
\begin{equation}\label{eq:powerless_AR}
\begin{split}
	\mathrm{AUC} & = 1/2,\\
	\mathrm{AR} & = 0.
\end{split}
\end{equation}
In case of a perfect classification system (i.e.\ there is a score value $s_0$ such that $F_D(s_0) =1, F_N(s_0)=0$)
we obtain if the distribution of the scores conditional on default is continuous
\begin{equation}\label{eq:perfect_AR}
\begin{split}
	\mathrm{AUC} & = 1, \\
	 \mathrm{AR} & = 1. 
\end{split}	 
\end{equation}
\end{subequations}
\end{corollary}

Relation \eqref{eq:powerless_AR} can obtain also in situations where $F_N \not= F_D$. 
For instance, \citet[][proposition 2.6]{ClaveroRasero} proved that \eqref{eq:powerless_AR}
applies in general when $F_N$ and $F_D$ have densities that are both symmetric with respect
to the same point.


\subsubsection{Example: Normally distributed scores}
\label{sec:normal}

Assume that the score distributions conditional on default and survival, respectively, are
normal:
\begin{equation}\label{eq:normal}
	S_D \sim \mathcal{N}(\mu_D, \sigma_D^2), \quad S_N \sim \mathcal{N}(\mu_N, \sigma_N^2).
\end{equation}
Formulas for ROC in the sense of \eqref{eq:ROC} and AUC in the sense of \eqref{eq:AUC} then easily
are derived:
\begin{equation}
	\begin{split}
	\mathrm{ROC}(u) & = \Phi\Bigl(\frac{\sigma_N\,\Phi^{-1}(u) + \mu_N - \mu_D}{\sigma_D}\Bigr), \quad 
	u \in [0,1]\\
	\mathrm{AUC} & = \Phi\biggl(\frac{\mu_N - \mu_D}{\sqrt{\sigma_N^2+\sigma_D^2}}\biggr).\label{eq:Satchell}
	\end{split}
\end{equation}
Note that \eqref{eq:Satchell} gives a closed form of AUC 
where \citet{Satchell&Xia} provided a formula involving
integration. See figure \ref{fig:1} for an illustration of \eqref{eq:normal} and \eqref{eq:Satchell}.

The unconditional score distribution $F$ can be derived from \eqref{eq:unconditional}. Under \eqref{eq:normal},
however, for $p \notin \{0,1\}$, $F$ is not a normal distribution function. Its inverse function $F^{-1}$ can be
evaluated numerically, but no closed-form representation is known. For plots of the CAP curve, therefore it is more
efficient to make use of representation \eqref{eq:CAPset}. The value of AR can be derived from the value
of AUC by proposition \ref{pr:AR_AUC}.


\subsubsection{Example: Density estimation with normal kernel}
\label{sec:kernel}

	Assume that there are samples $x_1, \ldots, x_{n_D}$ of scores of defaulted borrowers and 
	$y_1, \ldots, y_{n_N}$ of surviving borrowers. If the scores take on values on a continuous scale, 
	it makes sense to try and estimate densities of the defaulters' scores and survivors' scores, respectively.
	We consider here kernel estimation with a normal kernel as estimation approach \citep[see, e.g.][chapter 2]{Pagan99}.
	The resulting density estimates then are
	\begin{equation}\label{eq:dens_est}
	\begin{split}
	\widehat{f}_D(s) & = (n_D\,h_D)^{-1} \sum_{i=1}^{n_D} \varphi\Bigl(\dfrac{s-x_i}{h_D}\Bigr),\\
	\widehat{f}_N(s) & = (n_N\,h_N)^{-1} \sum_{i=1}^{n_N} \varphi\Bigl(\dfrac{s-y_i}{h_N}\Bigr),
	\end{split}
	\end{equation}
	where $h_D, h_N > 0$ denote appropriately selected \emph{bandwidths}. Silverman's rule of thumb 
	\citep[see, e.g.][equation (2.50)]{Pagan99} often yields reasonable results:
	\begin{equation}\label{eq:Silverman}
    h\ =\ 1.06\,\widehat{\sigma}\,T^{-1/5},
\end{equation}%
where $\widehat{\sigma}$  denotes the standard deviation of the sample $x_1, \ldots, x_{n_D}$ or 
$y_1, \ldots, y_{n_N}$, respectively. Equation \eqref{eq:dens_est}
immediately implies the following formulas for the corresponding estimated distribution functions:
	\begin{equation}\label{eq:dist_est}
	\begin{split}
	\widehat{F}_D(s) & = (n_D)^{-1} \sum_{i=1}^{n_D} \Phi\Bigl(\dfrac{s-x_i}{h_D}\Bigr),\\
	\widehat{F}_N(s) & = (n_N)^{-1} \sum_{i=1}^{n_N} \Phi\Bigl(\dfrac{s-y_i}{h_N}\Bigr).
	\end{split}
	\end{equation}
	ROC and CAP curves then can be drawn efficiently by taking recourse to \eqref{eq:ROCset} and
	\eqref{eq:CAPset}. An estimate of AUC (and then by
	proposition \ref{pr:AR_AUC} of AR) is given by a generalisation of \eqref{eq:Satchell}:	
\begin{equation}\label{eq:AUC_Satchell}
	\widehat{AUC} = (n_D\,n_N)^{-1} \sum_{i=1}^{n_D} \sum_{j=1}^{n_N} \Phi\Bigl(\dfrac{y_j - x_i}{\sqrt{h_N^2+h_D^2}}\Bigr).
\end{equation}
	See figure \ref{fig:2} for illustration.
	
\begin{remark}[Bias of kernel-based AUC-estimator]\label{rm:biased}\ \\
Assume that the samples $x_1, \ldots, x_{n_D}$ of scores of defaulted borrowers and 
$y_1, \ldots, y_{n_N}$ of scores of surviving borrowers are samples from normally distributed
score functions as in \eqref{eq:normal}. Then the expected value
of the AUC-estimator $\widehat{AUC}$ from \eqref{eq:AUC_Satchell} can be calculated as follows:
\begin{equation*}
	\mathrm{E}\bigl[\widehat{AUC}\bigr] \ = \ 
	\Phi\biggl(\frac{\mu_N - \mu_D}{\sqrt{h_N^2 + h_D^2+\sigma_N^2+\sigma_D^2}}\biggr).
\end{equation*}
Hence by \eqref{eq:Satchell}, the following observations apply:
\begin{align*}
	\bigl|\mathrm{E}\bigl[\widehat{AUC}\bigr]-1/2\bigr| &\ \le\ |\mathrm{AUC}-1/2|\\
	\mathrm{sign}\bigl(\mathrm{E}\bigl[\widehat{AUC}\bigr]-1/2\bigr) &\ =\ \mathrm{sign}(\mathrm{AUC}-1/2)\\
	\mu_D = \mu_N &\ \Leftrightarrow\ \mathrm{E}\bigl[\widehat{AUC}\bigr]\ =\ \mathrm{AUC}\\ 
	\mu_D = \mu_N &\ \Leftrightarrow\ \mathrm{AUC}\ =\ 1/2.
\end{align*}
In particular, in case $\mu_N > \mu_D$ the estimator $\widehat{AUC}$ on average underestimates 
the area under the curve while in case $\mu_N < \mu_D$ the area under the curve
is overestimated by $\widehat{AUC}$.
\end{remark}
To account for the potential bias of the AUC estimates by \eqref{eq:AUC_Satchell} as observed
in remark \ref{rm:biased}, in section \ref{sec:aspects} we will apply linear transformations 
to the density estimates \eqref{eq:dens_est}. These linear transformations make sure that the means
and variances of the estimated densities exactly match the empirical means and variances of 
the samples $x_1, \ldots, x_{n_D}$ and 	$y_1, \ldots, y_{n_N}$ respectively \citep[][section 3.4]{DavisonHinkley}.
Define
\begin{subequations}
\begin{align}
	b_D & = \sqrt{\frac{1/n \sum_{i=1}^{n_D} x_i^2 - \left(1/n \sum_{i=1}^{n_D} x_i\right)^2}
	{h_D^2 + 1/n \sum_{i=1}^{n_D} x_i^2 - \left(1/n \sum_{i=1}^{n_D} x_i\right)^2}}, &\quad
	a_D & = \frac{1-b_D}n \sum_{i=1}^{n_D} x_i,\\
	b_N & = \sqrt{\frac{1/n \sum_{j=1}^{n_N} y_j^2 - \left(1/n \sum_{j=1}^{n_N} y_j\right)^2}
	{h_N^2 + 1/n \sum_{j=1}^{n_N} y_j^2 - \left(1/n \sum_{j=1}^{n_N} y_j\right)^2}},
	 & \quad a_N & = \frac{1-b_N}n \sum_{j=1}^{n_N} y_j.
\end{align}
Replace then in equations \eqref{eq:dens_est}, \eqref{eq:dist_est}, and \eqref{eq:AUC_Satchell} 
\begin{equation}\label{eq:replace}
\begin{split}
	x_i \ \text{by}\ a_D + b_D\,x_i \quad \text{and}\quad h_D \ \text{by}\ b_D\,h_D,\\
	y_j \ \text{by}\ a_N + b_N\,y_j \quad \text{and}\quad h_N \ \text{by}\ b_N\,h_N,
\end{split}
\end{equation}
\end{subequations}
to reduce the bias from an application of \eqref{eq:AUC_Satchell} for AUC estimation. If, for instance,
in the right-hand panel of figure \ref{fig:2} the estimated ROC curve is based on the transformed
samples according to \eqref{eq:replace}, the resulting estimate of AUC is 71.2\%. Thus, at least in
this example, the ``transformed'' AUC estimate is closer to the true value of 71.6\% than the estimate
based on estimated densities without adjustments for mean and variance.

\subsection{Discontinuous score distributions}
\label{sec:discontinuous}

We have seen that in the case of continuous score distributions as considered in section \ref{sec:GenObsCont}
there are standard representations of ROC and CAP curves (theorem \ref{pr:functionROC}) 
that can be conveniently deployed to formally define the
area under the curve (AUC) and the accuracy ratio (AR) and to investigate some of their properties.
In this section, we will see that in a more general setting the use of 
the curve representations \eqref{eq:ROC} and \eqref{eq:CAP} can have counter-intuitive implications.
We then will look at modifications of \eqref{eq:ROC} and \eqref{eq:CAP} that avoid such
implications and show that these modifications are compatible with common interpolation approaches
to the ROC and CAP graphs as given by \eqref{eq:ROCset} and \eqref{eq:CAPset}. We will do so 
primarily with a view on the settings described in items 2) and 3) at the beginning of section
\ref{sec:framework}. For the sake of reference, 
the following two examples describe these settings in more detail.
\begin{example}[Rating distributions]\label{ex:dis}\ \\
Consider a \emph{rating system} with grades $1, 2, \ldots, n$ where $n$ stands for highest creditworthiness. 
The random variable $R$ which expresses a borrower's rating grade then is \emph{purely discontinuous} because
\begin{align*}
	\mathrm{P}[R = k] & \ge 0, \quad k \in \{1, 2, \ldots, n\}\\
	\mathrm{P}[R \notin \{1, 2, \ldots, n\}] & = 0.
\end{align*}
See the upper panel of figure \ref{fig:3} for illustration.
As in the case of score functions $S$, we write $R_D$ when considering $R$ on the sub-population of defaulters
and $R_N$ when considering $R$ on the sub-population of survivors.
\end{example}
\begin{example}[Sample-based empirical distributions]\label{ex:sample}\ \\
Assume -- as in section \ref{sec:kernel} -- that there are samples $x_1, \ldots, x_{n_D}$ of scores of defaulted borrowers and 
$y_1, \ldots, y_{n_N}$ of surviving borrowers. If there is no reason to believe that the samples were generated from
continuous score distributions, or if sample sizes are so large that kernel estimation becomes numerically inefficient, 
one might prefer to work with the empirical distributions of $S_D$ and $S_N$ as inferred from $x_1, \ldots, x_{n_D}$ and
$y_1, \ldots, y_{n_N}$, respectively:\\[1ex]
\begin{subequations}
For $w, z\in \mathbb{R}$ let
\begin{equation*}
	\delta_w(z) = \begin{cases}
	1, & z \le w\\
	0, & z > w.
\end{cases}
\end{equation*}
For $w \in \mathbb{R}$ define the \emph{empirical distribution function} for the sample $z_1, \ldots, z_n$ by
\begin{equation}\label{eq:EmpDist}
	\delta_w(z_1, \ldots, z_n) = 1/n \sum_{i=1}^n \delta_w(z_i).
\end{equation}
For $w, z\in \mathbb{R}$ let
\begin{equation*}
	\delta^\ast_w(z) = \begin{cases}
	1, & z < w\\
	1/2, & z = w\\
	0, & z > w.
\end{cases}
\end{equation*}
For $w \in \mathbb{R}$ define the \emph{modified empirical distribution function} for the sample $z_1, \ldots, z_n$ by
\begin{equation}\label{eq:EmpDistMod}
	\delta^\ast_w(z_1, \ldots, z_n) = 1/n \sum_{i=1}^n \delta^\ast_w(z_i).
\end{equation}
\end{subequations}
\end{example}
Of course, there is some overlap between examples \ref{ex:dis} and \ref{ex:sample}. The samples in example
\ref{ex:sample} could have been generated from rating distributions as described in example \ref{ex:dis} (see lower
panel of figure \ref{fig:3} for illustration).
Then example \ref{ex:sample} just would be a special case of example \ref{ex:dis}. The more interesting case
in example \ref{ex:sample} therefore is the case where $\{x_1, \ldots, x_{n_D}\}\cap 
\{y_1, \ldots, y_{n_N}\} = \emptyset$. This will occur with probability 1 when the two sub-samples are
generated from continuous score distributions. 

\textbf{Some consequences of discontinuity:}
\begin{itemize}
\item In the settings of examples \ref{ex:dis} and \ref{ex:sample} the CAP and ROC graphs
as defined by \eqref{eq:CAPset} and \eqref{eq:ROCset} consist of finitely many points.
\item CAP and ROC functions as defined by \eqref{eq:CAP} and \eqref{eq:ROC} 
are piecewise constant for rating grade variables $R$ as in example \ref{ex:dis} and
empirical distribution functions as in example \ref{ex:sample}.
See left panel of figure \ref{fig:4} for illustration.
\item Proposition \ref{pr:powerless} does not apply. In particular, the graphs of CAP and ROC functions
as defined by \eqref{eq:CAP} and \eqref{eq:ROC} for powerless score functions with discontinuities 
are not identical with the diagonal line.
See left panel of figure \ref{fig:4} for illustration.
\item Let $S$ be a random variable with a distribution that is concentrated on finitely many points as in
	example \ref{ex:dis} or \ref{ex:sample}. Let $S'$ be a random variable with the same distribution as $S$ 
	but independent of $S$. Then we have
\begin{equation}\label{eq:diag}
	\mathrm{P}[S = S'] > 0.
\end{equation}
\end{itemize}


\subsubsection{Observations on the general case}
\label{sec:GenObs}

In this section, we first look at what happens with corollary \ref{co:AUC_AR} if no continuity assumption obtains.

\begin{proposition}[AUC and AR in the general case]\label{pr:AUCgeneral}\ \\
Define AUC and AR by \eqref{eq:AUC} and \eqref{eq:AR}, respectively, with ROC and CAP as given in 
\eqref{eq:ROC} and \eqref{eq:CAP}. Let $S_D$ and $S_N$ denote independent random variables 
with distribution functions $F_D$ (score distribution conditional on default) and $F_N$ (score distribution
conditional on survival). Assume that $S'_D$ is an independent copy of $S_D$. Then 
\begin{align*}
\mathrm{AUC} & = \mathrm{P}[S_D \le S_N],\\
\mathrm{AR} & = 2\,\mathrm{P}[S_D \le S_N] - 1 + \frac{p}{1-p}\,\mathrm{P}[S_D = S'_D].
\end{align*}
\end{proposition}
\textbf{Proof.} The equation for AUC follows from application of Fubini's theorem to the right-hand side
of \eqref{eq:AUC}. Observe that in general
\begin{subequations}
\begin{align}\label{eq:decomp}
	2\,\mathrm{P}[S_D \le S'_D] & = 1 + \mathrm{P}[S_D = S'_D] \\
	\intertext{and therefore}
	\mathrm{P}[S_D \le S'_D] - 1/2 & = \mathrm{P}[S_D = S'_D]/2.\label{eq:diff}
\end{align}
\end{subequations}
Inserting this last identity into \eqref{eq:ARgeneral} yields the equation for AR. \hfill $\Box$

\begin{corollary}\label{co:AUC_AR_gen}
Define AUC and AR by \eqref{eq:AUC} and \eqref{eq:AR}, respectively, with ROC and CAP as given in 
\eqref{eq:ROC} and \eqref{eq:CAP}. Let $S_D$ and $S_{D'}$ denote independent random variables 
with distribution function $F_D$.
In case of a powerless classification system (i.e.\ $F_D = F_N$) 
we then have
\begin{equation}\label{eq:powerless_AR_gen}
\begin{split}
	\mathrm{AUC} & = 1/2 + \mathrm{P}[S_D = S'_D]/2,\\
	\mathrm{AR} & = \frac{\mathrm{P}[S_D = S'_D]}{1-p}.
\end{split}
\end{equation}
In case of a perfect classification system (i.e.\ there is a score value $s_0$ such that $F_D(s_0) =1, F_N(s_0)=0$)
we have
\begin{subequations}
\begin{align}
	\mathrm{AUC} & = 1 \label{eq:perfect_AUC_gen}\\
\intertext{and, if $p > 0$,} 
	 \mathrm{AR} & = 1 + \frac{p}{1-p}\,\mathrm{P}[S_D = S'_D]. \label{eq:perfect_AR_gen}
\end{align}
\end{subequations}
\end{corollary}
When corollary \ref{co:AUC_AR_gen} is compared to corollary \ref{co:AUC_AR}, it becomes clear 
that definitions \eqref{eq:ROC} and \eqref{eq:CAP} are unsatisfactory when it comes to 
calculate AUC and AR for powerless or perfect score functions with potential discontinuities. In 
particular, AUC and AR of powerless score functions then will not equal any longer 50\% and 0, respectively.
AR of a perfect score function can even be greater than 100\% when calculated for a score function with
discontinuities.

Definitions \eqref{eq:ROC} and \eqref{eq:CAP} of ROC and CAP curves, however, can be modified in a way
such that proposition \ref{pr:AR_AUC} and corollary \ref{co:AUC_AR} obtain without the assumption 
that the score function is continuous.

\begin{definition}[Modified ROC and CAP functions]\label{de:modROC}\ \\
\begin{subequations}
Denote by $F_N$ and $F_D$ the distribution functions of the survivor scores and the defaulter scores
respectively. Let $S_D$ be a random variable with distribution function $F_D$. The 
\emph{Modified Receiver Operating Characteristic function} $\mathrm{ROC}^\ast(u)$ then is defined by
\begin{align}\label{eq:ROC_mod}
	\mathrm{ROC}^\ast(u) & = \mathrm{P}\bigl[S_D < F_N^{-1}(u)\bigr] + \mathrm{P}\bigl[S_D = F_N^{-1}(u)\bigr]/2, \quad u \in [0,1].
\intertext{With $F$ denoting the unconditional distribution function of the scores, the 
\emph{Modified Cumulative Accuracy Profile function} $\mathrm{CAP}^\ast(u)$ is defined by}
\label{eq:CAP_mod}
	\mathrm{CAP}^\ast(u) & = \mathrm{P}\bigl[S_D < F^{-1}(u)\bigr] + \mathrm{P}\bigl[S_D = F^{-1}(u)\bigr]/2, \quad u \in [0,1].
\end{align}
\end{subequations}
\end{definition}
In general, we have 
\begin{equation*}
	\mathrm{ROC}^\ast(u) \le \mathrm{ROC}(u) \quad \text{and}\quad \mathrm{CAP}^\ast(u) \le \mathrm{CAP}(u), \quad
u \in [0,1].	
\end{equation*}
Compare the two panels of figure \ref{fig:4} for illustration.
If, however, the distribution function $F_D$ of the defaulter scores is continuous, 
\eqref{eq:ROC_mod} and \eqref{eq:ROC} are equivalent, and so are 
\eqref{eq:CAP_mod} and \eqref{eq:CAP} because
\begin{equation}
\begin{split}
	 \mathrm{ROC}(u) & = \mathrm{P}\bigl[S_D < F_N^{-1}(u)\bigr] + \mathrm{P}\bigl[S_D = F_N^{-1}(u)\bigr],\\
	 \mathrm{CAP}(u) & = \mathrm{P}\bigl[S_D < F^{-1}(u)\bigr] + \mathrm{P}\bigl[S_D = F^{-1}(u)\bigr].
\end{split}
\end{equation}
The following modified definitions of AUC and AR obviously coincide with the unmodified concepts of
AUC and AR from definition \ref{de:AUC&AR} when the underlying score distributions are continuous.
\begin{definition}[Modified area under the curve and modified accuracy ratio]\label{de:AUCmod}\ \\
For the function $\mathrm{ROC}^\ast$ given by \eqref{eq:ROC_mod} we define the \emph{modified area under the curve} 
$\mathrm{AUC}^\ast$ by
\begin{subequations}
\begin{align}\label{eq:AUC_mod}
\mathrm{AUC}^\ast & = \int_0^1 \mathrm{ROC}^\ast(u)\, d u.\\	
\intertext{For the function $\mathrm{CAP}^\ast$ given by \eqref{eq:CAP_mod} we define the 
\emph{modified accuracy ratio} $\mathrm{AR}^\ast$ by}
\mathrm{AR}^\ast & = \frac{2}{1-p}\, \Bigl(\int_0^1 \mathrm{CAP}^\ast(u) \, d u - 1/2\Bigr).\label{eq:AR_mod}
\end{align}
\end{subequations}
\end{definition}
Clearly, we have $\mathrm{AUC}^\ast \le \mathrm{AUC}$ and $\mathrm{AR}^\ast \le \mathrm{AR}$. 
The advantage of definition \ref{de:AUCmod} compared to definition \ref{de:AUC&AR} is that it gives
us versions of proposition \ref{pr:AR_AUC} and corollary \ref{co:AUC_AR} that obtain without
any continuity requirements on the score distributions.
\begin{proposition}\label{pr:AUCast}
Define $\mathrm{AUC}^\ast$ and $\mathrm{AR}^\ast$ by \eqref{eq:AUC_mod} and \eqref{eq:AR_mod}, respectively, with 
$\mathrm{ROC}^\ast$ and $\mathrm{CAP}^\ast$ as given in 
\eqref{eq:ROC_mod} and \eqref{eq:CAP_mod}. Let $S_D$ and $S_N$ denote independent random variables that
have the distribution of the scores conditional on default and on survival respectively. 
Then we obtain
\begin{subequations}
\begin{align}\label{eq:AUCstar}
\mathrm{AUC}^\ast & = \mathrm{P}[S_D < S_N] + \mathrm{P}[S_D = S_N]/2,\\[1ex]
\mathrm{AR}^\ast & = 2\,\mathrm{P}[S_D < S_N] + \mathrm{P}[S_D = S_N] - 1 = \mathrm{P}[S_D < S_N] - \mathrm{P}[S_D > S_N].
\end{align}
\end{subequations}
In particular, $\mathrm{AR}^\ast = 2\,\mathrm{AUC}^\ast - 1$ holds. 
\end{proposition}
\textbf{Proof.} By application of Fubini's theorem, obvious from the definitions 
of $\mathrm{AUC}^\ast$ and $\mathrm{AR}^\ast$. \hfill $\Box$

Note that \eqref{eq:AUCstar} by some authors \citep[e.g.][equation (12)]{Newson2001} is used
as definition of the area under the ROC curve.
\begin{corollary}\label{co:AUC_AR_mod}
In case of a powerless classification system (i.e.\ $F_D = F_N$) 
we have
\begin{equation}\label{eq:powerless_AR_mod}
\begin{split}
	\mathrm{AUC}^\ast & = 1/2,\\
	\mathrm{AR}^\ast & = 0.
\end{split}
\end{equation}
In case of a perfect classification system (i.e.\ there is a score value $s_0$ such that $F_D(s_0) =1, F_N(s_0)=0$)
we have
\begin{subequations}
\begin{align}
	\mathrm{AUC}^\ast & = 1 \label{eq:perfect_AUC_mod}\\
\intertext{and, if $p > 0$,} 
	 \mathrm{AR}^\ast & = 1. \label{eq:perfect_AR_mod}
\end{align}
\end{subequations}
\end{corollary}
Corollary \ref{co:AUC_AR_mod} gives a clear indication that for general score distributions
definition \ref{de:AUCmod} should be preferred to definition \ref{de:AUC&AR}. For the latter definition leads
to the results from corollary \ref{co:AUC_AR_gen} that are counter-intuitive in case of discontinuous
score distributions. In section \ref{sec:sample}, we will show that in the
settings of examples \ref{ex:dis} and \ref{ex:sample} definition \ref{de:AUCmod} also can be interpreted
in graphical terms.


\subsubsection{Examples: Rating distributions and empirical score distributions}
\label{sec:sample}

In this section, we look at examples \ref{ex:dis} and \ref{ex:sample} in more detail.
Observe first that both examples can be described in the same more general terms.

\textbf{Assumption G:} There is a finite number of states $z_1 < z_2 < \ldots < z_\ell$ such that
\begin{subequations}
\begin{equation}\label{eq:G}
\mathrm{P}[S_D \in \{z_1, \ldots,z_\ell\}] \ =\ 1\ =\ \mathrm{P}[S_N \in \{z_1, \ldots,z_\ell\}].
\end{equation}
Define for $i = 1, \ldots, \ell$
\begin{equation}\label{eq:probs}
	\begin{split}
		\mathrm{P}[S_D = z_i] & \ = \ \pi_i,\\
		\mathrm{P}[S_N = z_i] & \ = \ \omega_i.
	\end{split}
\end{equation}
\textbf{Convention:}
\begin{equation}
	z_0 \ = \ - \infty.
\end{equation}
To avoid redundancies in the notation we assume that 
\begin{equation}\label{eq:positive}
\pi_i + \omega_i > 0\ \text{for}\ i \ge 1.	
\end{equation}%
\end{subequations}

Choose $\ell = n$ and $z_i = i$ to see that \eqref{eq:G} (assumption G) is satisfied in the 
setting of example \ref{ex:dis}. Then it is obvious how to determine the probabilities $\pi_i$ and
$\omega_i$.

In case of example \ref{ex:sample} choose $\ell$ to be the 
number of elements of the set (combined sample) $\{x_1, \ldots, x_{n_D}, y_1, \ldots, y_{n_N}\}$
and $z_i$ as the $i$-th element of the ordered list of the different elements of the set. In this 
case we will have $1 \le \ell \le n_D + n_N$. The lower extreme case will occur when both the 
defaulter score sample and the survivor score sample are constant and have the same value. This 
seems unlikely to happen in practice. The greater limit for $\ell$ will be assumed when 
all the values in both the defaulter score and the survivor score samples are pairwise different.
This will occur even with probability one if both conditional score distributions are 
continuous.

For the probabilities $\pi_i$ and $\omega_i$ in \eqref{eq:probs}, in the setting of example 
\ref{ex:sample} we obtain
\begin{equation}\label{eq:probs_sample}
	\begin{split}
		\pi_i & \ = \ \delta_{z_i}(x_1, \ldots, x_{n_D}) - \delta_{z_{i-1}}(x_1, \ldots, x_{n_D}),\\
		\omega_i & \ = \ \delta_{z_i}(y_1, \ldots, y_{n_N}) - \delta_{z_{i-1}}(y_1, \ldots, y_{n_N}).
	\end{split}
\end{equation}

\paragraph{ROC, ROC$^\ast$, AUC, and AUC$^\ast$.}
Under \eqref{eq:G} (assumption G), the
ROC and ROC$^\ast$ functions according to \eqref{eq:ROC} and \eqref{eq:ROC_mod} can be described
more specifically as follows:
\begin{subequations}
\begin{align}
	\mathrm{ROC}(u) & = \begin{cases}
	0, & \text{if}\ 0 = u,\\
	\sum_{j=1}^i \pi_j, & \text{if}\
				\sum_{j=1}^{i-1} \omega_j < u \le 
				\sum_{j=1}^{i} \omega_j\\ 
	& \text{for}\ 1 \le i \le \ell.
			\end{cases}\label{eq:ROC_sample}\\
	\mathrm{ROC}^\ast(u) & = \begin{cases}
	0, & \text{if}\ 0 = u,\\
	\pi_i/2 + \sum_{j=1}^{i-1} \pi_j, & \text{if}\
				\sum_{j=1}^{i-1} \omega_j < u \le 
				\sum_{j=1}^{i} \omega_j\\ 
	& \text{for}\ 1 \le i \le \ell.\label{eq:ROCast_sample}
			\end{cases}
\end{align}
\end{subequations}
\begin{remark}\label{rm:redundant}
Observe that equations \eqref{eq:ROC_sample} and \eqref{eq:ROCast_sample}
can be become redundant to some extent in so far as the intervals on
their right-hand sides may be empty. This will happen in particular 
in the context of example \ref{ex:sample} whenever 
the samples $x_1, \ldots, x_{n_D}$ and $y_1, \ldots, y_{n_N}$ are disjoint.
Let 
$\widetilde{y}_1 < \ldots < \widetilde{y}_{k_N}$ be the ordered elements of the 
set $\{y_1, \ldots, y_{n_N}\}$ of survivor scores. Define $\widetilde{y}_0 = - \infty$.
More efficient versions of \eqref{eq:ROC_sample} and \eqref{eq:ROCast_sample} then can be
stated as
\begin{align*}
	\mathrm{ROC}(u) & = \begin{cases}
	0, & \text{if}\ 0 = u,\\
	\delta_{\widetilde{y}_k}(x_1, \ldots, x_{n_D}), & \text{if}\
				\delta_{\widetilde{y}_{k-1}}(y_1, \ldots, y_{n_N}) < u \le 
				\delta_{\widetilde{y}_k}(y_1, \ldots, y_{n_N})\\ 
	& \text{for}\ 1 \le k \le \ell.
			\end{cases}\\
	\mathrm{ROC}^\ast(u) & = \begin{cases}
	0, & \text{if}\ 0 = u,\\
	\delta^\ast_{\widetilde{y}_k}(x_1, \ldots, x_{n_D}), & \text{if}\
				\delta_{\widetilde{y}_{k-1}}(y_1, \ldots, y_{n_N}) < u \le 
				\delta_{\widetilde{y}_k}(y_1, \ldots, y_{n_N})\\ 
	& \text{for}\ 1 \le k \le \ell.
			\end{cases}
\end{align*}	 
\end{remark}		

Under \eqref{eq:G} (assumption G) we obtain for the set $\mathrm{gROC}$ from definition \ref{de:ROC&CAPcurves} 
\begin{equation}\label{eq:gROCsample}
	\mathrm{gROC} \ =\ \left\{\begin{pmatrix}0\\0\end{pmatrix}, 
	\begin{pmatrix}\omega_1\\ \pi_1\end{pmatrix}, \begin{pmatrix}\omega_1+\omega_2\\ \pi_1+\pi_2\end{pmatrix},
	\ldots, 
	\begin{pmatrix}\sum_{j=1}^{\ell-1} \omega_j\\ \sum_{j=1}^{\ell-1} \pi_j\end{pmatrix},
	\begin{pmatrix}1\\1\end{pmatrix}\right\}.
\end{equation}
Under assumption \eqref{eq:positive}, the points in $\mathrm{gROC}$ will be pairwise different. Hence there won't
be any redundancy in the representation \eqref{eq:gROCsample} of $\mathrm{gROC}$.  

As both the graphs of the $\mathrm{ROC}$ and the $\mathrm{ROC}^\ast$ functions 
as specified by \eqref{eq:ROC_sample} and \eqref{eq:ROCast_sample} can obviously be
discontinuous at $u = 0, u=\omega_1, \ldots$,
$u=\sum_{j=1}^{\ell-1} \omega_j$, in practice 
\citep[see, e.g.,][]{Newson2001, Fawcett2004, Engelmannetal03b} they are often replaced by
the linearly interpolated graph through the points of the set $\mathrm{gROC}$ as given by
\eqref{eq:gROCsample} (in the order of the points as listed there).
\begin{subequations}
\begin{proposition} Under \eqref{eq:G} (assumption G),
the area in the Euclidean plane enclosed by the x-axis, the vertical line through $x=1$ and
the graph defined by linear interpolation of the ordered point set $\mathrm{gROC}$ as given by
\eqref{eq:gROCsample} equals $\mathrm{AUC}^\ast$ as defined by \eqref{eq:AUC_mod} and
\eqref{eq:ROCast_sample}. Moreover, $\mathrm{AUC}^\ast$ can be calculated as
\begin{equation}\label{eq:AUCstar_gen}
	\mathrm{AUC}^\ast \ = \ 1/2 \sum_{i=1}^\ell \omega_i\,\pi_i + 
						\sum_{i=2}^\ell \omega_i \sum_{j=1}^{i-1} \pi_j.
\end{equation}
\end{proposition}
\textbf{Proof.}
\citet[][section III.1.2]{Engelmannetal03b} showed that the area under the interpolated ROC curve
equals $\mathrm{AUC}^\ast$ as represented by \eqref{eq:AUCstar}. Equation \eqref{eq:AUCstar_gen} 
follows immediately from \eqref{eq:AUCstar} and \eqref{eq:probs}. \hfill $\Box$

Still under \eqref{eq:G} (assumption G), it is easy to see that AUC from definition \ref{de:AUC&AR},
i.e.\ the ``continuous'' version of the area under the curve, can be calculated as
\begin{equation}\label{eq:AUC_gen}
	\mathrm{AUC} \ = \ \sum_{i=2}^\ell \omega_i \sum_{j=1}^{i} \pi_j\quad \ge\quad  \mathrm{AUC}^\ast.
\end{equation}
\end{subequations}
Observe that $\mathrm{AUC} = \mathrm{AUC}^\ast$ if and only if $\sum_{i=1}^\ell \omega_i\,\pi_i = 
\mathrm{P}[S_D = S_N] = 0$.
\begin{subequations}
\begin{remark}
In the specific setting of example \ref{ex:sample}, the representation of $\mathrm{ROC}^\ast(u)$
from remark \ref{rm:redundant} implies 
\begin{equation}\label{eq:MW}
	\mathrm{AUC}^\ast = (n_D\,n_N)^{-1} \sum_{i=1}^{n_D} \sum_{j=1}^{n_N} \delta^\ast_{y_j}(x_i).
\end{equation}
The right-hand side of \eqref{eq:MW} is up to the factor $n_D\,n_N$ 
identical to the statistic of the Mann-Whitney test on whether a distribution
is stochastically greater than another distribution \citep[see, e.g.,][]{Engelmannetal03b}. 
By means of the representation
of $\mathrm{ROC}(u)$ from remark \ref{rm:redundant}, it is not either hard to
show that
\begin{equation}\label{eq:AUCMW}
	\mathrm{AUC}\ =\ (n_D\,n_N)^{-1} \sum_{i=1}^{n_D} \sum_{j=1}^{n_N} \delta_{y_j}(x_i). 
\end{equation}
Clearly, $\mathrm{AUC}^\ast = \mathrm{AUC}$ if and only if the samples 
$x_1, \ldots, x_{n_D}$ and $y_1, \ldots, y_{n_N}$ are disjoint.
\end{remark}
\end{subequations}

\paragraph{CAP, CAP$^\ast$, AR, and AR$^\ast$.}
Recall from \eqref{eq:PDunconditional} that $p$ stands for the unconditional probability of default\footnote{%
\label{fn:p}{}In example \ref{ex:dis}, the value of $p$ is a model parameter that can be chosen as it is convenient.
In contrast, in example \ref{ex:sample} a natural (but not necessary) choice for the value of $p$ is $p = \frac{n_D}{n_D+n_N}$.}.
Under \eqref{eq:G} (assumption G), \eqref{eq:probs} therefore implies that $\mathrm{P}[S=z_i] =
p\,\pi_i + (1-p)\,\omega_i$. With this in mind, the following representations of $\mathrm{CAP}(u)$ and
$\mathrm{CAP}^\ast(u)$ are obvious:
\begin{subequations}
\begin{align}
	\mathrm{CAP}(u) & = \begin{cases}
	0, & \text{if}\ 0 = u,\\
	\sum_{j=1}^i \pi_j, & \text{if}\
				\sum_{j=1}^{i-1} \bigl(p\,\pi_j + (1-p)\,\omega_j\bigr) < u \le 
				\sum_{j=1}^{i} \bigl(p\,\pi_j + (1-p)\,\omega_j\bigr)\\ 
	& \text{for}\ 1 \le i \le \ell.
			\end{cases}\label{eq:CAP_sample}\\
	\mathrm{CAP}^\ast(u) & = \begin{cases}
	0, & \text{if}\ 0 = u,\\
	\pi_i/2 + \sum_{j=1}^{i-1} \pi_j, & \text{if}\
				\sum_{j=1}^{i-1} \bigl(p\,\pi_j + (1-p)\,\omega_j\bigr) < u \le 
				\sum_{j=1}^{i} \bigl(p\,\pi_j + (1-p)\,\omega_j\bigr)\\ 
	& \text{for}\ 1 \le i \le \ell.\label{eq:CAPast_sample}
			\end{cases}
\end{align}
\end{subequations}
Note that thanks to assumption \eqref{eq:positive} the redundancy issue mentioned in remark \ref{rm:redundant}
will not occur for representations\footnote{%
For more efficient calculations of $\mathrm{CAP}(u)$ or $\mathrm{CAP}^\ast(u)$ in the setting of
example \ref{ex:sample} nonetheless the observation might
be useful that $\sum_{j=1}^{i} \bigl(p\,\pi_j + (1-p)\,\omega_j\bigr) = 
\delta_{z_i}(x_1, \ldots, x_{n_D}, y_1, \ldots, y_{n_N})$ if $p$ is chosen as suggested in
footnote~\ref{fn:p}.} \eqref{eq:CAP_sample} and \eqref{eq:CAPast_sample}.

Under \eqref{eq:G} (assumption G) we obtain for the set $\mathrm{gCAP}$ from definition \ref{de:ROC&CAPcurves} 
\begin{equation}\label{eq:gCAPsample}
	\mathrm{gCAP} \ =\ \left\{\begin{pmatrix}0\\0\end{pmatrix}, 
	\begin{pmatrix}p\,\pi_1+(1-p)\,\omega_1\\ \pi_1\end{pmatrix}, 
	\ldots, 
	\begin{pmatrix}\sum_{j=1}^{\ell-1}\bigl(p\,\pi_j + (1-p)\,\omega_j\bigr)\\ \sum_{j=1}^{\ell-1} \pi_j\end{pmatrix},
	\begin{pmatrix}1\\1\end{pmatrix}\right\}.
\end{equation}
As the both the graphs of the $\mathrm{CAP}$ and the $\mathrm{CAP}^\ast$ functions 
as specified by \eqref{eq:CAP_sample} and \eqref{eq:CAPast_sample} are obviously
discontinuous at $u = 0, u=p\,\pi_1+(1-p)\,\omega_1, \ldots$,
$u=\sum_{j=1}^{\ell-1}\bigl(p\,\pi_j + (1-p)\,\omega_j\bigr)$, in practice 
\citep[see, e.g.,][]{Engelmannetal03b} they are often replaced by
the linearly interpolated graph through the points of the set $\mathrm{gCAP}$ as given by
\eqref{eq:gCAPsample} (in the order of the points as listed there).
\begin{subequations}
\begin{proposition}
Under \eqref{eq:G} (assumption G), the ratio of 1) the area in the Euclidean plane enclosed by 
the line $x=y$, the vertical line through $x=1$ and
the graph defined by linear interpolation of the ordered point set $\mathrm{gCAP}$ as given by
\eqref{eq:gCAPsample} and 2) the area enclosed by the line $x=y$, the vertical line through $x=1$ and
the CAP$^\ast$ curve of a perfect score function 
equals $\mathrm{AR}^\ast$ as defined by \eqref{eq:AR_mod} and
\eqref{eq:CAPast_sample}. Moreover, $\mathrm{AR}^\ast$ can be calculated as
\begin{equation}\label{eq:ARsample}
	\mathrm{AR}^\ast = \sum_{i=1}^\ell \omega_i\,\pi_i + 
						2\,\sum_{i=2}^\ell \omega_i \sum_{j=1}^{i-1} \pi_j - 1.
\end{equation}
\end{proposition}
\textbf{Proof.}
As in \citet[][section III.1.2]{Engelmannetal03b} one can show that the area under the interpolated CAP curve
equals $\mathrm{P}[S_D < S]+\mathrm{P}[S_D = S]/2$ where $S_D$ and $S$ are independent random variables with the
empirical distribution of the scores conditional on default and the unconditional empirical score distribution, respectively. 
If $S_N$ denotes a further independent random variable, with the distribution of the scores conditional on survival,
and $S'_D$ is an independent copy of $S_D$, 
this observation implies that 
\begin{align*}
	\text{Ratio of the areas 1) and 2)} & = \frac{\mathrm{P}[S_D < S]+\mathrm{P}[S_D = S]/2 -1/2}{1-p/2-1/2}\\
	& = \frac 2{1-p}\,\bigl(p\,\mathrm{P}[S_D < S'_D]+(1-p)\,\mathrm{P}[S_D < S_N]\\ 
	& \qquad\qquad + \mathrm{P}[S_D = S'_D]/2+
	(1-p)\,\mathrm{P}[S_D = S_N]/2\bigr) \\
	& = 2\,\mathrm{P}[S_D < S_N] + \mathrm{P}[S_D = S_N] -1.
\end{align*}
By proposition \ref{pr:AUCast}, this implies the first part of the assertion. 
\eqref{eq:ARsample} then is an immediate consequence of \eqref{eq:AUCstar_gen} and 
proposition \ref{pr:AUCast} once again. \hfill $\Box$

Still under \eqref{eq:G} (assumption G), by proposition \ref{pr:AUCgeneral} one can conclude 
that AR from definition \ref{de:AUC&AR},
i.e.\ the ``continuous'' version of the accuracy ratio, can be calculated as
\begin{equation}\label{eq:AR_gen}
	\mathrm{AR} \ = \ 2\,\sum_{i=2}^\ell \omega_i \sum_{j=1}^{i} \pi_j - 1 +
			\frac p{1-p}\,\sum_{i=1}^\ell \pi_i^2\quad >\quad  \mathrm{AR}^\ast.
\end{equation}
The ``$>$'' on the right-hand side of \eqref{eq:AR_gen} is implied by \eqref{eq:G} (i.e.\ at
least one $\pi_i$ is positive).
\end{subequations}

\begin{subequations}
\begin{remark}
In the specific setting of example \ref{ex:sample}, equation \eqref{eq:ARsample} is equivalent to
\begin{equation}
	\mathrm{AR}^\ast = \frac 2{n_D\,n_N} \sum_{i=1}^{n_D} \sum_{j=1}^{n_N} \delta^\ast_{y_j}(x_i) - 1.
\end{equation}
If $p = \frac{n_D}{n_D+n_N}$, by combining proposition \ref{pr:AUCgeneral} and \eqref{eq:AUCMW} one can also calculate AR for the setting of
example \ref{ex:sample}, i.e.\ a representation equivalent to \eqref{eq:AR_gen}:
\begin{equation}
	\mathrm{AR} \ = \ \frac 2{n_D\,n_N} \sum_{i=1}^{n_D} \sum_{j=1}^{n_N} \delta_{y_j}(x_i) - 1 + 
		\frac{1}{n_D\,n_N}\,\sum_{i=1}^{n_D} \sum_{j=1}^{n_D} \delta_{x_j}(x_i)\,\delta_{x_i}(x_j).
\end{equation}
Note that $\mathrm{AR} > \mathrm{AR}^\ast$ even if the samples $x_1, \ldots, x_{n_D}$ and $y_1, \ldots, y_{n_N}$ are disjoint.
This follows from $\sum_{i=1}^{n_D} \sum_{j=1}^{n_D} \delta_{x_j}(x_i)\,\delta_{x_i}(x_j) \ge \sum_{i=1}^{n_D} 1 = n_D > 0$.
\end{remark}
\end{subequations}


\section{Discriminatory power: Numerical aspects}
\label{sec:aspects}

\citet{Engelmannetal03a, Engelmannetal03b} compared for different sample sizes
approximate normality-based and bootstrap
	confidence intervals for AUC. 
	As they worked with a huge dataset of defaulter and non-defaulter scores,
	they treated the estimates on the whole dataset as ``true'' values -- an assumption confirmed by tight 
	confidence intervals. \citeauthor{Engelmannetal03a} then sub-sampled from the dataset to study the impact
	of smaller sample sizes. Their conclusion -- for scores both on continuous and discrete scales --
	was that even for defaulter samples of size ten the approximate
	and bootstrap intervals do not differ much and cover the ``true'' value.
	
	After having presented some general considerations on the impact on bootstrap performance by 
	sample size in sub-section \ref{sec:bootstrap}, in 
	sections \ref{sec:cont_experiments} and \ref{sec:finite} we supplement the observations of 
	\citeauthor{Engelmannetal03a} in a simulation study\footnote{%
	 Like \citet{Engelmannetal03a, Engelmannetal03b} we compare approximate normality-based and bootstrap
	confidence intervals for AUC. \citet{Newson2006} describes how jackknife methods can be applied
	to estimate confidence intervals for Somers'~D (and hence in particular for AUC).} where we
	sample from known analytical distributions. This way, we really know the true value of AUC and can
	determine whether or not the true value is covered by a confidence interval. Additionally, we study
	the impact of having an even smaller sample size of five defaulters. 
	
	Note that by proposition \ref{pr:AUCast} any conclusion on estimation uncertainty for AUC$^\ast$ also
	applies to AR$^\ast$.
	

\subsection{Bootstrap confidence intervals when the default sample size is small}
\label{sec:bootstrap}

\citet[][section 2.3]{DavisonHinkley} commented on the question of how large the sample size should be 
in order to generate meaningful bootstrap samples. \citeauthor{DavisonHinkley} observed that
if the size of the original sample is $n$ 
the number of different bootstrap samples that can be generated from this sample is no larger than 
$\binom{2\,n-1}{n}$. Table \ref{tab:1} shows the value of this term for the first eleven positive
integers.
When following the general recommendation by \citeauthor{DavisonHinkley} 
to generate at least 1000 bootstrap samples,
according to table \ref{tab:1} then beginning with $n=7$ it is possible not to have any identical (up to
permutations) samples. For sample size six and below the sample variation will be restricted 
for combinatorial reasons. This applies even more to samples on a 
discrete scale which in most cases include ties.
One should therefore expect that bootstrap intervals for AUC become less reliable when the size of the
defaulter score sample is six or less or when the sample includes ties.
A simple simulation experiment further illustrates this observation. For two samples of size $n\in\{1, \ldots, 11\}$ 
with $n$ different elements and $n-1$ different elements respectively, 
we run\footnote{%
All calculations for this paper were conducted with R version 2.6.2 \citep{Rsoftware}.} 100 
bootstrap experiments each with 1000 iterations. In each bootstrap experiment
we count how many of the generated samples are different. 

Table \ref{tab:2} indeed clearly demonstrates that the factual sample size from bootstrapping is 
significantly smaller than the nominal bootstrap sample size when the original sample has less
than nine elements. The impact of small size of the original sample is even stronger when the original
sample includes at least one tie (two identical elements). Observe, however, that the impact of 
diminished factual sample size is partially mitigated by the fact that for combinatorial reasons 
the frequencies of duplicated bootstrap samples will have some variation.

\paragraph{Bootstrap confidence intervals.}
In sections \ref{sec:cont_experiments} and \ref{sec:finite} we calculate \emph{basic bootstrap intervals} 
generated by \emph{nonparametric bootstrap} as described in section 2.4 of 
\citet{DavisonHinkley}. Technically speaking, if the original estimate of a parameter 
(e.g.\ of AUC$^\ast$) is $t$ and we 
have a bootstrap sample $t_1^\ast \le t_2^\ast \le \ldots \le t_n^\ast$ of estimates for the same
parameter, then the basic bootstrap interval $I$ at confidence level $\gamma \in (0,1)$ is given by
\begin{equation}\label{eq:basic_bootstrap}
	I \ = \ [2\,t - t^\ast_{n\,(1+\gamma)/2},\ 2\,t - t^\ast_{n\,(1-\gamma)/2}],
\end{equation}
where we assume that $(n+1)\,(1+\gamma)/2$ and $(n+1)\,(1-\gamma)/2$ are integers in the range from $1$ to $n$.
Our standard choice of $n$ and $\gamma$ in sections \ref{sec:cont_experiments} and \ref{sec:finite} is
$n = 999$ and $\gamma = 95\%$, leading to $(n+1)\,(1+\gamma)/2 = 975$ and $(n+1)\,(1-\gamma)/2 = 25$.

\paragraph{Approximate confidence intervals for AUC$^\ast$ based on the central limit theorem.} 
Additionally, in sections \ref{sec:cont_experiments} and \ref{sec:finite} 
we calculate \emph{approximate confidence intervals} 
for AUC according to \citet[][equation (12)]{Engelmannetal03b}.

\subsection{Simulation study: Continuous score distributions}
\label{sec:cont_experiments}

We consider the normal distribution
example from section \ref{sec:normal} with the following choice of parameters:
\begin{equation}\label{eq:parameters}
	\begin{split}
		\mu_D = 6.8, &\ \sigma_D = 1.96\\
		\mu_N = 8.5, &\ \sigma_N = 2
	\end{split}
\end{equation}
These parameters are chosen such as to match the first two moments of the binomial distributions
looked at in subsequent section \ref{sec:finite}.
According to \eqref{eq:Satchell}, under the normal assumption with 
parameters as in \eqref{eq:parameters} then we have $\mathrm{AUC} = \mathrm{AUC}^\ast = 71.615\%$.

For defaulter score sample sizes $n_D \in \{5, 10, 15, 20, 25, 30, 35, 40, 45, 50\}$ and constant survivor score sample 
size $n_N = 250$, we conduct $k=100$ times the following bootstrap experiment:
\begin{enumerate}
	\item Simulate a sample of size $n_D$ of independent normally distributed defaulter scores
	and a sample of size $n_N$ of independent normally distributed survivor scores, with parameters
	as specified in \eqref{eq:parameters}. 
	\item Based on the samples from step 1) calculate estimates 
	$\mathrm{AUC}_{\mathrm{kernel}}$ according to \eqref{eq:AUC_Satchell} and \eqref{eq:replace}
	and $\mathrm{AUC}_{\mathrm{emp}}$ according to \eqref{eq:MW} for AUC.
	\item Based on the samples from step 1) and $\mathrm{AUC}_{\mathrm{emp}}$ 
	calculate the normal 95\% confidence interval $I_{\mathrm{normal}}$ 
	\citep[as described by][equation (12)]{Engelmannetal03b}.
	\item Generate for each of the two samples from step 1) $r=999$ nonparametric bootstrap samples, thus
	obtaining $r=999$ pairs of bootstrap samples. 
	\item For each pair of bootstrap
	samples associated with bootstrap trial $i = 1, \ldots, r$ calculate 
	estimates $\widehat{AUC}_i$ according to \eqref{eq:AUC_Satchell}
	and \eqref{eq:replace} as well as $\widetilde{AUC}_i$ according to \eqref{eq:MW} for AUC. 
	\item Calculate basic bootstrap 95\% confidence intervals $I_{\mathrm{kernel}}$ and
	$I_{\mathrm{emp}}$ as described in \eqref{eq:basic_bootstrap} based on the estimate $\mathrm{AUC}_{\mathrm{kernel}}$ 
	and the sample $\bigl(\widehat{AUC}_i\bigr)_{i=1, \ldots, r}$
	and the estimate $\mathrm{AUC}_{\mathrm{emp}}$ and the sample $\bigl(\widetilde{AUC}_i\bigr)_{i=1, \ldots, r}$, respectively.
	\item Check whether or not
	\begin{align*}
		\mathrm{AUC} &\in I_{\mathrm{normal}}, &\ \mathrm{AUC} &\in I_{\mathrm{kernel}}, &\ \mathrm{AUC} &\in I_{\mathrm{emp}},\\
		50\% &\in I_{\mathrm{normal}}, &\ 50\% &\in I_{\mathrm{kernel}},&\ 50\% &\in I_{\mathrm{emp}}´.
	\end{align*}
\end{enumerate}

To give an impression of the variation encountered with the different confidence interval methodologies 
and the different sample
sizes, table \ref{tab:3} (for defaulter sample sizes $n_D=5$, $n_D=25$, and $n_D=45$) 
shows the AUC estimates from the original samples and the related 
confidence interval estimates for the first five experiments. Although it is clear from the tables that the estimates are
more stable and the confidence intervals are tighter for the larger defaulter score samples, it is nonetheless hard
to conclude from these results which of the estimation methods is most efficient.

Table \ref{tab:cover_continuous} and figure \ref{fig:cover_continuous} therefore provide 
information on how often the true AUC was covered by the confidence intervals and 
how often 50\% was an element of the confidence intervals. The check of the coverage of 50\% is of interest because as
long as 50\% is included in a 95\% confidence interval for AUC, one cannot conclude that the score function or rating
system under consideration has got any discriminatory power. 

According to table \ref{tab:cover_continuous} and figure \ref{fig:cover_continuous} coverage of the true AUC
is poor for defaulter sample size $n_D \le 15$ but becomes satisfactory for the larger defaulter sample sizes. 
At the same time, the values of coverage of 50\% indicate poor power 
for defaulter sample sizes $n_D \le 20$ and much better power for defaulter sample size $n_D=25$ and larger.

For all defaulter sample sizes the coverage differences both for true AUC and for 50\% are negligible in
case of the ``empirical'' confidence intervals 
and the kernel estimation-based confidence intervals. 
For the smaller defaulter sample sizes ($n_D \le 15$), coverage of true AUC by the normal 
confidence interval is clearly better than by the ``empirical'' confidence intervals 
and the kernel estimation-based confidence intervals but still less than the nominal level of
95\%. The better coverage of true AUC by the normal confidence intervals, however, comes at the
price of a much higher coverage of 50\% for defaulter samples sizes $n_D \le 20$ (type II error).
For defaulter sample sizes $n_D \ge 25$ differences in performance of the three approaches 
to confidence intervals seem to vanish.
\begin{remark}\label{rm:wilcox}
With a view on \eqref{eq:MW}, it follows from the duality of tests and confidence intervals
\citep[see, e.g.,][theorem 9.2.2]{Casella&Berger} that
the check of whether 50\% is covered by the AUC 95\% confidence interval is equivalent to 
conducting a Mann-Whitney test of whether the defaulter score distribution and the survivor score
distribution are equal (null hypothesis). The exact distribution of the Mann-Whitney test statistic
can be calculated with standard statistical software packages. Hence the 95\% confidence
interval coverage rates of 50\% 
reported in table \ref{tab:cover_continuous} can be double-checked against 
type~II error rates from
application of the two-sided Mann-Whitney test at 5\% type~I error level. 
\end{remark}
The type II error rates mentioned in remark \ref{rm:wilcox} are displayed in the second to last
column of table \ref{tab:cover_continuous}. For the sake of completeness, in the last column of
table \ref{tab:cover_continuous} type~II error rates from application of the two-sided Kolmogorov-Smirnov
test are presented, too. Comparison of the Mann-Whitney type~II error and the coverage of 
50\% by the AUC confidence intervals clearly indicates that for defaulter sample size $n_D \le 20$
the bootstrap confidence intervals are too narrow. With a view on table \ref{tab:2} this
observation does not come as a surprise for very small defaulter sample sizes but is slighly
astonishing for a defaulter sample size like $n_D = 15$. The confidence intervals based on
asymptotic normality, however, seem to perform quite well for sample size $n_D \ge 10$.
Comparing the last column of table \ref{tab:cover_continuous} to the second-last column moreover
shows that the Mann-Whitney test is clearly more powerful for smaller defaulter sample sizes
than the Kolmogorov-Smirnov test.

In summary, the simulation results suggest that in the continuous setting of this section
for defaulter sample size $n_D \ge 20$ the performance differences between the three approaches
to AUC confidence intervals considered are negligible. For defaulter sample size $n_D < 20$, however,
with a view on the coverage of the true AUC parameter it seems clearly preferable to deploy
the confidence interval approach based on asymptotic normality
\citep[as described, e.g., by][]{Engelmannetal03a, Engelmannetal03b}
because its coverage rates come closest to the nominal confidence level (but are still smaller). For
very small defaulter sample size $n_D \le 10$, poorer coverage of the true AUC parameter may come
together with a high type II error (high coverage of 50\%, indicating misleadingly that the score
function is powerless). 

On the basis of a more intensive simulation study that includes observations on coverage rates,
thus we can re-affirm and at the same time refine the conclusion by \citet{Engelmannetal03a, Engelmannetal03b}
that confidence intervals for AUC (and AR) based on asymptotic normality work reasonably well for sample data on 
a continuous scale, even for small defaulter sample size like $n_D = 10$ but not necessarily
for a very small  defaulter sample size like $n_D = 5$. Moreover, for defaulter sample sizes $n_D < 20$
the asymptotic normality confidence interval estimator out-performs bootstrap-based estimators.


\subsection{Simulation study: Discrete score distributions}
\label{sec:finite}

We consider the binomial distribution
example for 17 rating grades from figure \ref{fig:3} with probability parameter $p_D=0.4$ 
for the defaulter rating distribution and probability parameter $p_N = 0.5$ for 
the survivor rating distribution. As a consequence, the first two moments of 
the defaulter rating distribution match the first two moments of the defaulter
score distribution from section \ref{sec:cont_experiments} and the first two moments 
of the survivor rating distribution match the first two moments of the survivor
score distribution from section \ref{sec:cont_experiments}.
Moreover, also the discriminatory power of the fictitious rating system considered in this 
section is almost equal to the discriminatory power of the score function from 
section \ref{sec:cont_experiments} (AUC$^\ast$~71.413\% according to \eqref{eq:AUCstar_gen}
vs.\ AUC~71.615\%).

To assess the impact of the discreteness of the model, we conduct the same simulation 
exercise as in section \ref{sec:cont_experiments} but replace step 1) by step 1$^\ast$) which reads
\begin{itemize}
	\item[1$^\ast$)] Simulate a sample of size $n_D$ of independent binomially distributed defaulter ratings
	and a sample of size $n_N$ of independent binomially distributed survivor ratings, with 
	probability parameters $p_D=0.4$ and $p_N=0.5$ respectively.
\end{itemize}

As in section \ref{sec:cont_experiments}, to give an impression of the 
variation encountered with the different confidence interval methodologies 
and the different sample
sizes, table \ref{tab:4} (for defaulter sample sizes $n_D=5$, $n_D=25$, and $n_D=45$) 
shows the AUC estimates from the original samples and the related 
confidence interval estimates for the first five experiments. Although it is clear from the tables that the estimates are
more stable and the confidence intervals are tighter for the larger defaulter score samples, it is nonetheless hard
to conclude from these results which of the estimation methods is most efficient. Interesting is
also the result from experiment number 5 for sample size $n_D =5$ in table \ref{tab:4} which with
lower confidence bounds of 90.0\% and more looks very 
much like an outlier due to a defaulter sample concentrated at the bad end of the rating scale.

Table \ref{tab:cover_discrete} and figure \ref{fig:cover_discrete} provide 
information on how often the true AUC was covered by the confidence intervals and 
how often 50\% was an element of the confidence intervals. 
In contrast to table \ref{tab:cover_continuous} and figure \ref{fig:cover_continuous},
table \ref{tab:cover_discrete} and figure \ref{fig:cover_discrete} do not give a very
clear picture of the performance of the three AUC estimation approaches on the 
rating data. While coverage of 50\% (type II error) is high for defaulter sample
sizes smaller than $n_D = 30$, coverage of 50\% reaches very small values as in the continuous
case of section \ref{sec:cont_experiments} for larger defaulter sample sizes. Presumably due
to the relatively small number of 100 bootstrap experiments -- which already requires some 
hours of computation time --, according to figure \ref{fig:cover_discrete} 
there is some variation and not really a clear trend in
the level of coverage of the true AUC parameter. Even for a relatively high defaulter
sample size of $n_D = 40$ there is sort of a collapse of coverage of true AUC with 
percentages of 90\% or lower. For defaulter sample size of $n_D = 45$ or more there might
be some stabilisation at a satisfactory level.

As in the continuous case, 
for all defaulter sample sizes the coverage differences both for true AUC and for 50\% are negligible in
case of the ``empirical'' confidence intervals 
and the kernel estimation-based confidence intervals. 
For the smaller defaulter sample sizes ($n_D \le 20$), coverage of true AUC by the normal 
confidence interval is clearly better than by the ``empirical'' confidence intervals 
and the kernel estimation-based confidence intervals but still less than the nominal level of
95\%. The better coverage of true AUC by the normal confidence intervals, however, comes at the
price of a much higher coverage of 50\% for defaulter samples sizes $n_D \le 15$ (type II error).
For defaulter sample sizes $n_D \ge 25$ differences in performance of the three approaches 
to confidence intervals seem to vanish.
\begin{remark}\label{rm:fisher}
Remark \ref{rm:wilcox} essentially also applies to the setting of this section.
But take into account that in the presence of ties in the sample 
the equivalence between $\mathrm{AUC}^\ast$ as defined by \eqref{eq:AUC_mod} and
the Mann-Whitney statistic only holds when ranks for equal elements of the 
ordered total sample are assigned as \emph{mid-ranks}. With this in mind
we can double-check the 95\% confidence coverage rates of 50\% 
reported in table \ref{tab:cover_discrete} against type~II error rates from
application of the two-sided Mann-Whitney test at 5\% type~I error level 
in the same manner as we have done for remark \ref{rm:wilcox}.
\end{remark}
The type II error rates\footnote{%
Exact p-values for the Mann-Whitney test on samples with ties were calculated
with the function \emph{wilcox$\underline{\ }$test} from the R-software
package \emph{coin}.} mentioned in remark \ref{rm:fisher} are reported in the second to last
column of table \ref{tab:cover_discrete}. We have presented type~II error rates 
from application of the two-sided Kolmogorov-Smirnov test in the last column
of table \ref{tab:cover_continuous}. Due to the massive presence of ties in the  
discrete-case samples, however, application of the Kolmogorov-Smirnov test does not seem
appropriate in this section. Instead, we report type~II error rates 
from application of the two-sided exact Fisher test\footnote{%
The p-values of Fisher's
exact test have been calculated with the function \emph{fisher.test} (R-software package
\emph{stats}) in simulation mode due to too high memory and time requirements 
of the exact mode.} \citep[see, e.g.,][]{Weisstein}
in the last column of table \ref{tab:cover_discrete}. 

Again, comparison of the Mann-Whitney type~II error and the coverage of 
50\% by the AUC confidence intervals clearly indicates that for defaulter sample size $n_D \le 20$
the bootstrap confidence intervals are too narrow. With another view on table \ref{tab:2} this
is even less a surprise than in the continuous case. The confidence intervals based on
asymptotic normality, however, seem again to perform quite well for sample size $n_D \ge 10$.
Comparing the last column of table \ref{tab:cover_continuous} to the second-last column moreover
shows that the Mann-Whitney test is clearly more powerful for smaller and even some
moderate defaulter sample sizes than Fisher's test.

In summary, the simulation results suggest that in the discrete setting of this section
for defaulter sample size $n_D \ge 25$ the performance differences between the three approaches
to AUC confidence intervals considered are negligible. For defaulter sample size $n_D \le 20$, however,
with a view on the coverage of the true AUC parameter it seems clearly preferable to deploy
the confidence interval approach based on asymptotic normality
\citep[as described, e.g., by][]{Engelmannetal03a, Engelmannetal03b}
because its coverage rates come closest to the nominal confidence level (but are still smaller). For
very small defaulter sample size $n_D \le 10$, poorer coverage of the true AUC parameter may come
together with a high type II error (high coverage of 50\%, indicating misleadingly that the score
function is powerless). 

On the basis of this more intensive simulation study that includes observations on coverage rates,
thus we can re-affirm and at the same time refine 
the interesting conclusion by \citet{Engelmannetal03a, Engelmannetal03b} that confidence
intervals for AUC (and AR) based on asymptotic normality work reasonably (when
compared to other approaches) for sample data on 
a discrete scale, even for small defaulter sample size like $n_D = 10$ (but not necessarily
for smaller defaulter sample sizes). Bootstrap estimators are out-performed for
such small sample sizes by the asymptotic normality estimator. While the performance of
the three AUC confidence interval methods does not seem to be much worse for discrete scale
rating distributions than for continuous scale score distributions, there might be 
a higher likelihood of performance outliers -- as discussed at the beginning of section \ref{sec:bootstrap}.


\section{Determining PD curves by parametric estimation of ROC and CAP curves}
\label{sec:parametric}

In the context of portfolios with little default data, \citet{VanDerBurgt} suggested estimating the conditional
probability of default by fitting a one-parameter curve to the observed CAP curve, taking the derivative of 
the fitted curve and then calculating the conditional probabilities of default based on the
derivative. \Citeauthor{VanDerBurgt} did not provide much background information on why he chose the
specific one-parameter family of curves he used in the paper nor did he look more closely at
the properties of this family of curves.

In section \ref{sec:derivatives}, we provide some background information
on the implicit assumptions and implications when
working with parametric approaches for CAP and ROC functions.
In section \ref{sec:burgt}, we discuss \citeauthor{VanDerBurgt}'s approach in detail and introduce
three potential alternatives. In section \ref{sec:performance} we compare the performance of the four
approaches by looking at some numerical examples.

Parametric approaches (e.g.\ logit or van der Burgt's approaches) to PD curves are popular with practitioners because they can be designed such as to guarantee monotonicity of the conditional PD estimates. 
With an appropriate choice of the parametric
shape, it is also possible to replicate the exponential-like growth that is observed for corporate
default rates associated with agency ratings. However, as discussed in \citet[][section 4]{Tasche2008a},
monotonicity of PD curves must not be taken for granted. We will see in section \ref{sec:performance} that
both the erroneous assumption of monotonicity and the mis-specification of discriminatory power can
cause huge estimation errors when following a parametric approach to PD curve estimation. See \cite{Pluto&Tasche}
for a non-parametric approach that might be a viable alternative in particular 
when little default observation is available.

\subsection{Derivatives of CAP and ROC curves}
\label{sec:derivatives}

It is a well known fact that there is a close link between ROC and CAP curves on the one hand and conditional
probabilities of default on the other hand. Technically speaking, the link is based on the following
easy-to-prove (when making using of theorem \ref{pr:functionROC}) observation.

\begin{proposition}\label{pr:derivative}
Let $F_D$ and $F_N$ be distribution functions on an open interval $I \subset \mathbb{R}$. Assume that
$F_D$ has a density $f_D$ which is continuous on $I$ and that $F_N$ has a positive density $f_N$ 
that is continuous on $I$. Let $0 < p < 1$ be a fixed probability and define the mixed distribution
$F$ by \eqref{eq:unconditional}. Write $f$ for the density of $F$. Define $\mathrm{ROC}(u)$ and
$\mathrm{CAP}(u)$, $u \in (0,1)$ by \eqref{eq:ROC} and \eqref{eq:CAP}, respectively. Then both 
ROC and CAP are continuously differentiable for $u \in (0,1)$ with derivatives
\begin{subequations}
\begin{align}
	\mathrm{ROC}'(u) &\ = \ \frac{f_D\bigl(F_N^{-1}(u)\bigr)}{f_N\bigl(F_N^{-1}(u)\bigr)}, \\
	\mathrm{CAP}'(u) &\ = \ \frac{f_D\bigl(F^{-1}(u)\bigr)}{p\,f_D\bigl(F^{-1}(u)\bigr) + 
								(1-p)\,f_N\bigl(F^{-1}(u)\bigr)} \notag\\
					& \ =\ \frac{f_D\bigl(F^{-1}(u)\bigr)}{f\bigl(F^{-1}(u)\bigr)}. \label{eq:CAPderivative}	
\end{align}
\end{subequations}
\end{proposition}

Proposition \ref{pr:derivative} is of high interest in the context of individual default risk
analysis because -- in the notation of sections \ref{sec:model} and \ref{sec:framework}
-- the probability of default conditional on a score value $s$ is given by \eqref{eq:pd_continuous}.
Proposition \ref{pr:derivative} then immediately implies
\begin{subequations}
\begin{align}
	\mathrm{P}[D\,|\,S=s] &\ = \ \frac{p\,\mathrm{ROC}'\bigl(F_N(s)\bigr)}
			{p\,\mathrm{ROC}'\bigl(F_N(s)\bigr) + 1-p}\label{eq:condPD_ROC} \\
	 &\ = \ p\,\mathrm{CAP}'\bigl(p\,F_D(s)+(1-p)\,F_N(s)\bigr)\notag\\
	&\ = \ p\,\mathrm{CAP}'\bigl(F(s)\bigr). \label{eq:condPD_CAP}
\end{align}
\end{subequations}
Note that by \eqref{eq:condPD_CAP}, the derivative of a differentiable CAP curve for a borrower population
with unconditional probability of default $p>0$ is necessarily bounded from above by $1/p$. The following 
theorem shows on the one hand that this condition is not only a necessary but also a sufficient condition for a
distribution function on the unit interval to be a CAP curve. On the other hand, the theorem shows 
that a CAP curve
relates not only to one combination of conditional and unconditional score distributions but provides 
a link between conditional and unconditional score distributions which applies to an infinite number of
such combinations.

\begin{theorem}\label{th:chCAP}
Let $p \in (0,1)$ be a fixed probability. Let $f_D \ge 0$ be a density on $\mathbb{R}$ such that the 
set $I = \{f_D > 0\}$ is an open interval and $f_D$ is continuous in $I$. Denote by 
$F_D(s) = \int_{-\infty}^s f_D(v)\,d v$ the distribution function associated with $f_D$. Then the following two statements
are equivalent:
\begin{itemize}
	\item[(i)] The function $u \mapsto C(u)$, $u \in (0,1)$ is continuously differentiable in $u$ with 
	$\lim\limits_{u\to 0} C(u)= 0$, $\lim\limits_ {u \to 1} C(u) = 1$, 
	and $0 < C'(u) \le 1/p$, $u \in (0,1)$.
	\item[(ii)] There is a density $f_N \ge 0$ such that $\{f_N > 0\} = I$, $f_N$ is continuous in $I$
	and
\begin{equation}\label{eq:C}
	C(u) \ = \ F_D\bigl(F^{-1}(u)\bigr), \quad u \in (0,1),
\end{equation}
 	where $F(s) = p\,F_D(s) + (1-p)\,\int_{-\infty}^s f_N(v)\,d v$.
\end{itemize}
\end{theorem}
\textbf{Proof.}\\[1ex] 
\textbf{(i) $\Rightarrow$ (ii):} By assumption, $C$ maps $(0,1)$ onto $(0,1)$ and 
	the inverse $C^{-1}$ of $C$ exists. Define 
	$F(s) = C^{-1}\bigl(F_D(s)\bigr)$. Then $F$ is a  distribution function
	with $\lim\limits_{s \to \inf I} F(s)=0$, $\lim\limits_{s \to \sup I} F(s)=1$ and 
	density 
\begin{equation}\label{eq:f}
	f(s) \ = \ F'(s) \ = \ \frac{f_D(s)}{C'\bigl(F(s)\bigr)}, \quad s\in I.
\end{equation}
	Observe that $f(s)$ is positive and continuous in $I$. Hence the inverse $F^{-1}$ of $F$ exists. Let	
\begin{align}
	F_N(s) & \ = \ \frac{F(s) - p\,F_D(s)}{1-p}\quad s \in \mathbb{R}, \quad\text{and}\notag\\
	f_N(s) & \ = \ f_D(s)\,\frac{1/C'\bigl(F(s)\bigr)-p}{1-p}, \quad s\in I.
	\label{eq:non_default}
\end{align}
	By \eqref{eq:f}, then $f_N$ is the continuous derivative of $F_N$ and is positive in $I$ by assumption on $C'$
	and $f_D$. 
	This implies that
	$F_N$ is a distribution function with $\lim\limits_{s \to \inf I} F_N(s)=0$, 
	$\lim\limits_{s \to \sup I} F_N(s)=1$ and density $f_N$. By construction of $F$ and $F_N$, the functions
	$C$, $F_D$, and $F$ satisfy \eqref{eq:C}.\\[1ex]
\textbf{(ii) $\Rightarrow$ (i):} By construction, $F_D(s)$ and $F(s)$ are distribution functions which
	converge to 0 for $s \to \inf I$ and to 1 for $s \to \sup I$. This implies the limit statements for $C$. Equation
	\eqref{eq:C} implies that $C$ is continuously differentiable with derivative
\begin{equation*}
	0 \ < \ C'(u) \ =\ \frac{f_D\bigl(F^{-1}(u)\bigr)}{p\,f_D\bigl(F^{-1}(u)\bigr) + (1-p)\,f_N\bigl(F^{-1}(u)\bigr)}
	\ \le \ 1/p.
\end{equation*}
\rightline{$\Box$}
For the sake of completeness, we provide without proof the result corresponding to theorem \ref{th:chCAP} for ROC curves.
In contrast to the case of CAP curves, essentially every continuously differentiable and strictly 
increasing distribution function on the 
unit interval is the ROC curve for an infinite number of combinations of 
score distributions conditional of default and survival respectively.
\begin{proposition}\label{pr:chROC}
\textbf{}Let $f_D \ge 0$ be a density on $\mathbb{R}$ such that the 
set $I = \{f_D > 0\}$ is an open interval and $f_D$ is continuous in $I$. Denote by 
$F_D(s) = \int_{-\infty}^s f_D(v)\,d v$ the distribution function associated with $f_D$. Then the following two statements
are equivalent:
\begin{itemize}
	\item[(i)] The function $u \mapsto R(u)$, $u \in (0,1)$ is continuously differentiable in $u$ with 
	$\lim\limits_{u\to 0} R(u)= 0$, $\lim\limits_ {u \to 1} R(u) = 1$, 
	and $0 < R'(u)$, $u \in (0,1)$.
	\item[(ii)] There is a density $f_N \ge 0$ such that $\{f_N > 0\} = I$, $f_N$ is continuous in $I$
	and
\begin{equation}\label{eq:R}
	R(u) \ = \ F_D\bigl(F_N^{-1}(u)\bigr), \quad u \in (0,1),
\end{equation}
 	where $F_N(s) = \int_{-\infty}^s f_N(v)\,d v$.
\end{itemize}
\end{proposition}
The basic idea both with theorem \ref{th:chCAP} and proposition \ref{pr:chROC} is that 
if in the functional equation $f(x) = g(h^{-1}(x))$ two of the three functions $f$, $g$ and $h$ are given then
the third can be calculated by solving the equation for it. In the cases of ROC and CAP curves, matters can get
more complicated because the involved functions are not necessarily invertible. This would entail some technicalities
when trying to solve $f(x) = g(h^{-1}(x))$ for $g$ or $h$. However, to relate conditional probabilities of
default to ROC and CAP functions via \eqref{eq:condPD_ROC} and \eqref{eq:condPD_CAP} we need the existence of 
densities. This introduces some degree of smoothness as can be seen from 
theorem \ref{th:chCAP} and proposition \ref{pr:chROC}. Both the theorem and the proposition could also be stated with fixed 
distribution $F_N$ of the survivor scores. However, the survivor score distribution appears in the CAP function
only as a mixture with the defaulter score distribution. Therefore, stating theorem 
\ref{th:chCAP} with given survivor score distribution would no longer be  straight-forward
and the proof would involve the
implicit function theorem. As the additional insight by such a version of theorem \ref{th:chCAP} 
would be limited, in this paper the formulation of the theorem as provided above has been preferred.

\subsection{Van der Burgt's approach and alternatives}
\label{sec:burgt}

The one-parameter curve proposed by \citet{VanDerBurgt} for estimating CAP functions is
\begin{subequations}
\begin{equation}\label{eq:C.kappa}
	C_\kappa(u) \ = \ \frac{1-e^{-\kappa\,u}}{1- e^{-\kappa}}, \quad u \in [0,1],
\end{equation}
where $\kappa \in \mathbb{R}$ is the fitting parameter. The function $C_\kappa$ is obviously 
a distribution function on $[0,1]$. 
Moreover, for positive $\kappa$ the graph of $C_\kappa$ is concave as one might expect from the CAP curve of a score function
that assigns low scores to bad borrowers and high scores to good borrowers. For $\kappa \to 0$ the graph of $C_\kappa$
converges toward the diagonal line, i.e.\ the graph of a powerless score function.
The derivative of $C_\kappa$ and $\mathrm{AR}_\kappa$ associated with $C_\kappa$ according to \eqref{eq:AR} are easily computed as
\begin{align}
	C'_\kappa(u) & \ = \ \frac{\kappa\,e^{-\kappa\,u}}{1-e^{-\kappa}}, \label{eq:h_deriv}\\
	\mathrm{AR}_\kappa & \ = \ \frac 2{1-p}\left(\frac 1{1-e^{-\kappa}} - \frac 1 \kappa - 1/2\right).\label{eq:ARkappa}
\end{align}
\end{subequations}
In \eqref{eq:ARkappa} the parameter $p>0$ denotes the unconditional probability of default of the estimation sample in
the sense of section \ref{sec:approach}.
Observe from \eqref{eq:h_deriv} that for $\kappa > 0$
\begin{subequations}
\begin{equation}
	C'_\kappa(1) \ =\ \frac{\kappa}{1-e^{-\kappa}}\,e^{-\kappa} \ \le\ 
	C'_\kappa(u) \ \le \ \frac{\kappa}{1-e^{-\kappa}} \ =\ C'_\kappa(0), \quad u \in [0,1].
\end{equation}
Theorem \ref{th:chCAP} hence implies 
\begin{equation}
	\kappa \ < \ \frac{\kappa}{1-e^{-\kappa}} \ \le \ \frac 1 p.
\end{equation}
\end{subequations}
\begin{subequations}
Given a CAP curve $\mathrm{CAP}(u)$ to be approximated, in the setting of a 
continuous score function a natural approach to finding 
the best fit $\widehat{\kappa}$ would be a least squares procedure as the following
\begin{equation}\label{eq:ls}
\begin{split}
	\widehat{\kappa} & \ = \ \arg\min_{\kappa > 0} \int_0^1 \bigl(\mathrm{CAP}(u)-C_\kappa(u)\bigr)^2 d u \\
	& \ = \ \arg\min_{\kappa > 0} \mathrm{E}\left[\bigl(F_D(S)-C_\kappa(F(S))\bigr)^2\right].
\end{split}
\end{equation}
In practice, the integration in \eqref{eq:ls} would have to be replaced by a sample mean.
Alternatively,
\citet{VanDerBurgt} suggested inferring $\kappa$ by means of \eqref{eq:ARkappa} from an estimated\footnote{%
This requires that there is an estimate of $p$. \Citeauthor{VanDerBurgt} assumes $p \approx 0$ for the
purpose of estimating $\kappa$ but then, in a further step, makes use of the fact that $p$ is positive.} AR
(or via 
proposition \ref{pr:AR_AUC} from an estimated AUC). Assuming that an estimate of the unconditional 
probability $p$ is available, probabilities of default conditional on realised score values then can
be estimated via \eqref{eq:condPD_CAP}:
\begin{equation}\label{eq:VanDerBurgt_condPD}
		\mathrm{P}[D\,|\,S = s]  \ = \ \frac{p\,\kappa\,e^{-\kappa\,F(s)}}{1-e^{-\kappa}}.
\end{equation}
\end{subequations}
\Citet{VanDerBurgt}, however, applied the methodology to a discrete setting as described in example
\ref{ex:dis} (rating system with $n$ grades). 
In the notation of example \ref{ex:dis} van der Burgt's approach to finding the best fit parameter
$\widehat{\kappa}$ can be described as
\begin{subequations}
\begin{equation}\label{eq:unweighted}
\begin{split}
	\widehat{\kappa} & \ = \ \arg\min_{\kappa > 0} 
				\sum_{j=1}^n \bigl(\mathrm{P}[R_D \le j] - C_\kappa(\mathrm{P}[R \le j])\bigr)^2 \\
	& \ = \ \arg\min_{\kappa > 0} \mathrm{E}\left[\frac{\bigl(F_D(R)-C_\kappa(F(R))\bigr)^2}
		{\mathrm{P}[R=r]\bigm|_{r=R}}\right].
\end{split}
\end{equation}
Van der Burgt's approach hence can be regarded as sort of an unweighted regression in which the same
weights are given to rating grades with very few observations and grades with quite a lot of observations.
For calculating the estimates of the conditional probabilities of default \citet{VanDerBurgt} does not
deploy equation \eqref{eq:VanDerBurgt_condPD} but a modification that substitutes the 
unconditional score distribution function $F$ by the mean of its right and left continuous 
versions $(F + F(\,\cdot\, - 0))/2$:
\begin{equation}\label{eq:VanDerBurgt_mod}
		\mathrm{P}[D\,|\,R = j]  \ = \ \frac{p\,\kappa\,
		\exp\bigl(-\kappa\,(\mathrm{P}[R < j]+\mathrm{P}[R \le j])/2\bigr)}{1-e^{-\kappa}}.
\end{equation}
\end{subequations}
In his paper, \citet{VanDerBurgt} does not spend much time with explaining the why and how of his approach.
It is tempting to guess that the approach was more driven by the results than by theoretical considerations.
We observe that it is not obvious how to deploy van der Burgt's regression approach \eqref{eq:unweighted}
for a sample of scores from a score function with continuous scale. Therefore, 
for our example calculations in section \ref{sec:performance} we will make use of \eqref{eq:ls}
and \eqref{eq:VanDerBurgt_condPD} for the continuous setting and of \eqref{eq:unweighted} and \eqref{eq:VanDerBurgt_mod}
for the discrete setting of example \ref{ex:dis}.

In general, when choosing $C_\kappa$ for fitting a CAP curve, one should be aware that as a consequence 
of theorem \ref{th:chCAP} this choice implies
some structural links between the score distribution of the defaulters and the score distribution of 
the survivors. This is illustrated in figure \ref{fig:VanDerBurgtDensities} which shows for unconditional
probability of default $p = 0.01$ and different
values of $\kappa$ the survivor score densities that are implied by theorem \ref{th:chCAP} when the defaulter
score density is assumed to be standard normal. Clearly, for large $\kappa$ and, by \eqref{eq:ARkappa}, 
high discriminatory power the implied survivor score distributions are not normal as they are not symmetric.

This observation on the one hand might be considered not very appealing. On the other hand, it suggests
an alternative approach along the lines of section \ref{sec:normal} which provides in \eqref{eq:Satchell}
a two-parametric representation of the ROC function for the case of normally distributed defaulter
and survivor score distributions. 

As mentioned in section \ref{sec:normal}, no closed form is available
for the CAP function in case of normally distributed defaulter and survivor score distributions. This is one
reason why we consider in the following how to approximate general ROC curves (not CAP curves)
by the ROC function of the normal family as described in section \ref{sec:normal}. Another reason is 
that, in general, fitting ROC curves is conceptually sounder than fitting CAP curves 
because this way one can better separate the estimation of an unconditional probability of default
from the estimation of parameters of the fitting function.

By substituting in \eqref{eq:Satchell} the parameter $b> 0$ for $\sigma_N / \sigma_D$ and the parameter
$a\in\mathbb{R}$ for $\frac{\mu_N - \mu_D}{\sigma_D}$, we obtain a two-parametric family of ROC functions:
 \begin{subequations}
\begin{equation}\label{eq:normalROC}
	R_{a, b}(u) \ = \ \Phi\bigl(a + b\,\Phi^{-1}(u)\bigr), \quad u \in (0,1).
\end{equation}
From this, it follows readily that
\begin{align}\label{eq:Rab}
	R'_{a, b}(u) &\ = \ b\,\frac{\varphi\bigl(a + b\,\Phi^{-1}(u)\bigr)}{\varphi\bigl(\Phi^{-1}(u)\bigr)},\\
	\mathrm{AR}_{a, b} &\ = \ 2\,\Phi\bigl(\frac{a}{\sqrt{b^2+1}}\bigr) - 1.\label{eq:ARab}	
\end{align}
\end{subequations}
Clearly, a two-parameter family of functions will give better fits than a one-parameter family of functions as 
the one-parameter family proposed by \citet{VanDerBurgt}. In order to have a fair comparison, 
therefore, in the following we will focus on the one-parameter sub-family
of \eqref{eq:normalROC} specified by fixing $b$ at $b=1$. We simplify notation by writing $R_a$ for $R_{a, 1}$.
From section \ref{sec:normal} it follows that the one-parameter family of ROC functions $R_a$ includes, 
in particular, the ROC curves for normally distributed defaulter and survivor score functions when their variances
are equal. Equations \eqref{eq:Rab} and \eqref{eq:ARab} are simplified significantly for $R_a$:
\begin{subequations}
	\begin{align}\label{eq:Ra}
	R'_a(u) &\ = \ e^{-a\,\Phi^{-1}(u) - a^2/2},\\
	\mathrm{AR}_a &\ = \ 2\,\Phi\bigl(a/\sqrt{2}\bigr) - 1.\label{eq:ARa}	
\end{align}
\end{subequations}
When the unconditional probability of default $p$ is known, \eqref{eq:Ra} via \eqref{eq:condPD_ROC} implies
the following representation of the probability of default conditional on a realised score value:
\begin{subequations}
\begin{equation}\label{eq:PD_deriv}
	\mathrm{P}[D\,|\,S = s]  \ = \ \frac{1}
	{1 + \frac{1 - p}p\,\exp\bigl(a\,\Phi^{-1}(F_N(s))+a^2/2\bigr)}.
\end{equation}
Clearly, \eqref{eq:PD_deriv} can be rewritten as
\begin{align}\label{eq:PD_rewritten}
	\mathrm{P}[D\,|\,S = s]  &\ = \ \frac{1}
	{1 + \exp\bigl(\alpha + \beta\,\Phi^{-1}(F_N(s))\bigr)}\\
	\intertext{with}
	\alpha & \ = \ \log\bigl(\dfrac{1 - p}p\bigr) + a^2/2,\quad \beta \ = \ a.\notag
\end{align}
\end{subequations}
Thus, the conditional PDs derived from the one-parameter ROC approximation approach 
\eqref{eq:normalROC} with $b=1$ are the conditional PDs of a logit regression where
the default indicator is regressed on the explanatory variable $\Phi^{-1}(F_N(S))$.
In the special case where the score distribution conditional on survival is normal (i.e.\ 
$F_N(s) = \Phi\bigl((s-\mu)/\sigma\bigr)$ for some suitable constants
$\mu$ and $\sigma$), the right-hand side of 
equation \eqref{eq:PD_rewritten} coincides
with the conditional PDs of the common logit approach:
\begin{equation}\label{eq:PD_logit}
	\mathrm{P}[D\,|\,S = s]  \ = \ \frac{1}
	{1 + \exp\bigl(\alpha + \beta\,s\bigr)}.
\end{equation}
Thus \eqref{eq:PD_rewritten} can be considered a \emph{robust logit approach} that
replaces regression on the original score $S$ by regression on the transformed
score $\Phi^{-1}(F_N(S))$ to account for the fact that the score distribution might 
not be normal. As such, the suggestion by \citet{VanDerBurgt} leads to a potentially
quite useful modification of logit regression in the univariate case.

On an estimation sample in the sense of section \ref{sec:estimation}, parameters 
for logit-type raw conditional PDs as specified in equations \eqref{eq:PD_rewritten}
and \eqref{eq:PD_logit} can be estimated by maximum likelihood (MLE) procedures 
\citep[see, e.g.,][chapter 3]{Cramer2003}. In the case of \eqref{eq:PD_rewritten}, MLE will 
only work if $0 < F_N(x_i) < 1$ and $0 < F_N(y_j) < 1$ for all scores $x_i$ (defaulters)
and $y_j$ (survivors) in the estimation sample. This will not be the case if $F_N$ is
estimated as the empirical distribution function of the survivor sample $y_j$. To work around
this issue, the empirical distribution can be modified 
(as described in section \ref{sec:performance}). Another approach could be non-linear 
least squares estimation:
\begin{equation}
	(\widehat{\alpha}, \widehat{\beta}) \ = \ 
	\arg\min_{\alpha, \beta \in \mathbb{R}} \mathrm{E}\left[\bigl(\mathbf{1}_D-
	\bigl\{1 + \exp\bigl(\alpha + \beta\,\Phi^{-1}(F_N(S))\bigr)\bigr\}^{-1}\bigr)^2\right],
\end{equation}
where $\mathbf{1}_D = 1$ for defaulted borrowers and $\mathbf{1}_D = 0$ otherwise.

\Citet[][equation (5)]{VanDerBurgt} suggested inferring the value of parameter $\kappa$
specifying his CAP curve approximation \eqref{eq:C.kappa} from an estimate of AUC. This
idea can be used to derive another approach to the estimation of the parameters in 
\eqref{eq:PD_rewritten} or \eqref{eq:PD_logit}. To infer the values of two parameters, 
two equations are needed. A natural choice for the first equation is to equate a target value
$q$ for the unconditional PD and the mean of the conditional PDs:
\begin{subequations}
\begin{equation}\label{eq:first}
	q \ = \ \mathrm{E}\bigl[\mathrm{P}[D\,|\,S]\bigr].
\end{equation}
This equation can in general be used for the calibration of conditional PDs, see appendix 
\ref{sec:calibration} for details. A good choice for the second equation seems equating a target value $A$
for the area under the curve $\mathrm{AUC}^\ast$ and a 
representation of $\mathrm{AUC}^\ast$ in terms of the conditional
PDs:
\begin{equation}\label{eq:second}
	A \ = \ \frac{\mathrm{E}\left[ \left(\mathrm{E}\bigl[\mathrm{P}[D\,|\,S]\,\mathbf{1}_{\{S < s\}}\bigr]\bigm|_{s=S} 
	+ \mathrm{P}[S=s]\bigm|_{\{s=S\}} \mathrm{P}[D\,|\,S]/2\right)\,\bigl(1-\mathrm{P}[D\,|\,S]\bigr)\right]}
	{\mathrm{E}\bigl[\mathrm{P}[D\,|\,S]\bigr]\,\bigl(1-\mathrm{E}\bigl[\mathrm{P}[D\,|\,S]\bigr]\bigr)}
\end{equation}
\end{subequations}
This representation of $\mathrm{AUC}^\ast$ follows from proposition \ref{pr:AUCast}. 
Combining equations \eqref{eq:first} and \eqref{eq:second} for the inference of parameters
can be regarded as a \emph{quasi moment matching} approach. It is ``quasi'' moment matching because 
$\mathrm{AUC}^\ast$ is not a proper moment of the conditional PDs. The most natural alternative, 
the variance of the
conditional PDs, however, depends on the proportion of defaulters in the borrower population.
As this proportion clearly varies over time it would be difficult to determine an appropriate
target variance of the conditional PDs. In contrast, $\mathrm{AUC}^\ast$ by its definition 
does not depend on the proportion of defaulters in the borrower population. It is therefore
plausible to assume that discriminatory power displays less variation over time such that its
value can be inferred from a historical estimation sample and still applies to the current portfolio. 
The following example illustrates how the quasi moment matching approach works when the logit 
shape \eqref{eq:PD_logit} is supposed to apply to the conditional PDs.

\begin{example}[Quasi moment matching for PD curve calibration]\label{ex:QMM}\ \\
Let $s_1 \le s_2 \le \ldots \le s_n$ be a sorted calibration sample of credit scores in the sense of section 
\ref{sec:forecast} (possibly the scores of the current portfolio). Assume that the PDs conditional
on the score realisations can be described by \eqref{eq:PD_logit}. The sample versions of
equations \eqref{eq:first} and \eqref{eq:second} then read:
\begin{equation}\label{eq:quasi_sample}
\begin{split}
	q & \ =\ \frac 1 n \sum_{i=1}^n \frac 1 {1 + \exp\bigl(\alpha + \beta\,s_i\bigr)}, \\
	A & \ =\ \frac{\sum_{i=1}^n \frac{\exp(\alpha + \beta\,s_i)}{1 + \exp(\alpha + \beta\,s_i)} 
	\left(\frac 1 {2\,(1 + \exp(\alpha + \beta\,s_i))} + \sum_{j=1}^{i-1} \frac 1 {1 + \exp(\alpha + \beta\,s_j)}\right)}
					{\Big(\sum_{i=1}^n \frac 1 {1 + \exp(\alpha + \beta\,s_i)}\Big)
					\Big(\sum_{i=1}^n \frac{\exp(\alpha + \beta\,s_i)}{1 + \exp(\alpha + \beta\,s_i)}\Big)}.
\end{split}					
\end{equation}
Here $q$ is the \emph{target unconditional PD} which could be estimated for instance by econometric methods
\citep[see, e.g.,][section III]{Engelmann&Porath}. The variable $A$ stands for the \emph{target discriminatory power}, expressed
as area under the curve $\mathrm{AUC}^\ast$ which can be estimated from an estimation sample in the sense of
section \ref{sec:estimation} by means of \eqref{eq:MW}.\\
Solving the equation system \eqref{eq:quasi_sample} for the parameters $\alpha$ and $\beta$ then gives 
the \emph{quasi moment matching} coefficients for the logit approach to conditional PDs.
\end{example}


\subsection{Performance comparison}
\label{sec:performance}

To illustrate the operation of van der Burgt's approach and the three logit approaches introduced in 
section \ref{sec:burgt} and to compare their performance, we get back to the example from section 
\ref{sec:cont_experiments} for the continuous score distribution case and to the example from 
section \ref{sec:finite} for the case of a discrete rating distribution. The examples, together 
with some modifications, will show that none of the four approaches is uniformly superior to the others.
To see how the estimation methods work we conduct simulation experiments with five different 
scenarios.

The following scenarios are considered:
\begin{enumerate}
	\item Rating systems with discrete scales:	
\begin{itemize}
	\item Case 1: 17 grades, binomial distribution with probability parameter 0.4 for the defaulters' rating 
	distribution, binomial distribution with probability parameter 0.5 for the survivors' 
	rating distribution (as in section \ref{sec:finite}).
	\item Case 2: 7 grades, binomial distribution with probability parameter 0.3 for the defaulters' 
	rating distribution, binomial distribution with probability parameter 0.5 for the survivors'  
	rating distribution.
\end{itemize}
	\item Score functions with continuous scales:	
\begin{itemize}
	\item Case 3: Normal distribution with mean 6.8 and standard deviation 1.96 for the defaulters' score 
	distribution, normal distribution with mean 8.5 and standard deviation 2 for the survivors' score 
	distribution (as in section \ref{sec:cont_experiments}). Means and standard deviations are chosen such
	as to match those from the above discrete case 1.
	\item Case 4: Normal distribution with mean 2.1 and standard deviation 1.12 for the defaulters' score 
	distribution, normal distribution with mean 3.5 and standard deviation 1.22 for the survivors' score 
	distribution. Means and standard deviations are chosen such
	as to match those from the above discrete case 2.
	\item Case 5: Normal distribution with mean 0.0 and standard deviation 1.25 for the defaulters' score 
	distribution, normal distribution with mean 1.0 and standard deviation 1.0 for the survivors' score 
	distribution. Means and standard deviations here are chosen such as to have a larger difference in
	standard deviations than in cases 3 and 4 and to have the standard deviation for the defaulters' score 
	distribution exceeding the standard deviation for the survivors' score distribution.
\end{itemize}
\end{enumerate}
In cases 1 and 2, when the value of the unconditional PD is known, the true conditional PDs per rating grade
can be calculated according to \eqref{eq:pd_discrete}. In cases 3, 4, and 5 the true conditional PDs per 
score value can be calculated according to \eqref{eq:pd_continuous}.

For each scenario the following simulation experiment with 1000 iterations is conducted:
\begin{enumerate}
	\item Generate an estimation sample: Rating grades / scores of 25 (results in table \ref{tab:se})
	and 50 (results in table \ref{tab:se50}) defaulters and rating grades~/ scores of
	250 survivors.
	\item Based on the estimation sample, estimates are calculated for	
\begin{itemize}
	\item discriminatory power measured by $\mathrm{AUC}^\ast$ according to \eqref{eq:MW},
	\item parameters\footnote{%
	Here $p$ is actually a constant: $p = 25/(25 + 250) = 1/11$ and $p = 50/(25 + 250) = 1/6$
	respectively.} $p$ and $\kappa$ and 
	distribution function\footnote{%
	In the continuous cases 3, 4, and 5, an estimate of $F$ is calculated according to
	\eqref{eq:unconditional}. $F_D$ and $F_N$ in \eqref{eq:unconditional} are calculated
	from normal kernel density estimates with bias-correction as described in section \ref{sec:kernel}.
	In the discrete cases 1 and 2 the standard empirical distribution function is deployed for estimating $F$.} $F$ for 
	the raw conditional PDs suggested by \citet{VanDerBurgt}, 
	where the PDs are calculated 
	according to \eqref{eq:VanDerBurgt_condPD} in the continuous cases 3, 4, and 5, and  
	according to \eqref{eq:VanDerBurgt_mod} in the discrete cases 1 and 2,
	\item parameters $\alpha$ and $\beta$ and distribution function\footnote{%
	In the continuous cases 3, 4, and 5, $F_N$ is calculated from a normal 
	kernel density estimate with bias-correction as 
	described in section \ref{sec:kernel}. In the discrete cases 1 and 2, $F_N$ is estimated as the mean
	of the right-continuous and the left-continuous versions of the empirical distribution function. 
	Additionally, to avoid numerical issues when deploying maximum likelihood estimation, whenever the 
	result would be zero it is replaced by half of the minimum positive value of the modified 
	empirical distribution function.} $F_N$ for the raw conditional PDs
	according to the robust logit approach \eqref{eq:PD_rewritten},
	\item parameters\footnote{%
	$\alpha$ and $\beta$ are estimated by the standard logit MLE procedure 
	\citep[see, e.g.,][chapter 3]{Cramer2003}.} $\alpha$ and $\beta$ 
	for the raw conditional PDs according to the logit approach
	\eqref{eq:PD_logit}.
\end{itemize}
	\item Generate then a calibration sample: Rating grades / scores of 300 
	borrowers with unknown future solvency states.
	For each of the borrowers first a default / survival simulation with PD = 2.5\% is conducted. According to
	the result then a rating grade / score is drawn from the corresponding rating / score distribution. The
	simulated solvency state is not recorded.
	\item Based on the calibration sample, the raw PDs from step 2) are calibrated to an unconditional PD\footnote{%
	Hence, we implicitly assume that we have estimated exactly the true unconditional PD of 2.5\%. Of
	course, in practice this would be unlikely. For the purpose of comparing the performance of different
	estimators this assumption is nonetheless useful.} of
	2.5\%, as described in proposition \ref{pr:calibration} from appendix \ref{sec:calibration}.
	\item Based on the calibration sample, parameters $\alpha$ and $\beta$ 
	for PDs according to the quasi moment matching approach are inferred from an unconditional PD 
	2.5\% and the $\mathrm{AUC}^\ast$ estimate from step 2), as described in example \ref{eq:quasi_sample}.
	\item Based on the calibration sample, for each rating grade / score the differences between true conditional 
	PD and the PD estimates according to the four different approaches are calculated.
	\item Based on the four samples of PD differences, the standard error SE is calculated for each of the four
	approaches according to the generic formula
\begin{equation}\label{eq:se}
	 \mathrm{SE} \ = \ \sqrt{\frac 1 n \sum_{i=1}^n \bigl(\mathrm{P}[D\,|\,S=s_i]-
	 		\widehat{\mathrm{P}}[D\,|\,S=s_i]\bigr)^2},
\end{equation}
	 where $s_1, \ldots, s_n$ is the calibration sample, $\mathrm{P}[D\,|\,S=s_i]$ is the true conditional PD, and
	 $\widehat{\mathrm{P}}[D\,|\,S=s_i]$ is the estimated conditional PD.
\end{enumerate}
Actually, running the simulation as described in steps 1) to 7) only once would provide interesting insights but would 
not help too much when it comes to a comparison of the performances of the different estimators. For illustration,
have a look at figure \ref{fig:condPD} which displays the results (true conditional PDs and estimated conditional PDs)
of one simulation run of case 1 (17 rating grades scenario). 
All four estimates seem to fit well the true conditional PDs for rating grades
5 to 17. For rating grades 1 to 4 the fit seems much worse. The van der Burgt and the robust logit estimators even assign 
constant conditional PD estimates to rating grades 1 to 3. 

Note, however, that in this simulation run there were 
no observations of rating grades 1 to 3 and 16 and 17. In so far, on the one hand, 
it is questionable whether there should be at all any conditional PD estimates for grades 1 to 3 and 16 and 17. On the other
hand, it is not surprising that also for the logit and quasi moment matching estimators the fit at grades 1 to 3 and 16 and 17
is rather poor. Given the sizes of 300 or less of 
the estimation and calibration samples that are simulated, full coverage of the rating
scale by realised rating grades can only be expected for a rating scale with a significantly lower number of grades. 
In the following, therefore, we look also at the scenario case 2 -- a rating system with 7 grades only. The probability to
observe unoccupied rating grades when the sample size is about 250 is quite low under scenario case 2.

According to the single simulation run underlying figure \ref{fig:condPD} there might not be any dramatic differences of
the performances of the different estimators. 
More detailed information can be obtained from running a number of simulations. Tables \ref{tab:se} and
\ref{tab:se50} (for defaulter scores sample sizes 25 and 50 respectively in the estimation sample) show
quantiles of the distributions of the standard errors according to \eqref{eq:se} that were observed in 1000 
iterations of the experiment.

Observations from tables \ref{tab:se} and \ref{tab:se50}:
\begin{itemize}
	\item[(i)] When comparing the quantiles of the standard error distributions as displayed in table
	\ref{tab:se50} to the results from table \ref{tab:se}, it appears that the reductions in the low quantiles
	are moderate while the reductions in the higher quantiles are significantly larger. This observation
	indicates that the higher number of defaulter scores in the estimation sample mainly has an impact on
	the variance of the standard error distributions. Note that the distributional assumptions in cases 1 to 5
	have been chosen deliberately such that exact matches by one of the estimation approaches are not
	possible. Hence the standard error rates do not converge to zero for larger defaulter score samples. Rather
	the variances of their distributions will be diminished.
	\item[(ii)] For cases 1 to 3 the logit estimator is best according to the quantiles observed
	at levels 75\% or lower. In case 4, there is no clear picture. In case 5, the
	robust logit estimator is best. In cases 1 to 4, the variance of the standard error distribution of the
	van der Burgt estimator is clearly the least.
	\item[(iii)] The error magnitude in case 5 is much higher than in the other cases. 
	This might be due to the fact that the true conditional PD curve is not monotonous as a consequence of the
	fact that there is a relatively large difference between the variances of the two conditional score distributions.
	\item[(iv)] The van der Burgt estimator is less volatile than the other estimators (exception in case 5) but has also a much
	higher minimum error level. Actually, case 5 has been defined deliberately with a higher variance of
	the defaulters score distribution in order to challenge the performance of the van der Burgt estimator. 
	For figure \ref{fig:VanDerBurgtDensities} indicates that the van der Burgt estimator can adapt to the case where
	the survivors score distribution has larger variance than the defaulters score distribution but not 
	necessarily to the opposite case. 
	\item[(v)] Performance of the quasi moment matching estimator is inferior to the performance of the logit 
	estimator but the difference is not large.
\end{itemize}
\Citet[][section 5]{VanDerBurgt} described an approach to exploring the sensitivity of the conditional PD curves
estimated by means of estimator \eqref{eq:unweighted} with regard to uncertainty in the estimation sample. 
This approach is based on the potential variation of the
``concavity'' parameter $\kappa$. Observation (v) indicates that an analogous approach can be applied to
the logit estimator by exploring the sensitivities of the  quasi moment estimates 
with respect to $\mathrm{AUC}^\ast$ and the unconditional PD. 
\begin{remark}[Use of the quasi moment matching estimator for sensitivity analysis]\label{rm:sensitivity}
We have seen in section \ref{sec:aspects} how to construct confidence intervals for the area under the curve (and
equivalently for the accuracy ratio) even in case of defaulter scores sample sizes as small as five. 
By applying the quasi moment matching estimator, we can then 
generate conditional PD curves from different values for $\mathrm{AUC}^\ast$ as indicated by
the confidence interval. Similarly, one can vary the unconditional PD which is the other input to
the quasi moment matching estimator in order to investigate the sensitivity of the conditional 
PD curves with respect to the unconditional PD. 
\end{remark}
Table \ref{tab:prob.least} displays the results of another approach to the performance comparison of the conditional PD 
estimates. The table shows for both defaulter score sample sizes of 25 and
50 and all the five scenarios introduced earlier the frequencies (in 1000 simulation iterations) 
with which the four estimation approaches produced the least standard error. Hence the entries for the different
estimators in a row of table \ref{tab:prob.least} add up to 100\%. The results from 
table \ref{tab:prob.least} re-affirm observations (ii) and (v) made on tables \ref{tab:se} and \ref{tab:se50}
in so far as they also show dominance of the logit or quasi moment matching estimators in cases 1 to 3
and of the robust logit estimator in case 5. Table \ref{tab:prob.least} however, indicates a clear
superiority of the van der Burgt estimator in case 4 where the 
results of tables \ref{tab:se} and \ref{tab:se50} are less clear.

Also shown in table \ref{tab:prob.least} (last column) are the ratios of the conditional score distribution standard
deviations for the five considered scenarios. This helps to explain the performance results.
\begin{itemize}
	\item The logit and quasi moment matching estimators stand out in cases 1 and 3 because then the standard deviations are
	nearly equal and therefore \eqref{eq:PD_logit} describes an almost exact fit of the conditional PD curve 
	\cite[][section 6.1]{Cramer2003}. Note from tables \ref{tab:se} and \ref{tab:se50} that nevertheless 
	the estimation error realised with a logit or quasi moment matching estimator can be quite large. This can
	be explained with a sensitivity analysis as described in remark \ref{rm:sensitivity}. Figure \ref{fig:QMM}
	illustrates that matching a wrong AUC-specification has quite a dramatic impact on the shape of the estimated
	conditional PD curve, with the consequence of a high standard error. Although misspecification of the target unconditional PD has
	a much weaker impact, this observation clearly reveals significant vulnerability of parametric approaches to 
	conditional PD curve estimation by their dependence on assumptions on the shape of the conditional PD curve. 
	\item The van der Burgt estimator stands out in case 4 because it adapts best to a situation where the survivor score
	variance is significantly larger than the defaulter score variance. See figure \ref{fig:VanDerBurgtDensities} for
	a graphical demonstration of this adaptation property.
	\item With a view on case 4, it is surprising that the van der Burgt estimator does not stands out in case 2 
	although the survivor score variance is also significantly larger than the defaulter score variance.
	The different approaches to the estimation of $\kappa$ that we apply in cases 2 and 4 -- \eqref{eq:unweighted}
	vs.\ \eqref{eq:ls} -- might explain this observation. Weighted least squares as in \eqref{eq:ls} presumably
	comply better with the standard error definition \eqref{eq:se} which includes implicit weighting similar
	to \eqref{eq:ls}.
	\item The robust logit estimator stands out in case 5 because it adapts best to a situation where the 
	survivor score variance is significantly smaller than the defaulter score variance. The robust logit
	estimator, however, cannot represent non-monotonous conditional PDs either. That is why the fit even by 
	this estimator in case 5 is quite poor (as shown in tables \ref{tab:se} and \ref{tab:se50}).
\end{itemize}
 
As the final observation in this simulation study table \ref{tab:corr} shows Spearman rank correlations
between the absolute errors of the $\mathrm{AUC}^\ast$ estimates on the estimation sample and
the standard errors of the conditional PD estimates on the calibration sample. Again the last column of
the table displays the ratios of the conditional score distribution standard
deviations for the five considered scenarios. Table \ref{tab:corr} demonstrates that there is a clear
relation between the two estimation errors if the variances of the conditional score distributions
are approximately equal. The less equal the variances of the conditional score distributions are, the
weaker the relation seems to be. However, the almost vanishing correlations in case~5 could also be caused by
the rather high estimation errors observed for the conditional PDs in this case. Hence, it seems
premature to draw a firm conclusion from this limited evidence.


\section{Conclusions}
\label{sec:conclusions}

In this paper, we have treated some topics that are not very closely related at first glance:
\begin{enumerate}
	\item In section \ref{sec:framework} we have looked in detail at the question of
	how to define and calculate consistently discriminatory power in terms of area under the curve (AUC)
	and accuracy ratio (AR). We have seen that there are good reasons to base the definitions of AUC
	and AR on definition \ref{de:modROC} of modified ROC and CAP curves. Section \ref{sec:sample} 
	provides reasy-to-use formulas for the estimation of AUC and AR from samples. 
	\item In section \ref{sec:aspects} we have reported the results of a simulation study which
	refined related studies by \citet{Engelmannetal03a, Engelmannetal03b} on the performance of
	confidence interval estimators for AUC. We have confirmed that the asymptotic normality confidence 
	interval estimator is most reliable. However, not surprisingly even this estimator performs 
	not very well when applied to defaulter score samples of size ten or less.
	\item In section \ref{sec:parametric} we have discussed a proposal by \citet{VanDerBurgt} to derive
	PD curve estimates by a one-parameter approach to the estimation of CAP curves. By providing background
	information, we have shown that there are some quite natural logit-related alternatives to van der Burgt's proposal.
	We have then investigated the performance of the different estimators by another simulation study. 
	The results of this study are mixed in that they demonstrate on the one hand that none of the discussed 
	estimation methods is
	uniformly best and on the other hand that, in general, by following a parametric approach one risks
	huge estimation errors caused by the implicit structural assumptions of the estimators.
\end{enumerate}
The common theme in this list is the fact that all the estimation concepts and techniques can be deployed
in an implementation of the two-phases approach to PD curve estimation as described in section \ref{sec:approach}.
In the first phase of this approach one estimates shape parameters that are essentially equivalent to
discriminatory power as expressed by AUC or AR (\citeauthor{VanDerBurgt}'s concavity parameter $\kappa$ or the parameter
$\beta$ in the logit curves from section \ref{sec:burgt}). In the second phase of the approach the 
\emph{raw PD curve} from the first phase is calibrated on the current portfolio such that the resulting
unconditional probability of default fits an independently determined target unconditional PD. The technical details
of this calibration step are described in appendix~\ref{sec:calibration}.



\appendix

\section{Appendix: Calibration of a PD curve to a target unconditional PD}
\label{sec:calibration}

We assume that for each score or rating grade $s$ an estimate $p_{D,p}(s) = \mathrm{P}_p[D\,|\,S=s]$ 
of the PD conditional on the score has been
made. The index $p$ indicates that these PDs depend on the unconditional 
PD $p = \frac{n_D}{n_D+n_N}$ where $n_D$ is the size of the defaulter estimation sample 
$x_1, \ldots, x_{n_D}$ and $n_N$ stands for the
size of the survivor estimation sample $y_1, \ldots, y_{n_N}$. The PDs $p_{D,p}(s)$ are called
\emph{raw PDs}. In section \ref{sec:parametric} we look at some parametric approaches to
the estimation of such raw PDs.

It is, however, unlikely that the unknown defaulter proportion (the actual unconditional PD) 
in a given calibration sample $s_1, \ldots,
s_n$ (possibly the current portfolio) is $p$. We assume that instead there is an estimate $\pi \not= p$ of
this unknown unconditional PD. The aim is then to find a transformation of the raw PDs evaluated on
the sample $s_1, \ldots,
s_n$ such that their mean equals $\pi$. By \eqref{eq:def_condPD} this is a necessary condition for
having unbiased estimates of the conditional PDs.

In the special cases of a rating system with a fixed number of grades $k$ (i.e.\ $S$ is a discrete 
random variable) and of a score function
with conditional score densities $f_N$ and $f_D$, we know from equations \eqref{eq:pd_discrete}
and \eqref{eq:pd_continuous} that the unconditional PD $q$ and the corresponding conditional PDs 
$p_{D,q}(s)$ satisfy the following equation:
\begin{align}\label{eq:lik.ratio}
	\lambda(s) & \ = \ \frac q {1-q}\,\frac{1-p_{D,q}(s)}{p_{D,q}(s)},\\
\intertext{where the \emph{likelihood ratio} $\lambda$ is defined as}
	\lambda(s) & \ = \ \begin{cases}
	\dfrac{f_N(s)}{f_D(s)}, & $S$ \ \text{continuous},\\[2ex]
	\dfrac{\mathrm{P}[S=s\,|\,N]}{\mathrm{P}[S=s\,|\,D]}, & $S$\ \text{discrete}.
\end{cases}\notag
\end{align}
%
As mentioned in section \ref{sec:forecast}, 
we assume that the conditional score distributions are the same in the estimation
and in the calibration sample. Then also the likelihood ratios are the same in the estimation
and in the calibration sample. Hence \eqref{eq:lik.ratio} applies both to the raw PDs with 
unconditional PD $p$ and to the conditional PDs $p_{D,\pi}(s)$ corresponding to the unconditional PD $\pi$. This 
observation implies\footnote{%
This approach seems to be common knowledge \citep[see, e.g.,][section 5.3]{OeNB&FMA}.}
\begin{align}
	p_{D,\pi}(s) &\ =\  \frac 1{1 + \frac{1-\pi}{\pi}\,\lambda(s)}\notag\\
	 &\ =\ \frac 1{1 +
	\frac{1-\pi}{\pi}\,\frac{p}{1-p}\,\frac{1-p_{D,p}(s)}{p_{D,p}(s)}}.\label{eq:pd_transform}
\end{align}
The PDs from \eqref{eq:pd_transform} often will not have the required 
property that
their mean equals $\pi$, even if the conditional score distributions for the estimation and 
the calibration samples are really the same:
\begin{equation}\label{eq:unequal}
	\pi \ \not= \ \frac 1 n \sum_{i=1}^n p_{D,\pi}(s_i).
\end{equation}
This is due to the facts 
\begin{itemize}
	\item that $\pi$ is unlikely to be an \emph{exact} forecast of the unconditional default rate and
	\item that the sample $s_1, \ldots, s_n$ is the result of randomly sampling from a
	mixture of the unconditional score 
	distributions. Hence the empirical distribution of the sample is 
	likely to be somewhat different to the theoretical unconditional score 
	distribution as presented in \eqref{eq:unconditional}.
\end{itemize}
Depending upon how much different are the conditional score distributions underlying the estimation sample
and the calibration sample respectively and how good a forecast for the true unconditional PD the
estimate $\pi$ is, the difference between the left-hand and the right-hand sides of \eqref{eq:unequal}
can be of quite different magnitudes. It can become quite large in particular if $\pi$ is not a 
point-in-time forecast of the unconditional PD but rather an estimate of a through-the-cycle central
tendency.

Whatever the magnitude of the difference is, it may be desirable to obtain equality of the both sides of
\eqref{eq:unequal} by adjusting the conditional PDs on its right-hand side. The obvious approach to this
would be to apply a constant multiplier to each of the $p_{D,\pi}(s_i)$. This approach, however, on the
one hand lacks a theoretical foundation and, on the other hand, has the disadvantage that conditional PD
values higher than 100\% may be the consequence of multiplication with a constant factor that is
possibly greater than 100\%.

In a more sophisticated approach the $p_{D,\pi}(s_i)$ on the right-hand side of \eqref{eq:unequal} are
replaced by $p_{D,q}(s_i)$ where $q$ is chosen in such a way as to match the left-hand side of \eqref{eq:unequal}
with its right-hand side. In this approach $p_{D,q}(s_i)$ is specified by \eqref{eq:pd_transform} (with $\pi$
substituted by $q$ and $s$ substituted by $s_i$). Hence $q$ is a solution of the equation
\begin{subequations}
\begin{equation}\label{eq:q_root}
	\pi \ = \ \frac 1 n \sum_{i=1}^n \frac 1{1 +
	\frac{1-q}{q}\,\frac{p}{1-p}\,\frac{1-p_{D,p}(s_i)}{p_{D,p}(s_i)}}.
\end{equation}
Recall that the raw PDs $p_{D,p}(s_i)$ are assumed to be known. It it not difficult to see that the actual
value of $p$ in the fraction $\frac{p}{1-p}$ in \eqref{eq:q_root} 
does not matter for the values of the transformed PDs because
the transformed PDs depend on $q$ and $p$ only through the term $\frac{1-q}{q}\,\frac{p}{1-p}$. 
Hence it
is sufficient to consider the simplified (case $p=1/2$ in the fraction $\frac{p}{1-p}$) equation
\begin{equation}\label{eq:unique}
	\pi \ = \ \frac 1 n \sum_{i=1}^n \frac 1{1 +
	\frac{1-q}{q}\,\frac{1-p_{D,p}(s_i)}{p_{D,p}(s_i)}}.
\end{equation}
The right-hand side of this equation is continuous and strictly increasing in $q$ and
converges toward 0 for $q \to 0$ and toward 1 for $q \to 1$. Therefore there is a unique
solution $q$ for \eqref{eq:unique}.
\end{subequations}
\begin{proposition}\label{pr:calibration}
Let $s_1, \ldots, s_n$ be a sample of scores or rating grades. Assume that an estimate $\pi \in (0,1)$ of 
the unconditional PD in $s_1, \ldots, s_n$ is given and that there is a set of raw conditional
PDs $p_D(s_1), \ldots, p_D(s_n)$ associated with $s_1, \ldots, s_n$. If at least one of
the $p_D(s_i)$ is greater than 0 and less than 1, then there is a unique
solution $q = q(\pi) \in (0,1)$ to equation \eqref{eq:unique}. The numbers
\begin{equation}
	\pi_i \ = \ \frac 1{1 +
	\frac{1-q(\pi)}{q(\pi)}\,\frac{1-p_{D}(s_i)}{p_{D}(s_i)}}, \quad i = 1, \ldots, n
\end{equation}
are called \emph{$\pi$-calibrated conditional PDs associated with the sample $s_1, \ldots, s_n$}.
\end{proposition}


\newpage

\refstepcounter{figure}
%
 Figure \thefigure:
  \emph{Score densities and ROC and CAP curves in the case of normal conditional score densities (see section 
  \ref{sec:normal}). Parameter values as in \eqref{eq:parameters}. Unconditional PD 10\%.}
\label{fig:1}
\begin{center}
\ifpdf
    \resizebox{\height}{4.0in}{\includegraphics[width=12.0in]{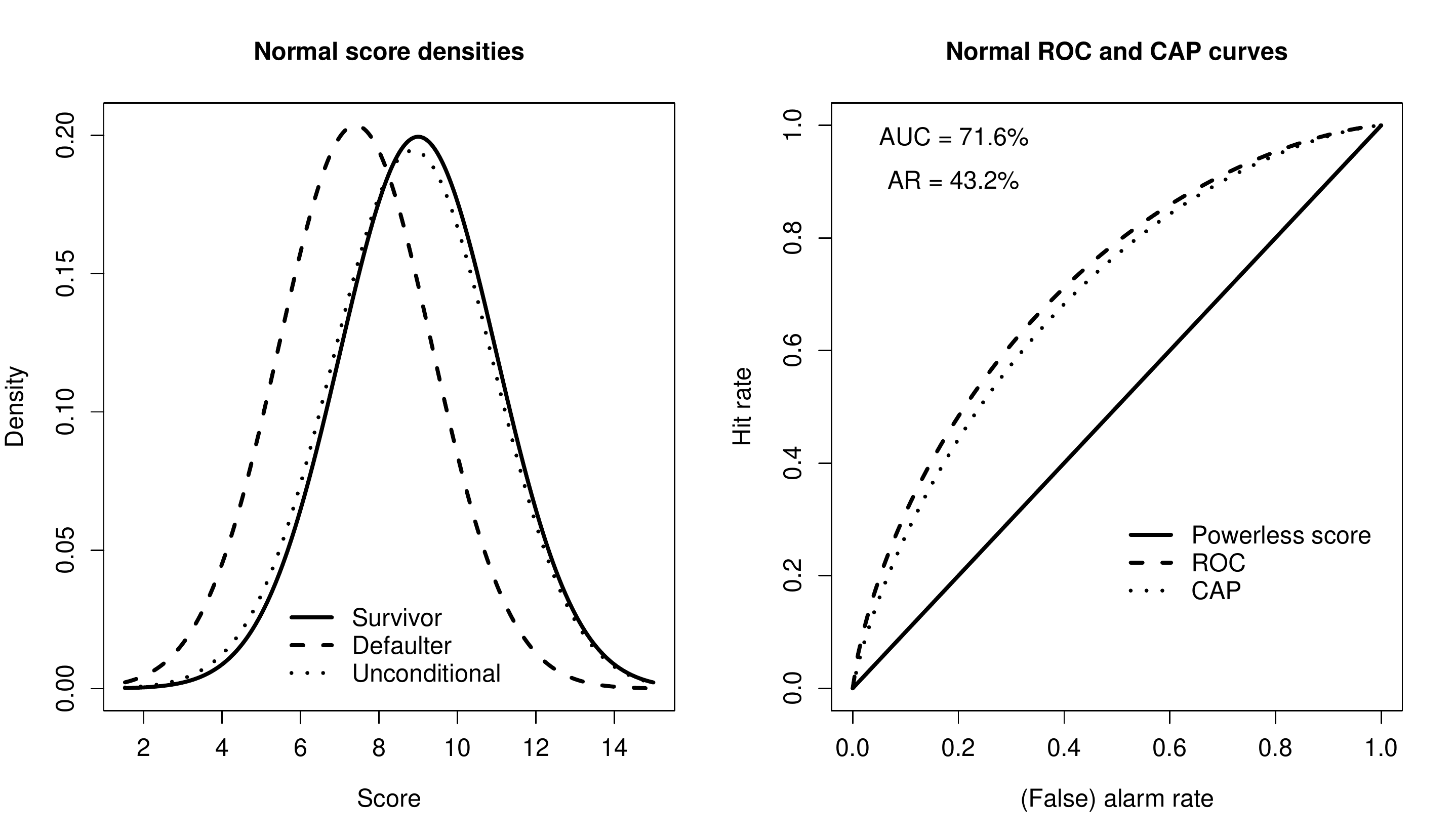}}
\else
\begin{turn}{270}
\resizebox{\height}{7.0in}{\includegraphics[width=2.5in]{NormalExample.ps}}
\end{turn}
\fi
\end{center}

\newpage
\refstepcounter{figure}
%
  Figure \thefigure:
  \emph{Non-parametric estimates (with normal kernels) of conditional score densities and ROC curve. Samples of size $n_D=5$ and
  $n_N = 250$ from normal densities as in figure \ref{fig:1}.}
\label{fig:2}
\begin{center}
\ifpdf
    \resizebox{\height}{4.0in}{\includegraphics[width=12.0in]{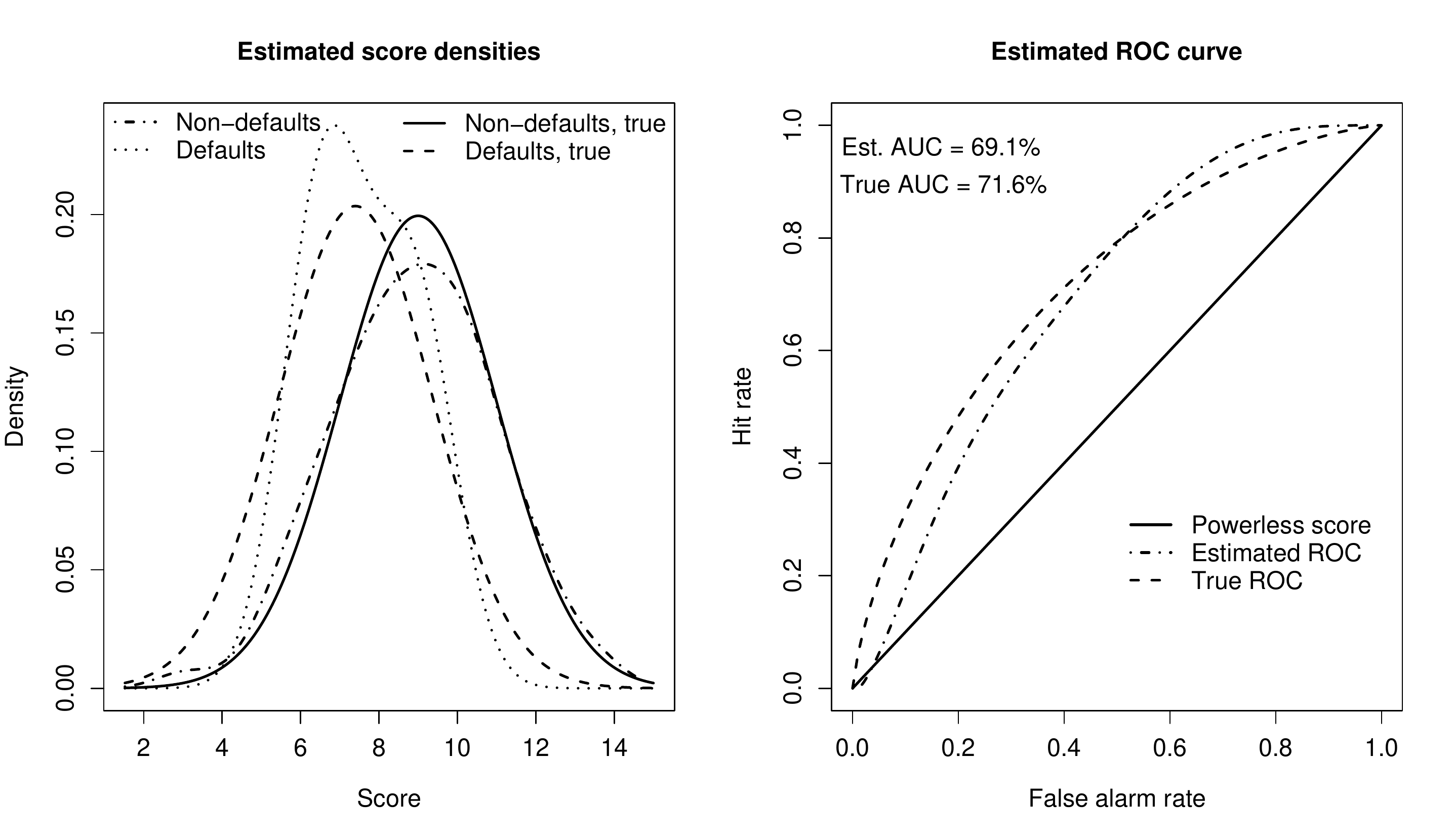}}
\else
\begin{turn}{270}
\resizebox{\height}{7.0in}{\includegraphics[width=2.5in]{DensityEstimationExample.ps}}
\end{turn}
\fi
\end{center}

\newpage
\refstepcounter{figure}
%
  Figure \thefigure:
  \emph{Fictitious conditional rating distributions for a rating system with 17 grades.\\ 
  Upper panel: Defaulters'
  distribution is binomial with success probability 40\%; survivors' distribution is binomial
  with success probability 50\%.\\
  Lower panel: Defaulters'
  distribution by sampling 5 times from defaulters' distribution from upper panel. 
  Survivors'
  distribution by sampling 250 times from survivors' distribution from upper panel. \\
  Note the different scaling of the y-axis in the two panels.}
\label{fig:3}
\begin{center}
\ifpdf
    \resizebox{\height}{6in}{\includegraphics[width=6.0in]{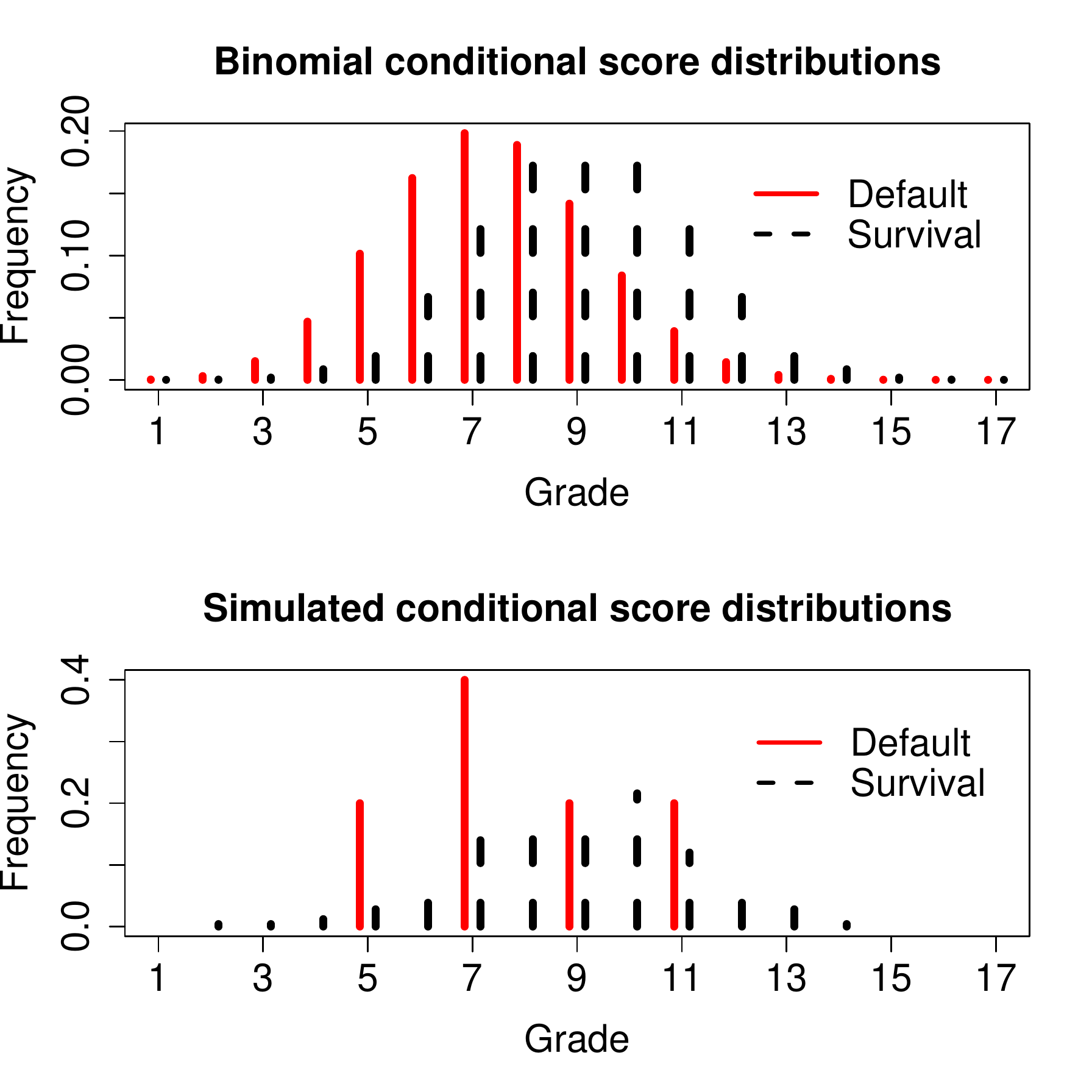}}
\else
\begin{turn}{270}
\resizebox{\height}{6in}{\includegraphics[width=6in]{BinomialScores.ps}}
\end{turn}
\fi
\end{center}

\newpage
\refstepcounter{figure}
%
  Figure \thefigure:
  \emph{Discrete and modified ROC curves. Conditional rating distributions as in upper panel of figure \ref{fig:3}.}
\label{fig:4}
\begin{center}
\ifpdf
    \resizebox{\height}{4.0in}{\includegraphics[width=12.0in]{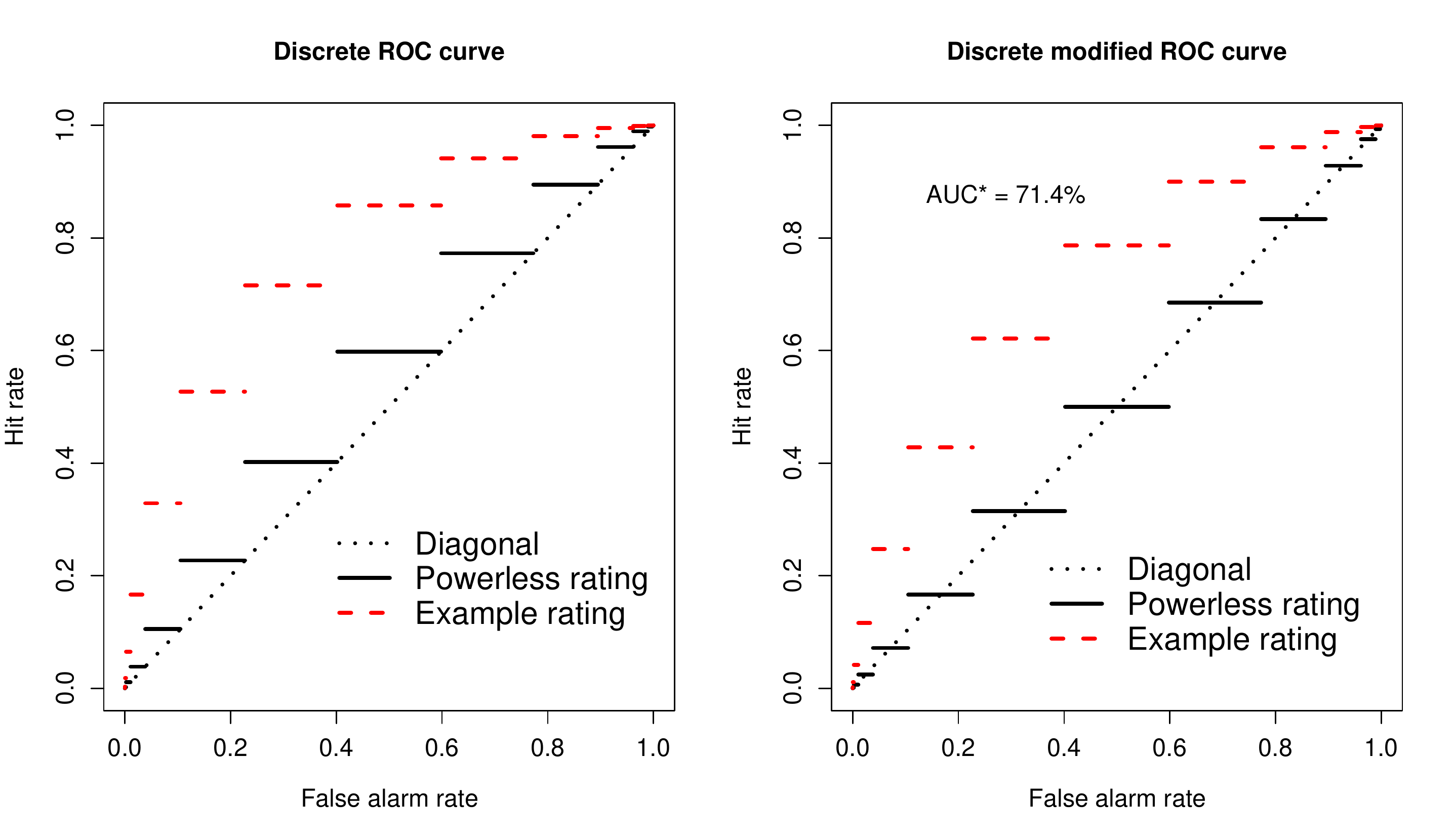}}
\else
\begin{turn}{270}
\resizebox{\height}{7.0in}{\includegraphics[width=2.5in]{DiscreteExample.ps}}
\end{turn}
\fi
\end{center}

\newpage
\refstepcounter{figure}
%
  Figure \thefigure:
  \emph{Modified and interpolated ROC curves. Conditional rating distributions as in lower panel of figure \ref{fig:3}.}
\label{fig:5}
\begin{center}
\ifpdf
    \resizebox{\height}{4.0in}{\includegraphics[width=12.0in]{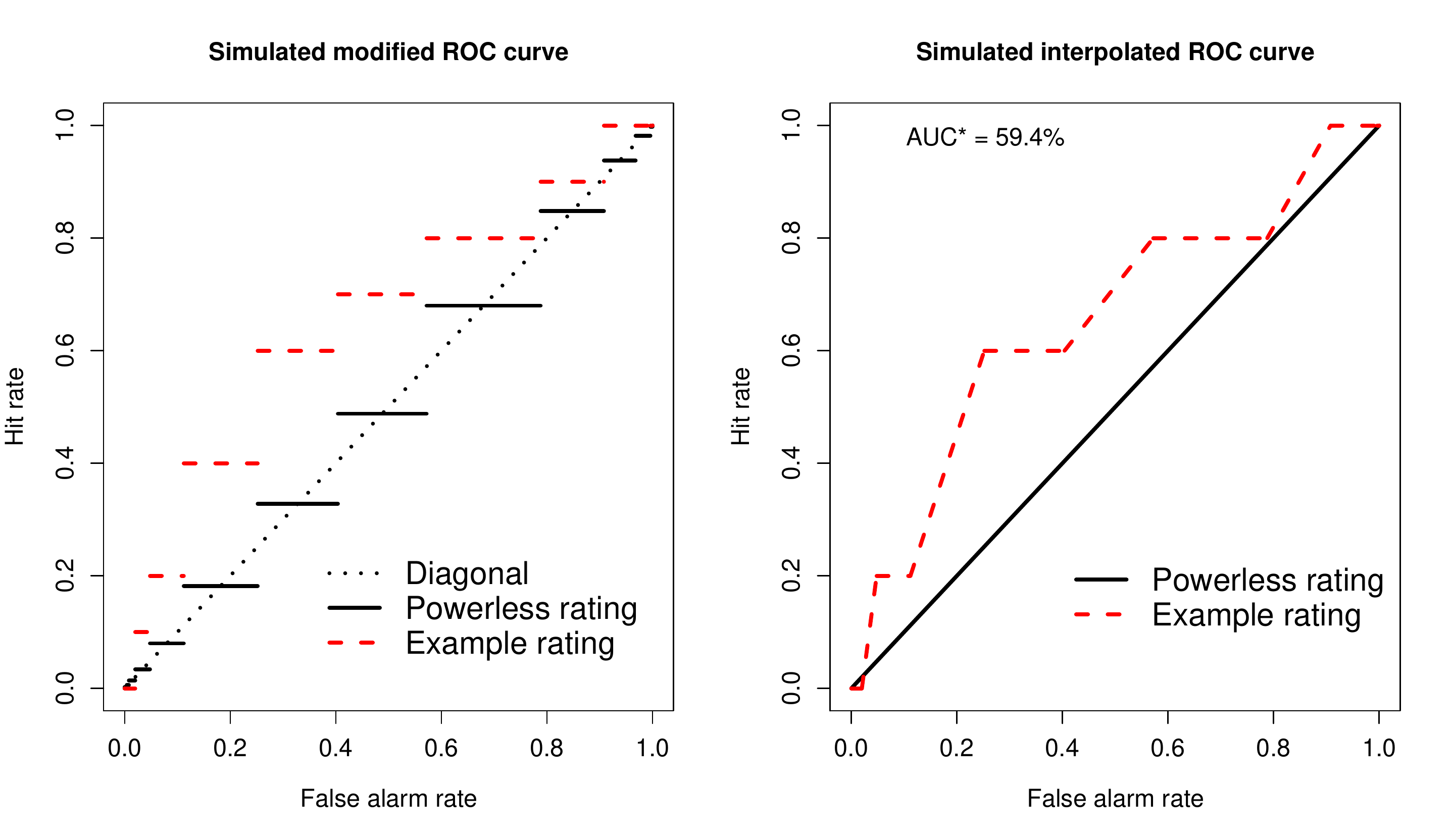}}
\else
\begin{turn}{270}
\resizebox{\height}{7.0in}{\includegraphics[width=2.5in]{SimulatedExample.ps}}
\end{turn}
\fi
\end{center}

\newpage
\refstepcounter{figure}
%
  Figure \thefigure:
  \emph{Example from section \ref{sec:cont_experiments}. Coverage of true AUC and 50\% by
95\% confidence intervals as function of sample size $n_D$ 
of defaulter scores. Differentiation according to estimation method. 
Total hits in 100 experiments. Exact results in table \ref{tab:cover_continuous}.}
\label{fig:cover_continuous}
\begin{center}
\ifpdf
    \resizebox{\height}{6.0in}{\includegraphics[width=6.0in]{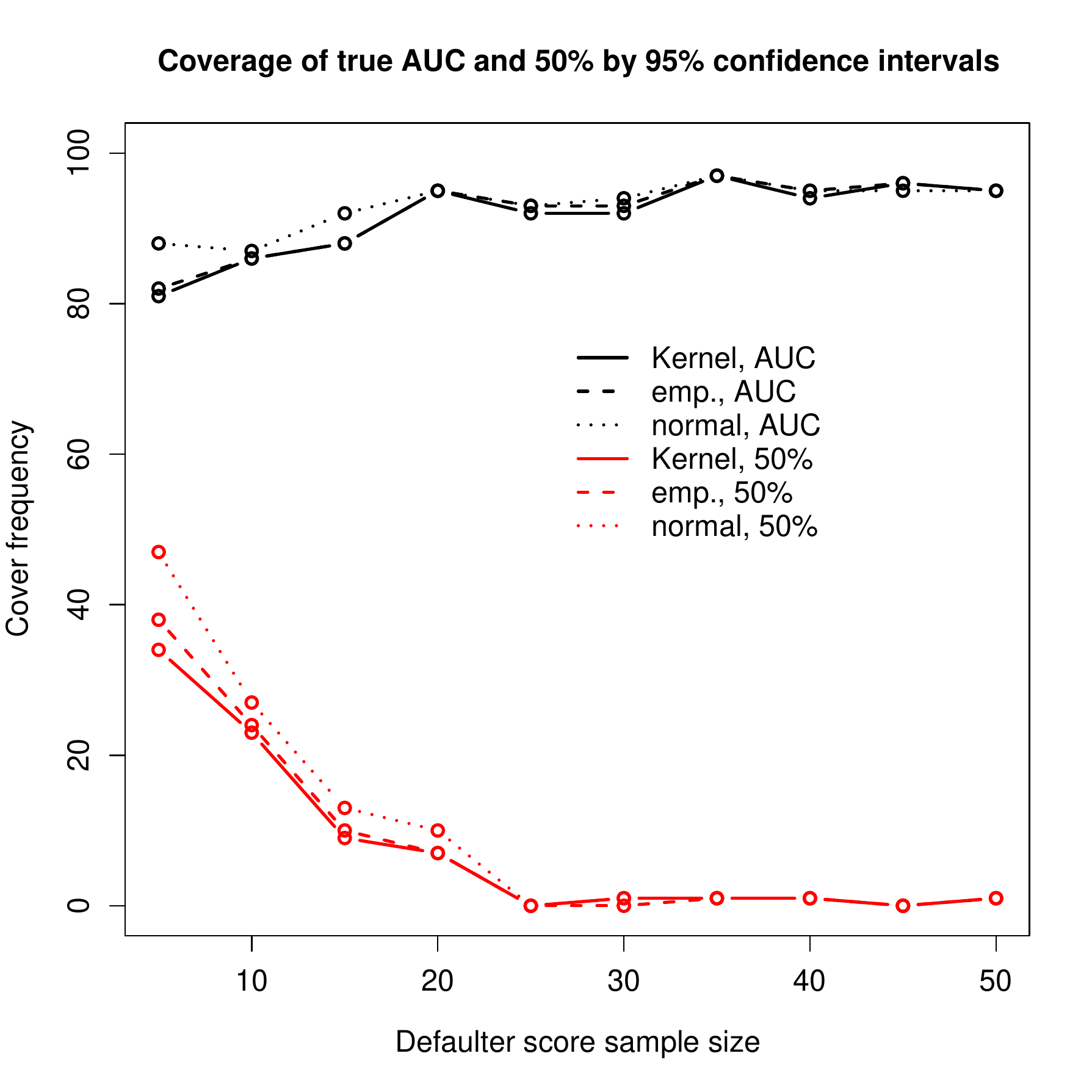}}
\else
\begin{turn}{270}
\resizebox{\height}{6.0in}{\includegraphics[width=6.0in]{Cover_continuous.ps}}
\end{turn}
\fi
\end{center}

\newpage
\refstepcounter{figure}
%
  Figure \thefigure:
  \emph{Example from section \ref{sec:finite}. Coverage of true AUC and 50\% by
95\% confidence intervals as function of sample size $n_D$ 
of defaulter scores. Differentiation according to estimation method. Total hits in 100 experiments. 
Exact results in table \ref{tab:cover_discrete}.}
\label{fig:cover_discrete}
\begin{center}
\ifpdf
    \resizebox{\height}{6.0in}{\includegraphics[width=6.0in]{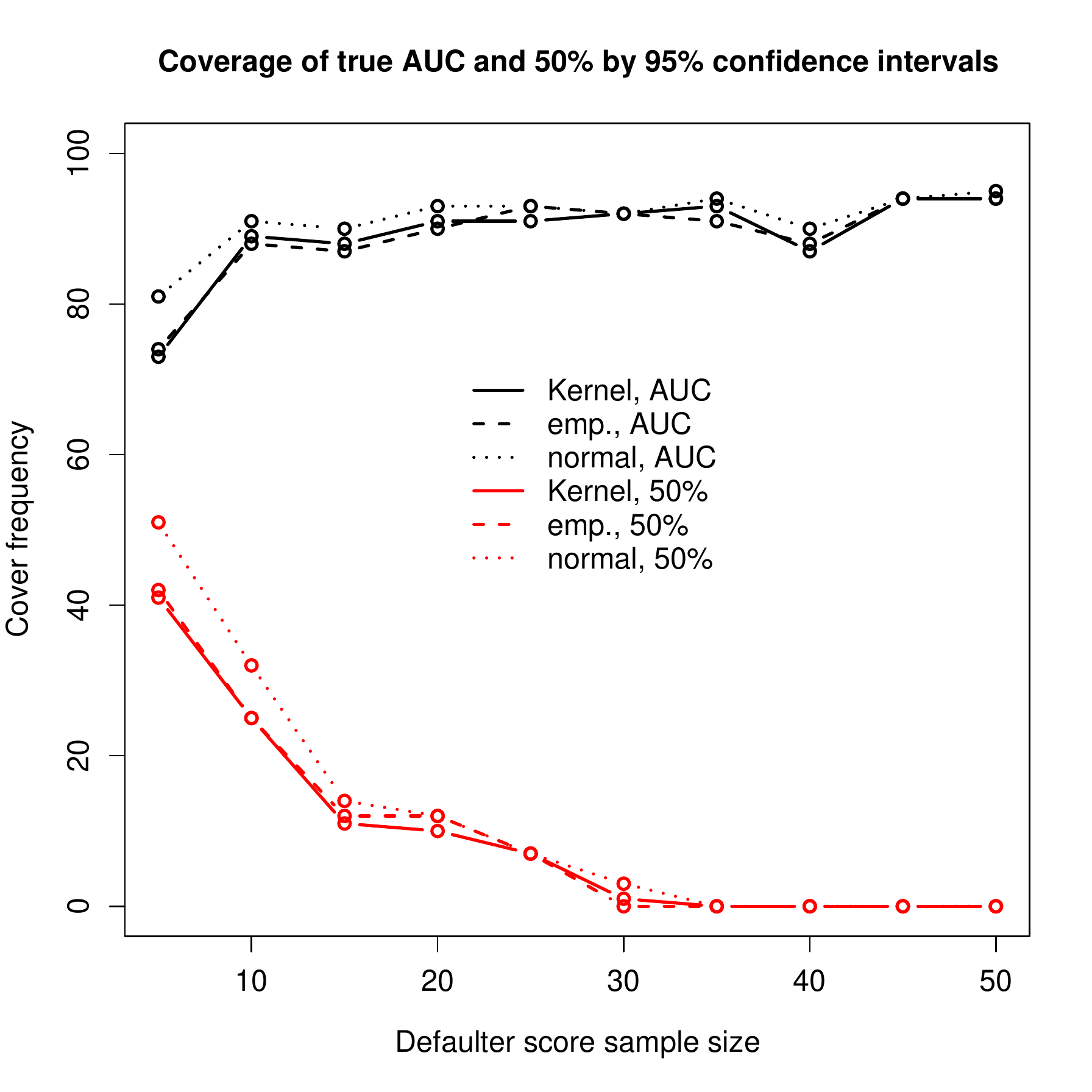}}
\else
\begin{turn}{270}
\resizebox{\height}{6.0in}{\includegraphics[width=6.0in]{Cover_Discrete.ps}}
\end{turn}
\fi
\end{center}

\newpage
\refstepcounter{figure}
%
  Figure \thefigure:
  \emph{Score densities implied by van der Burgt's parametric approach to CAP curves \eqref{eq:C.kappa} when
  the default score distribution is standard normal. The non-default score densities are calculated
  according to \eqref{eq:non_default}, with unconditional probability of default $p = 0.01$.}
\label{fig:VanDerBurgtDensities}
\begin{center}
\ifpdf
    \resizebox{\height}{6.0in}{\includegraphics[width=6.0in]{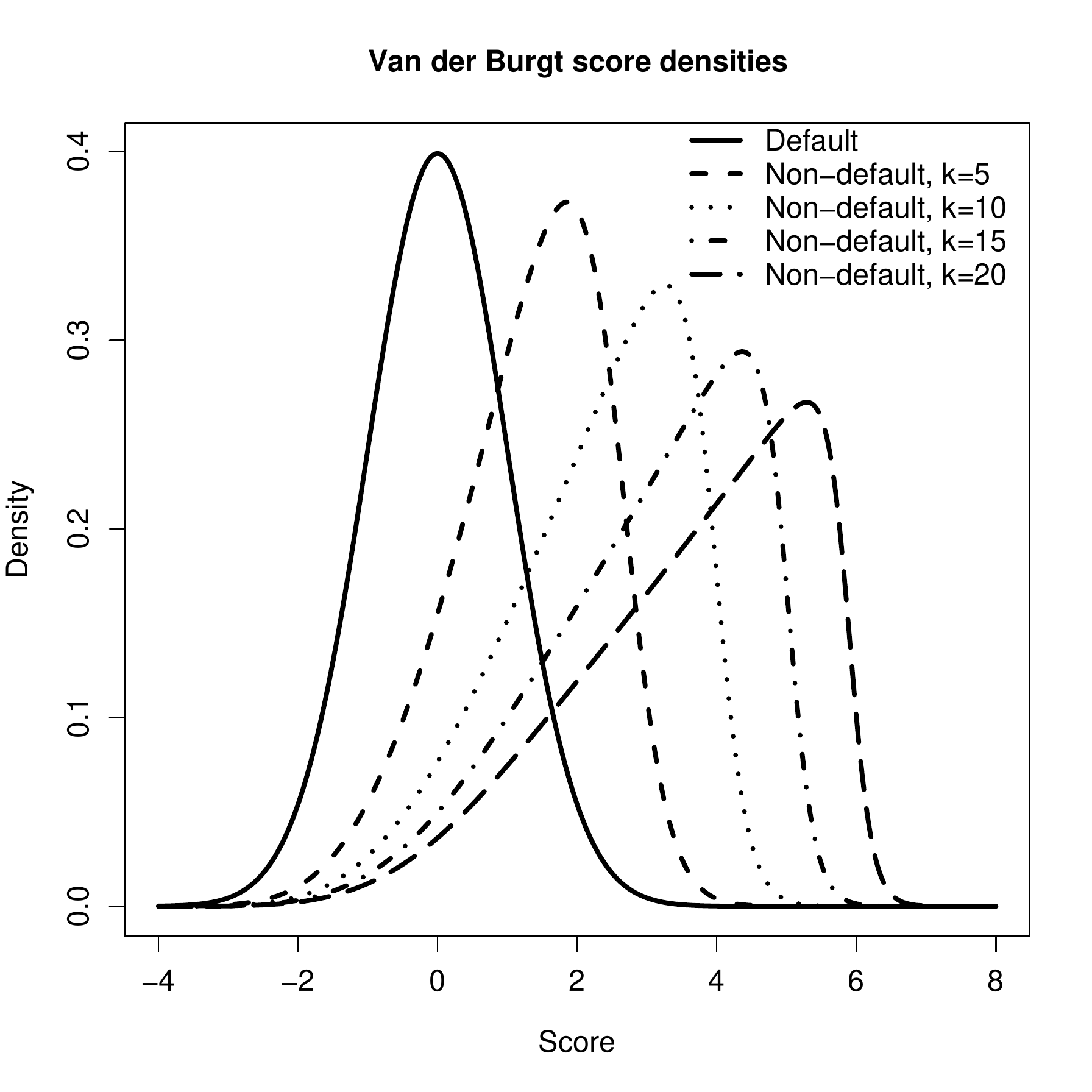}}
\else
\begin{turn}{270}
\resizebox{\height}{6.0in}{\includegraphics[width=6.0in]{VanDerBurgtDensities.ps}}
\end{turn}
\fi
\end{center}

\newpage
\refstepcounter{figure}
%
  Figure \thefigure:
  \emph{True conditional PDs and estimated conditional PDs for the case 1 scenario (rating system with 17 grades)
  from section \ref{sec:performance}. 
  Defaulter scores sample size 25 in estimation sample.}
\label{fig:condPD}
\begin{center}
\ifpdf
    \resizebox{\height}{6.0in}{\includegraphics[width=6.0in]{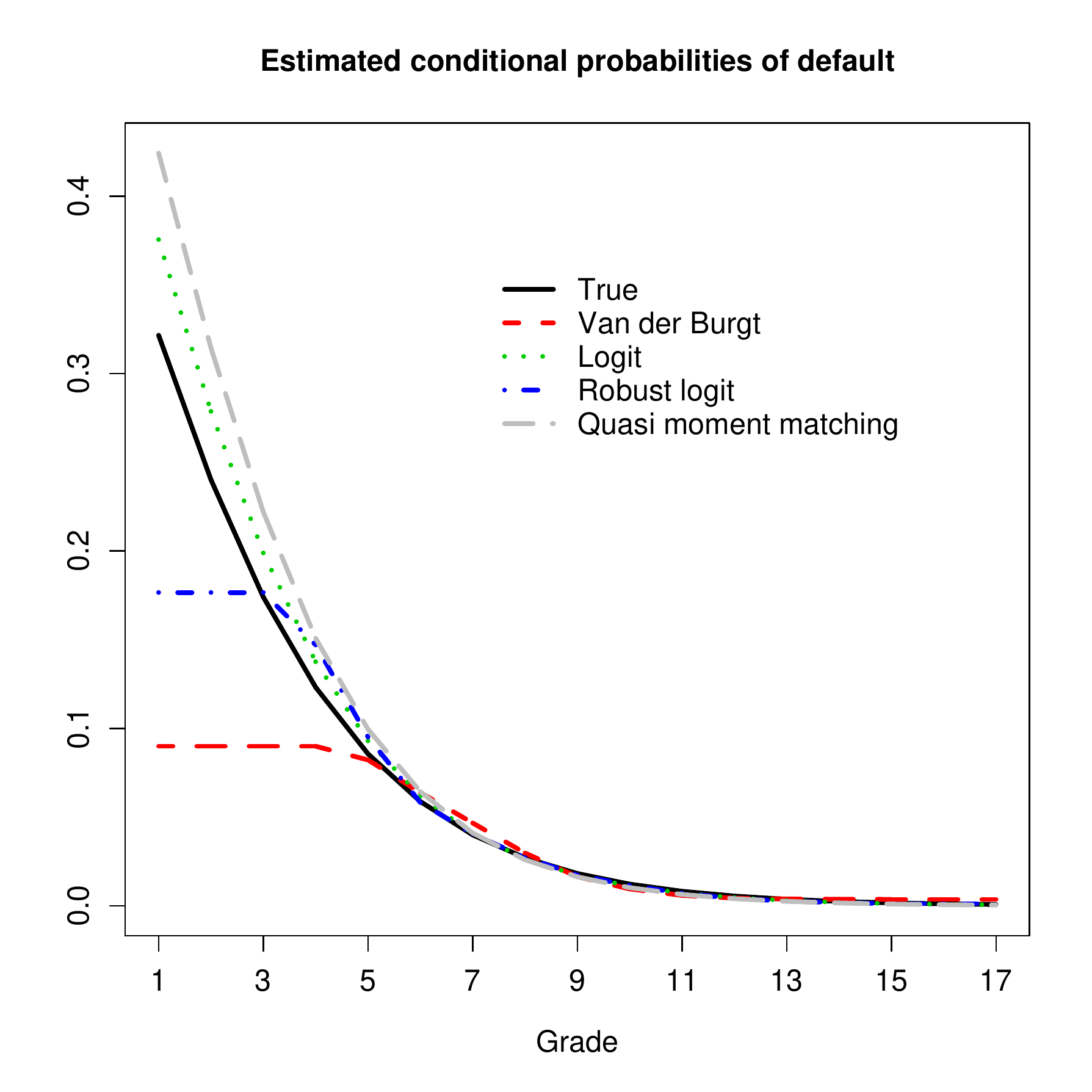}}
\else
\begin{turn}{270}
\resizebox{\height}{6.0in}{\includegraphics[width=6.0in]{Fig_condPD.ps}}
\end{turn}
\fi
\end{center}

\newpage
\refstepcounter{figure}
%
  Figure \thefigure:
  \emph{True conditional PDs and conditional PDs estimates by the quasi moment 
  matching estimator (example \ref{ex:QMM}) for the case 3 scenario (continuous
  score function with nearly equal conditional score distribution variances)
  from section \ref{sec:performance}. 
  Estimates based on calibration sample of size 300. Estimates are matched to the true
  unconditional PD and the true AUC (``best fit''), to a too small unconditional PD and
  the true AUC (``lower PD''), and to the true unconditional PD and a too high AUC
  (``higher AUC'').}
\label{fig:QMM}
\begin{center}
\ifpdf
    \resizebox{\height}{6.0in}{\includegraphics[width=6.0in]{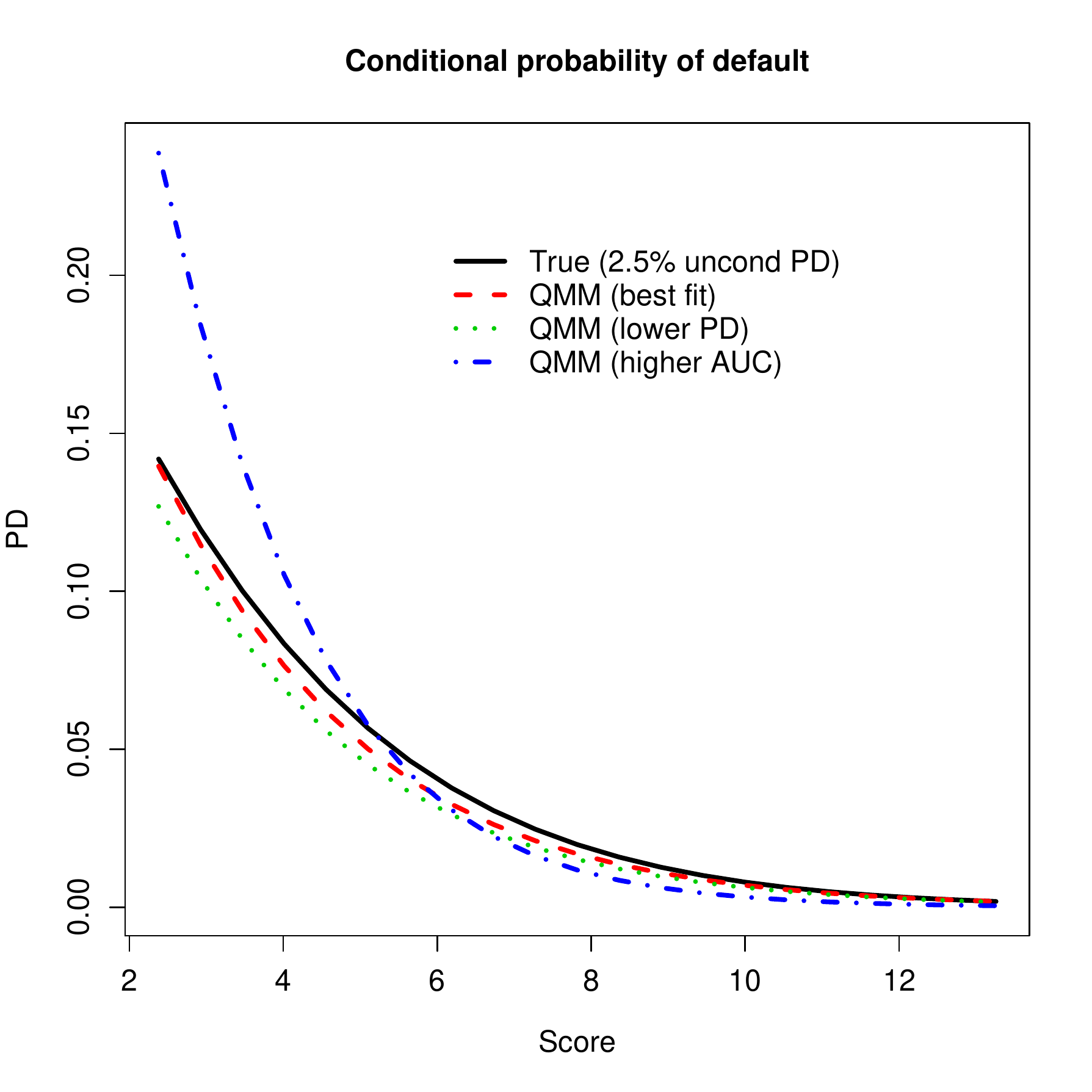}}
\else
\begin{turn}{270}
\resizebox{\height}{6.0in}{\includegraphics[width=6.0in]{QMM.ps}}
\end{turn}
\fi
\end{center}

\newpage

\refstepcounter{table}
    \begin{center}
\label{tab:1}
\parbox{15cm}{Table \thetable: \emph{Maximum number of different bootstrap samples as function of size $n$
of original sample.}}\\[2ex]
\begin{tabular}{|c||c|c|c|c|c|c|c|c|c|c|c|}
\hline
$n$ & 1 & 2 & 3 & 4 & 5 & 6 & 7 & 8 & 9 & 10 & 11\\ \hline\hline
$\binom{2\,n-1}{n}$ & 1 & 3 & 10 & 35 & 126 & 462 & 1716 & 6435 & 24310 & 92378 & 352716\\ \hline
\end{tabular}
\end{center}

\refstepcounter{table}
    \begin{center}
\label{tab:2}
\parbox{15cm}{Table \thetable: \emph{Estimated (from 100 simulation experiments) 
mean numbers $\mu_n$ and $\nu_n$ of different (after sorting) samples in 1000 bootstrap iterations
of size $n$ with $n$ different elements and $n-1$ different elements,
respectively.}}\\[2ex]
\begin{tabular}{|c||c|c|c|c|c|c|c|c|c|c|c|}
\hline
$n$ & 1 & 2 & 3 & 4 & 5 & 6 & 7 & 8 & 9 & 10 & 11\\ \hline\hline
$\mu_n$ & 1.0  & 3.0  & 10.0  & 35.0  & 117.0  & 323.0  & 620.2  & 844.6  & 945.8  & 983.2  & 995.1\\ \hline
$\nu_n$ & NA  & 1.0  & 4.0  & 15.0  & 52.0  & 160.1  & 389.6  & 679.8  & 873.0  & 957.6  & 987.0\\ \hline
\end{tabular}
\end{center}

\refstepcounter{table}
    \begin{center}
\label{tab:3}
\parbox{15cm}{Table \thetable: \emph{Results of first five bootstrap experiments as described in
section \ref{sec:cont_experiments}. Confidence intervals at 95\% level. Default sample sizes $n_D = 5$,
$n_D = 25$, and $n_D = 45$.}}\\[2ex]
\begin{tabular}{|c||c|c|c|c|c|c|c|c|}
\hline
\multicolumn{9}{|c|}{True AUC = $71.615\%$}    \\\hline
Exp.~no. & $\mathrm{AUC}_{\mathrm{kernel}}$ & \multicolumn{2}{c|}{$I_{\mathrm{kernel}}$} &
$\mathrm{AUC}_{\mathrm{emp}}$ & \multicolumn{2}{c|}{$I_{\mathrm{emp}}$} &   
\multicolumn{2}{c|}{$I_{\mathrm{normal}}$} \\\hline \hline
& \multicolumn{8}{|c|}{$n_D = 5$}   \\ \hline
1 & 56.95\% & 37.19\% & 80.26\% & 56.80\% & 36.96\% & 81.28\% & 32.88\% & 80.72\% \\ \hline
2 & 68.67\% & 47.30\% & 92.75\% & 68.56\% & 47.12\% & 94.08\% & 42.19\% & 94.93\% \\ \hline
3 & 62.14\% & 42.19\% & 84.07\% & 62.32\% & 42.08\% & 84.64\% & 37.52\% & 87.12\% \\ \hline
4 & 75.37\% & 54.99\% & 97.46\% & 73.52\% & 52.00\% & 95.92\% & 48.88\% & 98.16\% \\ \hline
5 & 62.19\% & 37.26\% & 86.07\% & 63.12\% & 38.48\% & 88.96\% & 34.53\% & 91.71\% \\ \hline\hline
& \multicolumn{8}{|c|}{$n_D = 25$}    \\ \hline
1 & 65.67\% & 55.21\% & 77.19\% & 65.89\% & 55.25\% & 77.74\% & 54.42\% & 77.36\% \\ \hline
2 & 65.55\% & 55.18\% & 76.22\% & 65.60\% & 54.93\% & 76.85\% & 54.59\% & 76.61\% \\ \hline
3 & 62.44\% & 51.86\% & 73.42\% & 62.18\% & 51.62\% & 73.52\% & 51.04\% & 73.31\% \\ \hline
4 & 70.44\% & 59.02\% & 81.74\% & 71.28\% & 60.91\% & 82.62\% & 59.80\% & 82.76\% \\ \hline
5 & 66.62\% & 55.78\% & 77.62\% & 66.62\% & 56.05\% & 78.27\% & 55.63\% & 77.62\% \\ \hline\hline
& \multicolumn{8}{|c|}{$n_D = 45$}    \\ \hline
1 & 68.63\% & 61.30\% & 77.42\% & 68.47\% & 61.12\% & 77.08\% & 60.45\% & 76.49\% \\ \hline
2 & 72.07\% & 63.34\% & 81.71\% & 71.74\% & 62.66\% & 82.15\% & 62.63\% & 80.86\% \\ \hline
3 & 75.00\% & 68.27\% & 83.28\% & 74.95\% & 68.24\% & 83.38\% & 67.35\% & 82.55\% \\ \hline
4 & 71.45\% & 63.67\% & 79.81\% & 71.15\% & 63.04\% & 79.70\% & 62.26\% & 80.03\% \\ \hline
5 & 67.31\% & 59.77\% & 75.60\% & 67.06\% & 59.34\% & 75.52\% & 59.16\% & 74.95\% \\ \hline
\end{tabular}
\end{center}

\newpage
\refstepcounter{table}
    \begin{center}
\label{tab:cover_continuous}
\parbox{15cm}{Table \thetable: \emph{Example from
section \ref{sec:cont_experiments}. Coverage of true AUC and 50\% by
95\% confidence intervals. With differentiation according to estimation method and sample size $n_D$ 
of defaulter scores. Total hits in 100 experiments.
MW means Mann-Whitney test, KS means Kolmogorov-Smirnov test.}}\\[2ex]
\begin{tabular}{|c||c|c|c||c|c|c||c|c|}
\hline
 & \multicolumn{3}{c||}{True AUC in interval} &  \multicolumn{3}{c||}{50\% in interval} & 
 \multicolumn{2}{c|}{Type II error rate} \\\hline
Method: & Kernel & emp. & normal & Kernel & emp. & normal & MW & KS \\ \hline\hline
$n_D=5$ & 81 & 82 & 88 & 34 & 38 & 47 & 57 & 68\\ \hline
$n_D=10$ & 86 & 86 & 87 & 23 & 24 & 27 & 29 & 39\\ \hline
$n_D=15$ & 88 & 88 & 92 & 9 & 10 & 13 & 13 & 19\\ \hline
$n_D=20$ & 95 & 95 & 95 & 7 & 7 & 10 & 10 & 14\\ \hline
$n_D=25$ & 92 & 93 & 93 & 0 & 0 & 0 & 0 & 10\\ \hline
$n_D=30$ & 92 & 93 & 94 & 1 & 0 & 1 & 0 & 5 \\ \hline
$n_D=35$ & 97 & 97 & 97 & 1 & 1 & 1 & 1 & 2 \\ \hline
$n_D=40$ & 94 & 95 & 95 & 1 & 1 & 1 & 1 & 0 \\ \hline
$n_D=45$ & 96 & 96 & 95 & 0 & 0 & 0 & 0 & 0 \\ \hline
$n_D=50$ & 95 & 95 & 95 & 1 & 1 & 1 & 1 & 1\\ \hline
\end{tabular}
\end{center}

\refstepcounter{table}
    \begin{center}
\label{tab:4}
\parbox{15cm}{Table \thetable: \emph{Results of first five bootstrap experiments as described in
section \ref{sec:finite}. Confidence intervals at 95\% level. Default sample sizes $n_D = 5$,
$n_D = 25$, and $n_D = 45$.}}\\[2ex]
\begin{tabular}{|c||c|c|c|c|c|c|c|c|}
\hline
\multicolumn{9}{|c|}{True AUC = $71.413\%$}    \\\hline
Exp.~no. & $\mathrm{AUC}_{\mathrm{kernel}}$ & \multicolumn{2}{c|}{$I_{\mathrm{kernel}}$} &
$\mathrm{AUC}_{\mathrm{emp}}$ & \multicolumn{2}{c|}{$I_{\mathrm{emp}}$} &   
\multicolumn{2}{c|}{$I_{\mathrm{normal}}$} \\\hline \hline
& \multicolumn{8}{|c|}{$n_D = 5$}   \\ \hline
1 & 69.04\% & 48.88\% & 94.86\% & 69.12\% & 49.64\% & 95.96\% & 43.18\% & 95.06\% \\ \hline
2 & 63.97\% & 36.87\% & 92.31\% & 62.80\% & 35.04\% & 92.76\% & 30.07\% & 95.53\% \\ \hline
3 & 68.52\% & 45.21\% & 96.57\% & 65.08\% & 40.00\% & 89.64\% & 36.47\% & 93.69\% \\ \hline
4 & 69.53\% & 50.58\% & 92.35\% & 68.28\% & 49.16\% & 89.92\% & 45.61\% & 90.95\% \\ \hline
5 & 95.41\% & 91.81\% & 100.00\% & 95.20\% & 91.40\% & 100.00\% & 90.00\% & 100.00\% \\ \hline\hline
& \multicolumn{8}{|c|}{$n_D = 25$}    \\ \hline
1 & 68.10\% & 57.62\% & 79.41\% & 68.90\% & 58.78\% & 79.82\% & 57.91\% & 79.90\% \\ \hline
2 & 69.10\% & 59.85\% & 79.31\% & 68.71\% & 59.56\% & 79.29\% & 58.36\% & 79.06\% \\ \hline
3 & 69.98\% & 60.66\% & 79.51\% & 69.62\% & 60.10\% & 79.43\% & 59.74\% & 79.49\% \\ \hline
4 & 66.33\% & 55.05\% & 77.66\% & 66.31\% & 55.30\% & 77.65\% & 54.94\% & 77.69\% \\ \hline
5 & 80.26\% & 71.77\% & 89.89\% & 79.73\% & 71.11\% & 89.89\% & 70.27\% & 89.19\% \\ \hline\hline
& \multicolumn{8}{|c|}{$n_D = 45$}    \\ \hline
1 & 72.00\% & 64.56\% & 81.00\% & 71.64\% & 64.25\% & 80.51\% & 63.75\% & 79.53\% \\ \hline
2 & 73.34\% & 66.93\% & 80.65\% & 73.03\% & 66.70\% & 80.32\% & 66.20\% & 79.85\% \\ \hline
3 & 73.22\% & 65.62\% & 81.36\% & 73.14\% & 65.70\% & 81.26\% & 65.36\% & 80.93\% \\ \hline
4 & 74.13\% & 66.46\% & 82.05\% & 73.92\% & 66.33\% & 81.98\% & 66.17\% & 81.68\% \\ \hline
5 & 76.63\% & 70.11\% & 82.89\% & 76.32\% & 69.84\% & 82.68\% & 69.74\% & 82.89\% \\ \hline
\end{tabular}
\end{center}

\newpage
\refstepcounter{table}
    \begin{center}
\label{tab:cover_discrete}
\parbox{15cm}{Table \thetable: \emph{Example from section \ref{sec:finite}. Coverage of true AUC and 50\% by
95\% confidence intervals. With differentiation according to estimation method and sample size $n_D$ 
of defaulter scores. Total hits in 100 experiments.
MW means Mann-Whitney test, Fisher means Fisher's exact test.}}\\[2ex]
\begin{tabular}{|c||c|c|c||c|c|c||c|c|c|}
\hline
 & \multicolumn{3}{c||}{True AUC in interval} &  \multicolumn{3}{c||}{50\% in interval} & 
 \multicolumn{2}{c|}{Type II error rate} \\\hline
Method: & Kernel & emp. & normal & Kernel & emp. & normal & MW & Fisher \\ \hline\hline
$n_D=5$ & 73 & 74 & 81 & 41 & 42 & 51 & 63 & 79\\ \hline
$n_D=10$ & 89 & 88 & 91 & 25 & 25 & 32 & 32 & 61\\ \hline
$n_D=15$ & 88 & 87 & 90 & 11 & 12 & 14 & 15 & 49\\ \hline
$n_D=20$ & 91 & 90 & 93 & 10 & 12 & 12 & 10 & 37\\ \hline
$n_D=25$ & 91 & 93 & 93 & 7 & 7 & 7 & 6 & 30\\ \hline
$n_D=30$ & 92 & 92 & 92 & 1 & 0 & 3 & 1 & 19\\ \hline
$n_D=35$ & 93 & 91 & 94 & 0 & 0 & 0 & 0 & 9\\ \hline
$n_D=40$ & 87 & 88 & 90 & 0 & 0 & 0 & 1 & 8\\ \hline
$n_D=45$ & 94 & 94 & 94 & 0 & 0 & 0 & 0 & 7\\ \hline
$n_D=50$ & 94 & 94 & 95 & 0 & 0 & 0 & 0 & 5\\ \hline
\end{tabular}
\end{center}

\newpage
\refstepcounter{table}
    \begin{center}
\label{tab:se}
\parbox{15cm}{Table \thetable: \emph{Standard errors according to \eqref{eq:se} for different
approaches to estimation of conditional probabilities of default.
Measured in simulation experiment as described in section \ref{sec:performance}.
Defaulter scores sample size 25 in estimation sample.}}\\[2ex]
\begin{tabular}{|c||c|c|c|c|}
\hline
Quantile level & \multicolumn{4}{c|}{Standard errors} \\\hline\hline
& Quasi moment matching & Logit & Robust logit & Van der Burgt \\\hline\hline
 & \multicolumn{4}{c|}{Case 1 (17 rating grades)} \\\hline
5\% & 0.097\% & 0.088\% & 0.185\% & 0.483\% \\ \hline
25\% & 0.268\% & 0.269\% & 0.339\% & 0.656\% \\ \hline
50\% & 0.528\% & 0.496\% & 0.553\% & 0.82\% \\ \hline
75\% & 0.9\% & 0.861\% & 0.91\% & 1.062\% \\ \hline
95\% & 1.62\% & 1.584\% & 1.7\% & 1.479\% \\ \hline
 & \multicolumn{4}{c|}{Case 2 (7 rating grades)} \\\hline\hline
5\% & 0.115\% & 0.116\% & 0.213\% & 0.58\% \\ \hline
25\% & 0.32\% & 0.317\% & 0.392\% & 0.717\% \\ \hline
50\% & 0.577\% & 0.543\% & 0.657\% & 0.86\% \\ \hline
75\% & 0.942\% & 0.915\% & 1.089\% & 1.12\% \\ \hline
95\% & 1.628\% & 1.67\% & 2.022\% & 1.593\% \\ \hline\hline 
 & \multicolumn{4}{c|}{Case 3 (continuous $\sim$ 17 grades)} \\\hline\hline
 5\% & 0.121\% & 0.123\% & 0.169\% & 0.45\% \\ \hline
25\% & 0.285\% & 0.27\% & 0.372\% & 0.621\% \\ \hline
50\% & 0.523\% & 0.517\% & 0.644\% & 0.791\% \\ \hline
75\% & 0.933\% & 0.908\% & 1.111\% & 1.027\% \\ \hline
95\% & 1.75\% & 1.761\% & 2.518\% & 1.512\% \\ \hline\hline
 & \multicolumn{4}{c|}{Case 4 (continuous $\sim$ 7 grades)} \\\hline\hline
5\% & 0.288\% & 0.28\% & 0.261\% & 0.425\% \\ \hline
25\% & 0.471\% & 0.454\% & 0.533\% & 0.599\% \\ \hline
50\% & 0.735\% & 0.715\% & 0.886\% & 0.794\% \\ \hline
75\% & 1.224\% & 1.185\% & 1.662\% & 1.046\% \\ \hline
95\% & 2.235\% & 2.315\% & 3.387\% & 1.559\% \\ \hline\hline
 & \multicolumn{4}{c|}{Case 5 (continuous, different variances)} \\\hline\hline
5\% & 0.646\% & 0.634\% & 0.493\% & 1.133\% \\ \hline
25\% & 0.902\% & 0.875\% & 0.782\% & 1.66\% \\ \hline
50\% & 1.348\% & 1.24\% & 1.164\% & 2.282\% \\ \hline
75\% & 2.18\% & 1.913\% & 1.74\% & 3.238\% \\ \hline
95\% & 3.7\% & 3.398\% & 3.028\% & 4.994\% \\ \hline 
\end{tabular}
\end{center}

\newpage
\refstepcounter{table}
    \begin{center}
\label{tab:se50}
\parbox{15cm}{Table \thetable: \emph{Standard errors according to \eqref{eq:se} for different
approaches to estimation of conditional probabilities of default.
Measured in simulation experiment as described in section \ref{sec:performance}.
Defaulter scores sample size 50 in estimation sample.}}\\[2ex]
\begin{tabular}{|c||c|c|c|c|}
\hline
Quantile level & \multicolumn{4}{c|}{Standard errors} \\\hline\hline
& Quasi moment matching & Logit & Robust logit & Van der Burgt \\\hline\hline
 & \multicolumn{4}{c|}{Case 1 (17 rating grades)} \\\hline
5\% & 0.083\% & 0.085\% & 0.17\% & 0.414\% \\ \hline
25\% & 0.216\% & 0.237\% & 0.31\% & 0.561\% \\ \hline
50\% & 0.418\% & 0.4\% & 0.483\% & 0.696\% \\ \hline
75\% & 0.684\% & 0.697\% & 0.777\% & 0.885\% \\ \hline
95\% & 1.168\% & 1.183\% & 1.302\% & 1.243\% \\ \hline\hline
 & \multicolumn{4}{c|}{Case 2 (7 rating grades)} \\\hline
5\% & 0.101\% & 0.097\% & 0.182\% & 0.426\% \\ \hline
25\% & 0.265\% & 0.267\% & 0.36\% & 0.545\% \\ \hline
50\% & 0.477\% & 0.475\% & 0.594\% & 0.68\% \\ \hline
75\% & 0.802\% & 0.781\% & 0.943\% & 0.879\% \\ \hline
95\% & 1.289\% & 1.288\% & 1.754\% & 1.295\% \\ \hline\hline 
 & \multicolumn{4}{c|}{Case 3 (continuous $\sim$ 17 grades)} \\\hline\hline
5\% & 0.114\% & 0.117\% & 0.163\% & 0.356\% \\ \hline
25\% & 0.244\% & 0.245\% & 0.336\% & 0.524\% \\ \hline
50\% & 0.437\% & 0.44\% & 0.557\% & 0.676\% \\ \hline
75\% & 0.741\% & 0.722\% & 0.928\% & 0.898\% \\ \hline
95\% & 1.299\% & 1.275\% & 2.222\% & 1.264\% \\ \hline\hline
 & \multicolumn{4}{c|}{Case 4 (continuous $\sim$ 7 grades)} \\\hline\hline
5\% & 0.281\% & 0.281\% & 0.259\% & 0.307\% \\ \hline
25\% & 0.456\% & 0.426\% & 0.5\% & 0.466\% \\ \hline
50\% & 0.695\% & 0.681\% & 0.852\% & 0.643\% \\ \hline
75\% & 1.095\% & 1.05\% & 1.458\% & 0.855\% \\ \hline
95\% & 1.829\% & 1.92\% & 3.214\% & 1.24\% \\ \hline\hline
 & \multicolumn{4}{c|}{Case 5 (continuous, different variances)} \\\hline\hline
5\% & 0.635\% & 0.634\% & 0.47\% & 0.968\% \\ \hline
25\% & 0.872\% & 0.867\% & 0.745\% & 1.485\% \\ \hline
50\% & 1.273\% & 1.257\% & 1.082\% & 2.12\% \\ \hline
75\% & 2.008\% & 1.923\% & 1.594\% & 3.055\% \\ \hline
95\% & 3.351\% & 3.07\% & 2.688\% & 4.638\% \\ \hline
\end{tabular}
\end{center}

\newpage
\refstepcounter{table}
    \begin{center}
\label{tab:prob.least}
\parbox{15cm}{Table \thetable: \emph{Probabilities to produce least standard error for different
approaches to estimation of conditional probabilities of default.
Measured in simulation experiment as described in section \ref{sec:performance}.
QMM means ``Quasi moment matching''. $\sigma_D$ and $\sigma_N$
are the standard deviations of the defaulter score distribution
and survivor score distribution respectively.}}\\[2ex]
\begin{tabular}{|c||c|c|c|c||c|}
\hline
Case & \multicolumn{4}{c||}{Probability to produce least standard error} & $\sigma_D / \sigma_N$\\\hline\hline
& QMM & Logit & Robust logit & Van der Burgt & \\\hline\hline
 & \multicolumn{4}{c||}{Default sample size 25} & \\\hline
1 (17 grades) & 40.5\% & 33.1\% & 16.6\% & 9.8\% & 98\%\\ \hline
2 (7 grades) & 39\% & 27.8\% & 21.7\% & 11.5\% & 91.7\%\\ \hline
3 (continuous $\sim$ 17 grades) & 35.2\% & 27\% & 16.9\% & 20.9\% & 98\%\\ \hline
4 (continuous $\sim$ 7 grades) & 29.5\% & 16.3\% & 16.7\% & 37.5\% & 91.7\%\\ \hline
5 (different variances) & 25\% & 21.1\% & 52.5\% & 1.4\% & 125\%\\ \hline\hline
 & \multicolumn{4}{c||}{Default sample size 50} & \\\hline
1 (17 grades) & 43.7\% & 31\% & 14.8\% & 10.5\% & 98\%\\ \hline
2 (7 grades) & 35.8\% & 28.5\% & 20.9\% & 14.8\% & 91.7\%\\ \hline
3 (continuous $\sim$ 17 grades) & 35\% & 28\% & 14.3\% & 22.7\% & 98\%\\ \hline
4 (continuous $\sim$ 7 grades) & 26\% & 14.4\% & 12.9\% & 46.7\% & 91.7\%\\ \hline
5 (different variances) & 25\% & 17.8\% & 56\% & 1.2\% & 125\%\\ \hline
\end{tabular}
\end{center}

\refstepcounter{table}
    \begin{center}
\label{tab:corr}
\parbox{15cm}{Table \thetable: \emph{Spearman correlations of absolute error of
AUC estimate and standard error of conditional PD curve estimate
for different
approaches to estimation of conditional probabilities of default.
Measured in simulation experiment as described in section \ref{sec:performance}.
QMM means ``Quasi moment matching''. $\sigma_D$ and $\sigma_N$
are the standard deviations of the defaulter score distribution
and survivor score distribution respectively.}}\\[2ex]
\begin{tabular}{|c||c|c|c|c||c|}
\hline
Case & \multicolumn{4}{c||}{Spearman correlation} & $\sigma_D / \sigma_N$\\\hline\hline
& QMM & Logit & Robust logit & Van der Burgt & \\\hline\hline
 & \multicolumn{4}{c||}{Default sample size 50} & \\\hline
1 (17 grades) & 87.2\% & 86\% & 78.1\% & 55.6\% & 98\% \\ \hline
2 (7 grades) & 80.7\% & 79.5\% & 65.6\% & 54.6\% & 91.7\%\\ \hline
3 (continuous $\sim$ 17 grades) & 85.4\% & 85.5\% & 70.8\% & 59.3\% & 98\% \\ \hline
4 (continuous $\sim$ 7 grades) & 51\% & 46.9\% & 38.6\% & 62.9\% & 91.7\%\\ \hline
5 (different variances) & 18.7\% & 12.4\% & 25.5\% & 9.2\% & 125\%\\ \hline\hline
 & \multicolumn{4}{c||}{Default sample size 50} & \\\hline
1 (17 grades) & 81.9\% & 81.8\% & 71.8\% & 46.8\% & 98\%\\ \hline
2 (7 grades) & 74.4\% & 75.5\% & 60.4\% & 49.6\% & 91.7\%\\ \hline
3 (continuous $\sim$ 17 grades) & 79.5\% & 78.8\% & 59.7\% & 49\% & 98\%\\ \hline
4 (continuous $\sim$ 7 grades) & 36.4\% & 30.1\% & 28.2\% & 54.8\% & 91.7\%\\ \hline
5 (different variances) & 6.9\% & 4\% & 11.8\% & 7.2\% & 125\%\\ \hline
\end{tabular}
\end{center}

\end{document}